\documentclass[a4paper,fleqn]{cas-dc}

\usepackage{subfiles}
\usepackage[numbers,sort&compress]{natbib}
\usepackage[labelfont=bf]{caption}
\usepackage[labelfont=bf]{subcaption}
\usepackage{float}
\usepackage{threeparttable}
\usepackage[nameinlink]{cleveref}
\Crefformat{appendix}{#2App.~#1#3}
\Crefformat{figure}{#2Fig.~#1#3}
\Crefrangeformat{figure}{Figs.~#3#1#4--#5\crefstripprefix{#1}{#2}#6}
\crefrangelabelformat{figure}{#3#1#4--#5\crefstripprefix{#1}{#2}#6}
\Crefmultiformat{figure}%
  {\edef\mycrefprefixinfo{#1}Figs.~#2#1#3}%
  { and~#2\crefstripprefix{\mycrefprefixinfo}{#1}#3}%
  {, #2\crefstripprefix{\mycrefprefixinfo}{#1}#3}%
  {, and~#2\crefstripprefix{\mycrefprefixinfo}{#1}#3}

\makeatletter
\newcommand\appendix@section[1]{%
  \refstepcounter{section}%
  \orig@section*{Appendix \@Alph\c@section: #1}%
  \addcontentsline{toc}{section}{Appendix \@Alph\c@section: #1}%
}
\let\orig@section\section
\g@addto@macro\appendix{\let\section\appendix@section}
\makeatother

\def\tsc#1{\csdef{#1}{\textsc{\lowercase{#1}}\xspace}}
\tsc{WGM}
\tsc{QE}
\tsc{EP}
\tsc{PMS}
\tsc{BEC}
\tsc{DE}

\graphicspath{{Template/}{Template/figs/}{Figures/}{Figs/}}

\usepackage{color}

\begin{document}
\let\WriteBookmarks\relax
\def\floatpagepagefraction{1}
\def\textpagefraction{.001}
\shorttitle{Multilayer network analysis of \textit{C. elegans}}
\shortauthors{T Maertens et~al.}

\title [mode = title]{Multilayer network analysis of \textit{C. elegans}: \\ Looking into the locomotory circuitry}                      
%
%

\author[1]{Thomas Maertens}
\ead{thomas.maertens@gmx.net}
%
\credit{conceptualized this study, performed simulations, analyzed the results, contributed to writing the manuscript}
\address[1]{Institut f\"ur Theoretische Physik, Technische Universit\"at Berlin, Hardenbergstra{\ss}e 36, 10623 Berlin, Germany}

\author[1,2]{Eckehard Sch\"oll}
\ead{schoell@physik.tu-berlin.de}
\ead[URL]{http://www.itp.tu-berlin.de/schoell}
\credit{contributed to the research question, discussion of the results and writing the manuscript}

\author[1,2]{Jorge Ruiz}
\ead{jorge.ruiz@bccn-berlin.de}
%
\credit{data curation, contributed to the research question, discussion of the results and writing the manuscript}
\address[2]{Bernstein Center for Computational Neuroscience Berlin, Humboldt-Universit{\"a}t zu Berlin, Berlin, Germany}

\author[3,1,2]{Philipp H\"ovel}[%
    orcid=0000-0002-1370-4272
]
\cormark[1]
\ead{philipp.hoevel@ucc.ie}
\ead[URL]{http://publish.ucc.ie/researchprofiles/D019/philipphoevel}
\credit{contributed to the research question, discussion of the results and writing the manuscript}
\address[3]{School of Mathematical Sciences, University College Cork, Western Road, Cork T12 XF64, Ireland}

\cortext[cor1]{Corresponding author}




\newcommand{\elegans}[0]{\textit{C. elegans}}

\newcommand{\keyword}[1]{\textbf{#1}}


\newcommand{\vecvar}[1]{\ensuremath{\boldsymbol{#1}}}

\newcommand{\vecvel}[1]{\ensuremath{\boldsymbol{\dot{#1}}}}

\newcommand{\vecacc}[1]{\ensuremath{\boldsymbol{\ddot{#1}}}}


\newcommand{\specialcell}[2][c]{%
  \begin{tabular}[#1]{@{}l@{}}#2\end{tabular}} 

\newcommand{\nextline}{\setlength{\emergencystretch}{1em}}


\begin{abstract}
We investigate how locomotory behavior is generated in the brain focusing on the paradigmatic connectome of nematode \textit{Caenorhabditis elegans} (\elegans) and on neuronal activity patterns that control forward locomotion.
We map the neuronal network of the worm as a multilayer network that takes into account various neurotransmitters and neuropeptides.
Using logistic regression analysis, we predict the neurons of the locomotory subnetwork. Combining Hindmarsh-Rose equations for neuronal activity with a leaky integrator model for muscular activity, we study the dynamics within this subnetwork and predict the forward locomotion of the worm using a harmonic wave model.
The application of time-delayed feedback control reveals synchronization effects that contribute to a coordinated locomotion of \elegans.
Analyzing the synchronicity when the activity of certain neurons is silenced informs us about their significance 
for a coordinated locomotory behavior. Since the information processing is the same in humans and \elegans, the study of the locomotory circuitry provides new insights for understanding how the brain generates motion behavior.

\end{abstract}


\begin{highlights}
\item New perspective: \elegans\ connectome as multilayer network
\item Prediction of the locomotory subnetwork using logistic regression analysis
\item Investigation of patterns of neuronal activity that drive forward locomotion 
\item Proof of a network-based central pattern generator
\item Modeling forward locomotion based on harmonic waves
\item Enhancing synchronicity of harmonic waves by time-delayed feedback control
\item Determining significance of neurons for a coordinated locomotion behavior
\end{highlights}


\maketitle


\section{Introduction}
\label{sec:intro}

The human brain is a complex neuronal network with hundreds of billions of neurons arranged in a multitude of interconnected circuits.
A major goal of neuroscience research is to understand how mental processes and behavior are controlled by the brain \cite[Chapter 1]{Haken.1996}. 
In order to understand how behavior is generated, it is essential to understand the structure and the function of the individual circuits, along with the underlying patterns of neuronal activity \cite{Cazemier.2016,Chen.2007}.
Therefore, it is necessary to identify the neurons and their connections within each circuit.
In view of the complexity of the brain, simpler models need to be considered.

The nematode \elegans\ is an important animal model for almost all areas of experimental biology. The tiny creature provides a complete description of a development lineage, a nervous system, and a genome \cite{Gjorgjieva.2014}. Although the genome of the worm is very simple, it has remarkable similarities to the human genome \cite{Lai.2000}. The neurons of \elegans\ and humans are almost identical \cite{Hobert.2013}. The nerve ring located at the head region of the worm is an equivalent counterpart to the human central nervous system 
\cite{Chiu.2011}. 
Compared to the complexity of the brain, the nervous system of the adult \elegans\ hermaphrodite can be treated more easily because it consist of a fixed number of only $302$.
It serves as a prototype for investigations of neuronal networks since all synaptic connections between the neurons have been completely mapped by electron microscopy with reasonable accuracy at the cellular level \cite{White.1986,Albertson.1976,Chen.2007,Varshney.2011,Bentley.2016,Cook.2019}.

The locomotion is the most important behavior of \elegans\ and is made possible through $95$ body wall muscles \cite{Altun.2009}.
For example, the worm must be able to move in order to search for food, conspecifics, or improved conditions \cite{Gjorgjieva.2014} but also to react to certain environmental influences, such as a gentle body touch. 
The \elegans\ nervous system includes three main types of neurons: sensory neurons, interneurons, and motor neurons. 
In case of a body touch, information processing starts with sensory neurons that transmit information to interneurons.
The latter in turn stimulate motor neurons that stimulate muscle cells so that the worm reacts with locomotion \cite[Section II]{Driscoll.1997}.
In general, information processing of external influences in humans does not function differently even if the human nervous system is much more elaborate.
For example, humans can also perceive touches via receptors in their skin \cite{Owens.2014} and react to them with movements of corresponding body parts.
Therefore, studying the locomotory subnetwork of \elegans\ could provide new insights for understanding how the brain generates motion behavior.

In this study, we map the somatic nervous system of \elegans\ as a multilayer network including neurons and body wall muscles.
For a better understanding of properties and processes of the network, the underlying neurotransmitters and neuropeptides in chemical synapses are taken into account. 
These represent different types of interactions between the nodes and define the multilayer network \cite{Bentley.2016}.
We perform a logistic regression analysis of this network in order to identify the locomotory subnetwork that comprises all neurons involved in locomotion behavior. 
Moreover, we focus the patterns of neuronal and muscular activity 
for a circuit of the locomotory subnetwork that 
initiates forward and backward locomotion in response to a touch stimulus on the head or tail.
In order to describe the locomotion of the worm, we develop a harmonic wave model.
Applying time-delayed feedback control, synchronization effects emerge that contribute to a coordinated locomotion behavior.
This approach also allows us to perform additional synchronicity analyses from which conclusions can be drawn about the significance of neurons within the circuit.



\section{Construction of multilayer network}
\label{sec:multilayer_network}

In this section, we introduce a multilayer network of \elegans\ based on several datasets (see \Cref{tab:overview_datasets}). For the neurons, the information of class affiliation and neuron type is taken into account (ID~1). Neuronal connectivity describes how neurons are connected to each other, either through chemical synapses or gap junctions 
(ID~2). The neurons in the network are extended by 95 body wall muscle cells as these are essential for analysis of locomotion. Motor neurons act on muscle cells through neuromuscular junctions (ID~3). To better understand the interactions of chemical synapses, they are separated into various neurotransmitters and neuropeptides. This requires transmitter and receptor information for the individual neurons and information about which transmitters and receptors couple together (ID~4). 
By considering different neurotransmitters and neuropeptides, the network receives many extra layers that provide a deeper knowledge of the \elegans\ connectome and biological networks in general. 
\begin{table}[!htbp]
\centering
                \captionof{table}[Datasets with network specific information.]{\keyword{Datasets with network specific information.} The neuron and connectivity data (ID~1 -- ID~3) can be found on WormAtlas \cite{wormatlas.2020}. The transmitter and receptor data (ID~4) are partly provided by \cite{Bentley.2016} and have been supplemented by own work.}
                \label{tab:overview_datasets}%
                \resizebox{\columnwidth}{!}{%
               \begin{tabular}{lll}
\toprule
\textbf{ID} & \textbf{Dataset} & \textbf{Content} \\
\midrule
1   & Neurons & Name, class, and neuron type \\
\midrule
\multirow{2}[0]{*}{2} & Neuronal  & Synapse type with number of synapses \\
    & connectivity & between presynaptic and postsynaptic partners \\
\midrule
\multirow{2}[0]{*}{3} & Neuromuscular & Number of synapses between neurons and \\
    & connectivity & body wall muscle cells \\
\midrule
\multirow{2}[1]{*}{4} & Transmitters & Neurotransmitter, neuropeptide, and \\
    & and receptors & neuroreceptor information for the neurons \\
\bottomrule
\end{tabular}%
}
\end{table}
\begin{figure*}[htbp] 
\centering
	\begin{subfigure}{.43\textwidth}
	  \centering
	  \includegraphics[width=.95\linewidth]{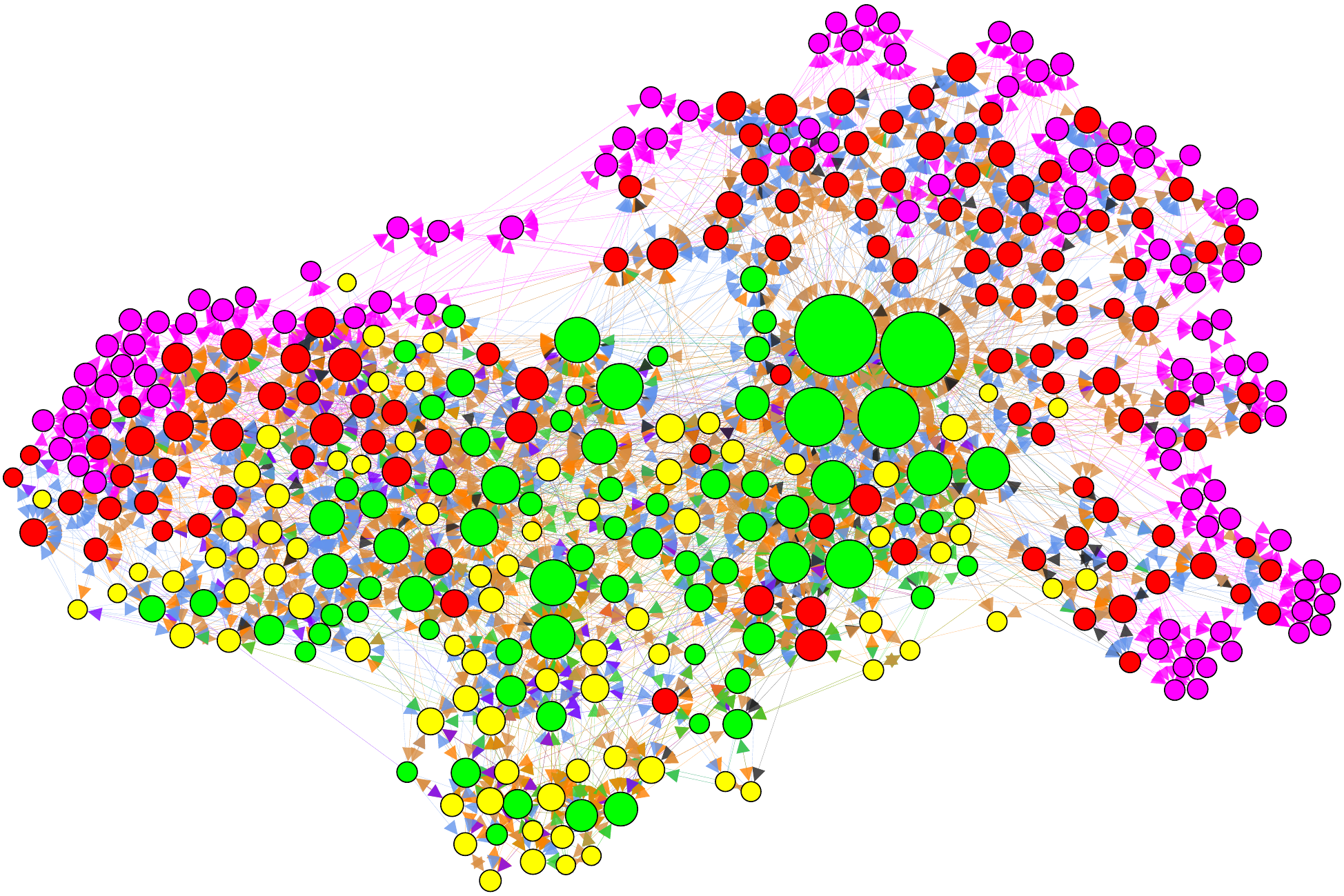} 
	  \caption{Multilayer network}
	  \label{fig:multilayer_network_all}
	\end{subfigure}
	\begin{subfigure}{.43\textwidth}
	  \centering
	  \includegraphics[width=.95\linewidth]{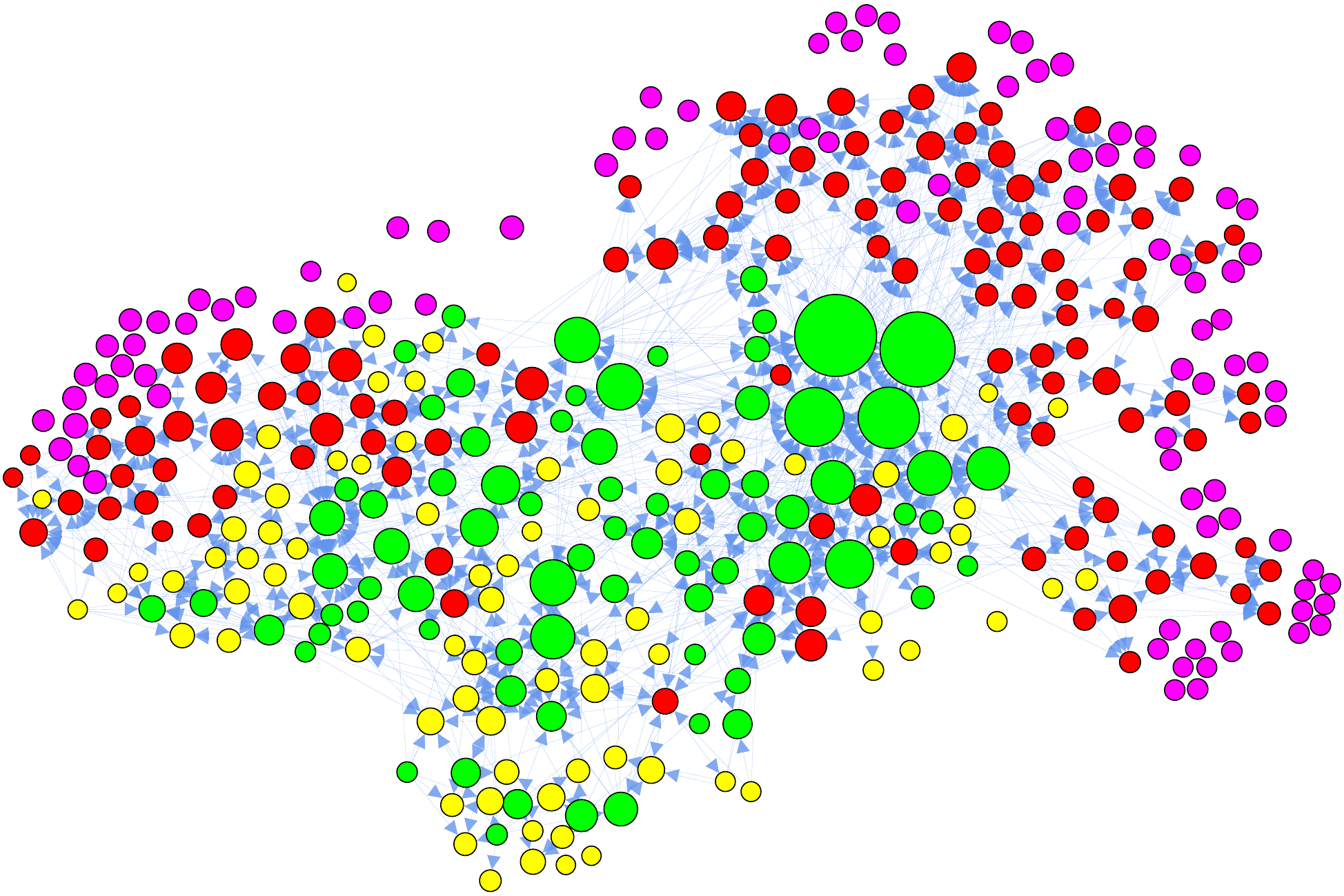} 
	  \caption{Acetylcholine layer ($33.4\%$)}
	  \label{fig:multilayer_network_ach}
	\end{subfigure}
		\newline
	\begin{subfigure}{.43\textwidth}
	  \centering
	  \includegraphics[width=.95\linewidth]{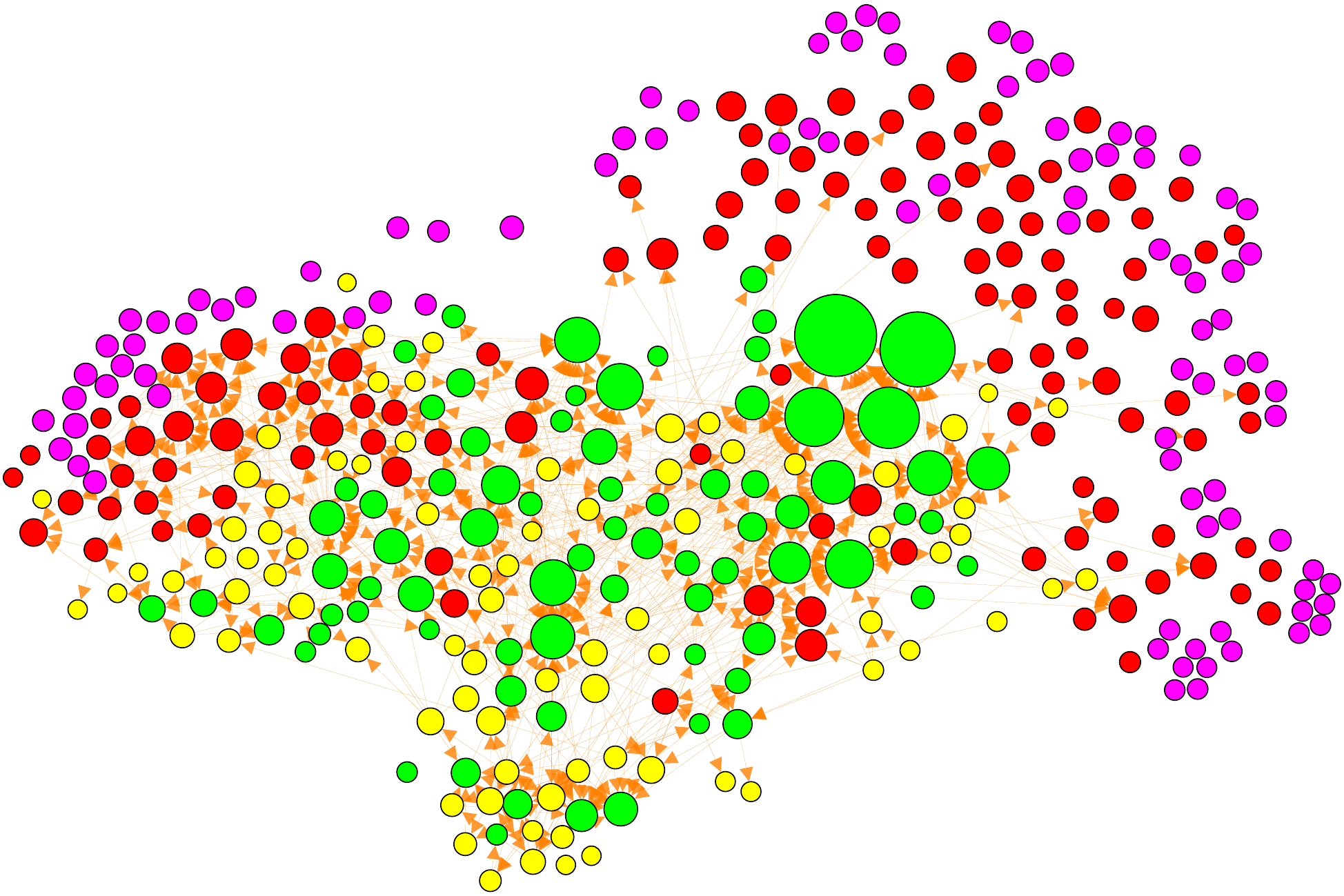} 
	  \caption{Glutamate layer ($20.5\%$)}
	  \label{fig:multilayer_network_glu}
	\end{subfigure}
	\begin{subfigure}{.43\textwidth}
	  \centering
	  \includegraphics[width=.95\linewidth]{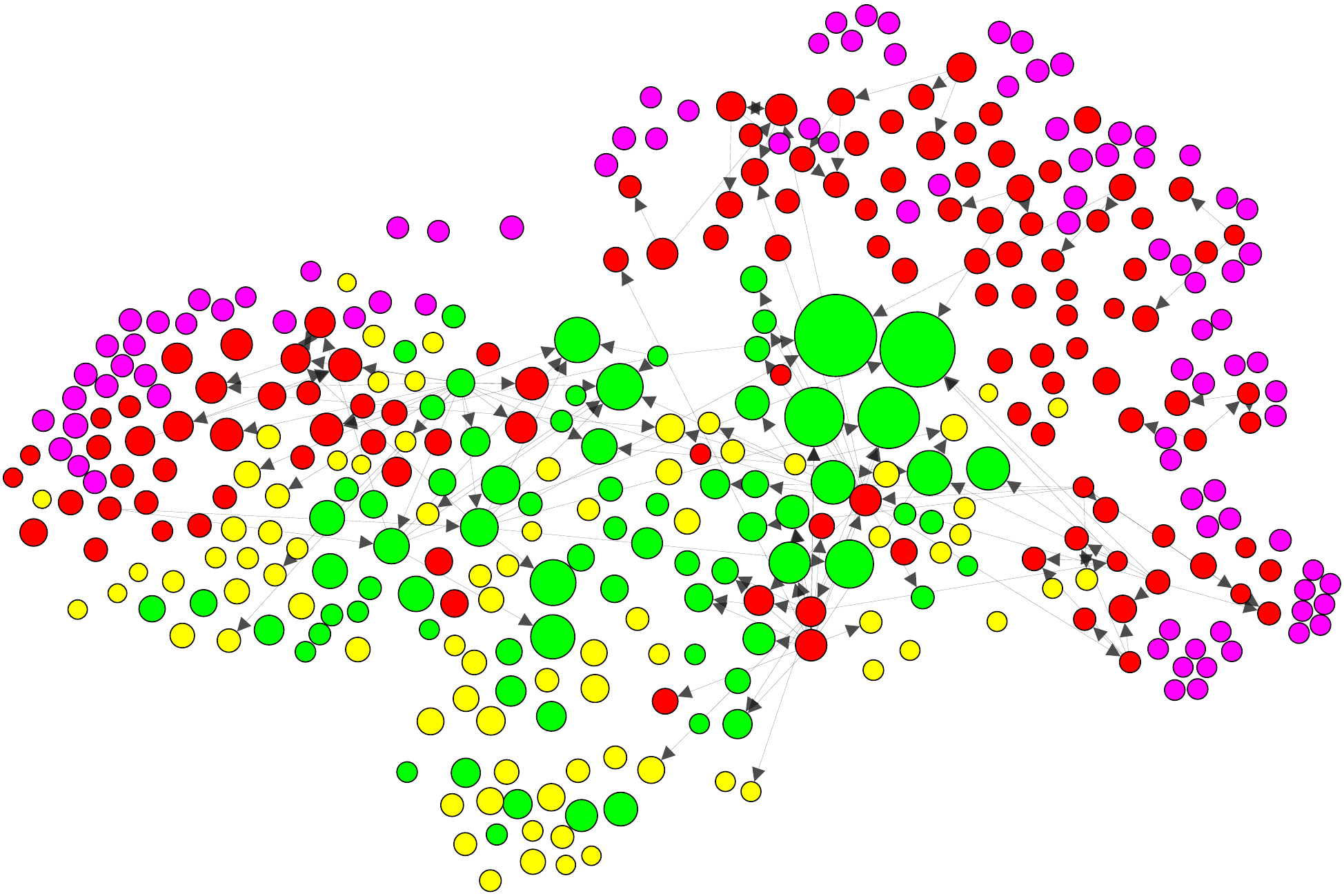}
	  \caption{\textit{gamma}-Aminobutyric acid layer ($3.8\%$)}
	  \label{fig:multilayer_network_gaba}
	\end{subfigure}
			\newline
	\begin{subfigure}{.43\textwidth}
	  \centering
	  \includegraphics[width=.95\linewidth]{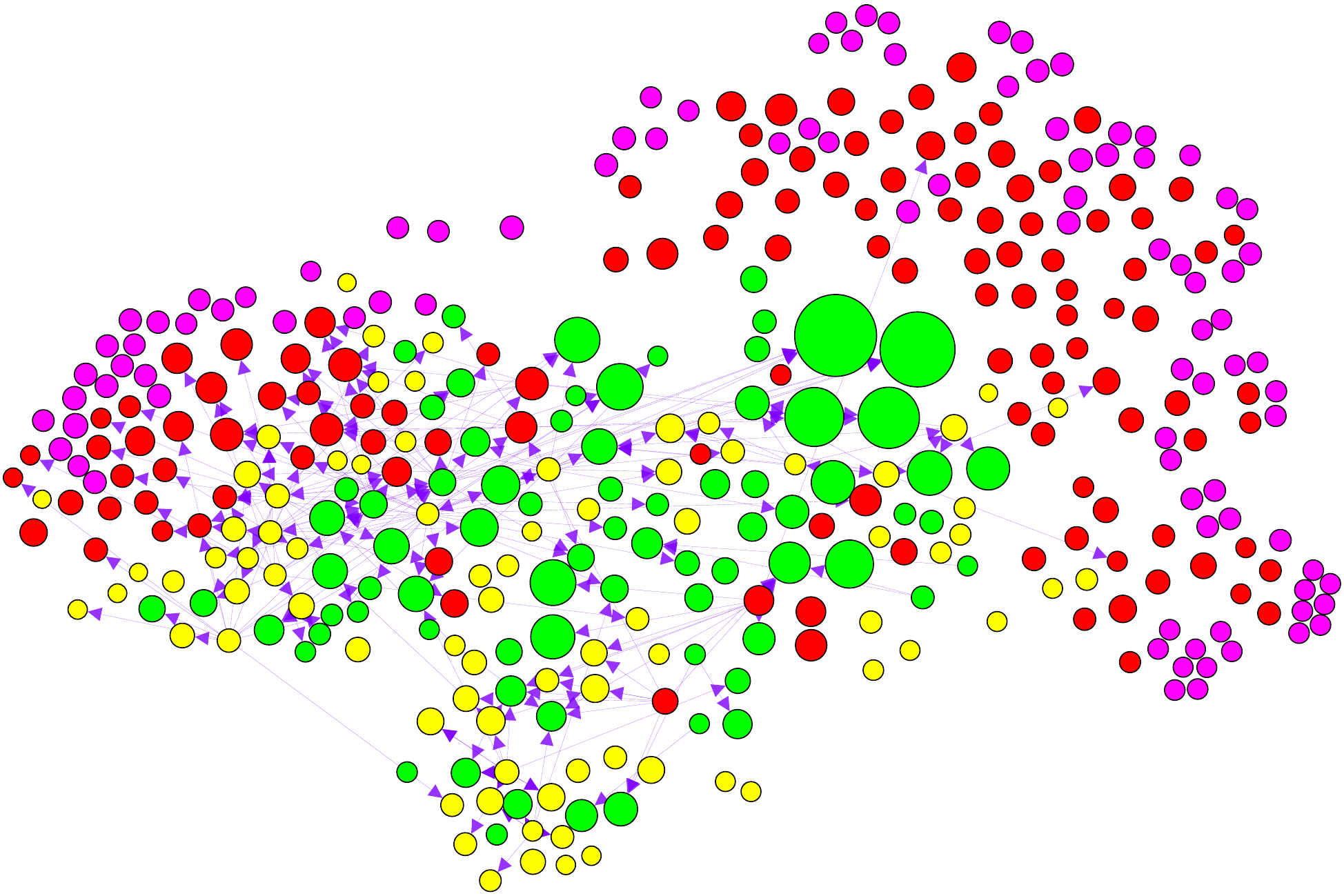}  
	  \caption{Monoamine layer ($5.8\%$)}
	  \label{fig:multilayer_network_ma}
	\end{subfigure}
	\begin{subfigure}{.43\textwidth}
	  \centering
	  \includegraphics[width=.95\linewidth]{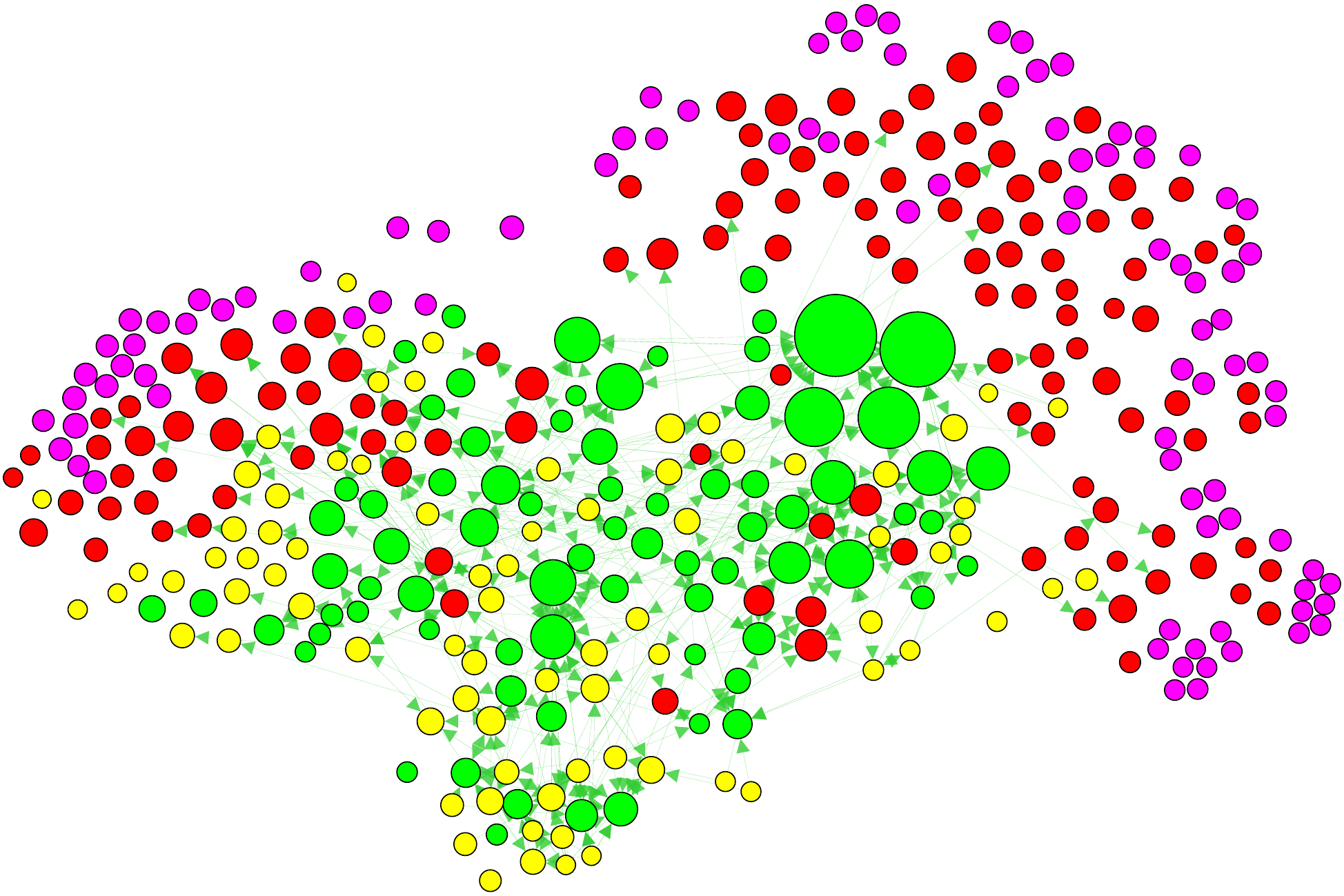} 
	  \caption{Peptide layer ($11.8\%$)}
	  \label{fig:multilayer_network_pep}
	\end{subfigure}
	\newline
	\begin{subfigure}{.43\textwidth}
	  \includegraphics[width=.95\linewidth]{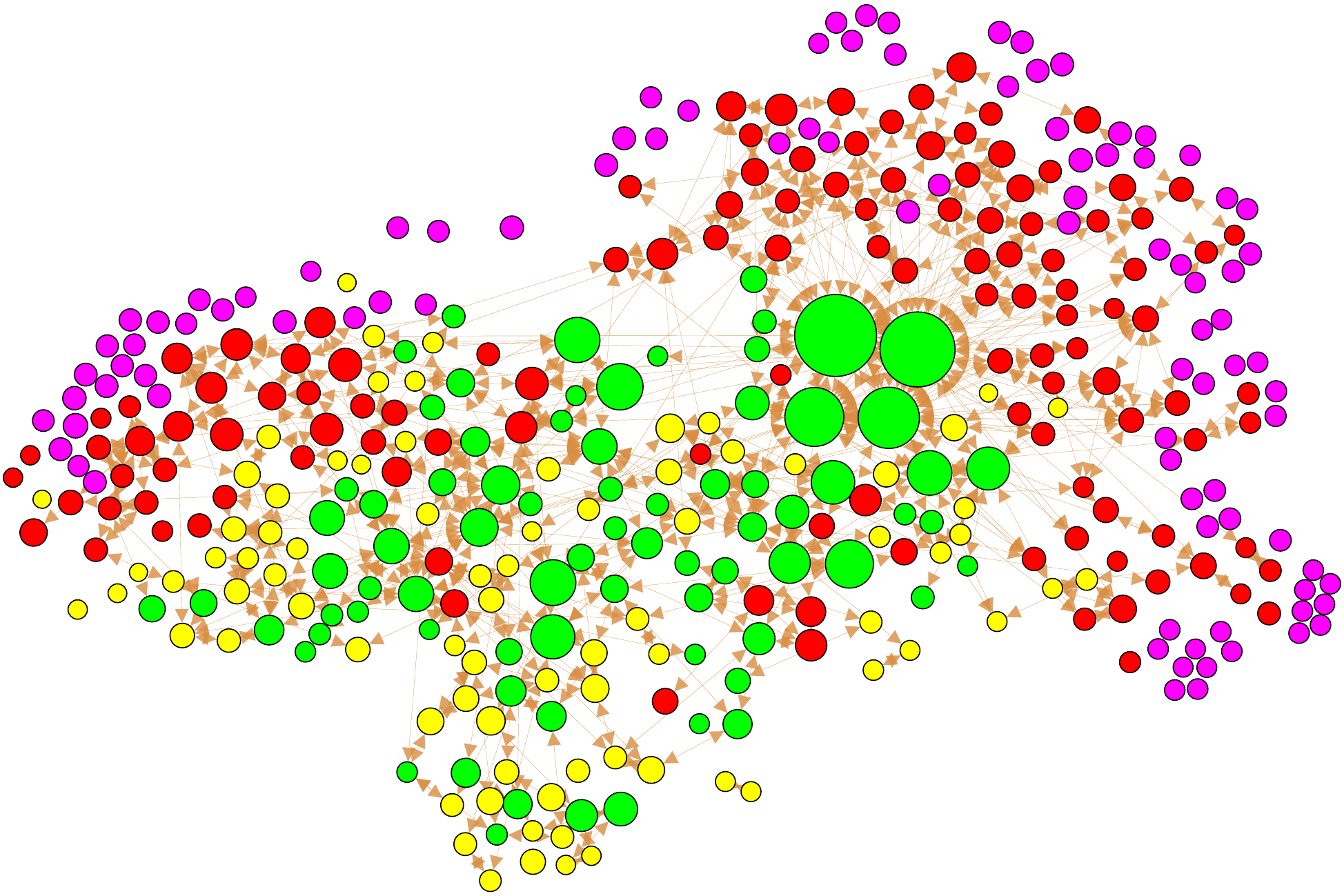}
	  \caption{Electrical layer ($29.1\%$)}
	  \label{fig:multilayer_network_el}
	\end{subfigure}
	\begin{subfigure}{.43\textwidth}
	  \includegraphics[width=.95\linewidth]{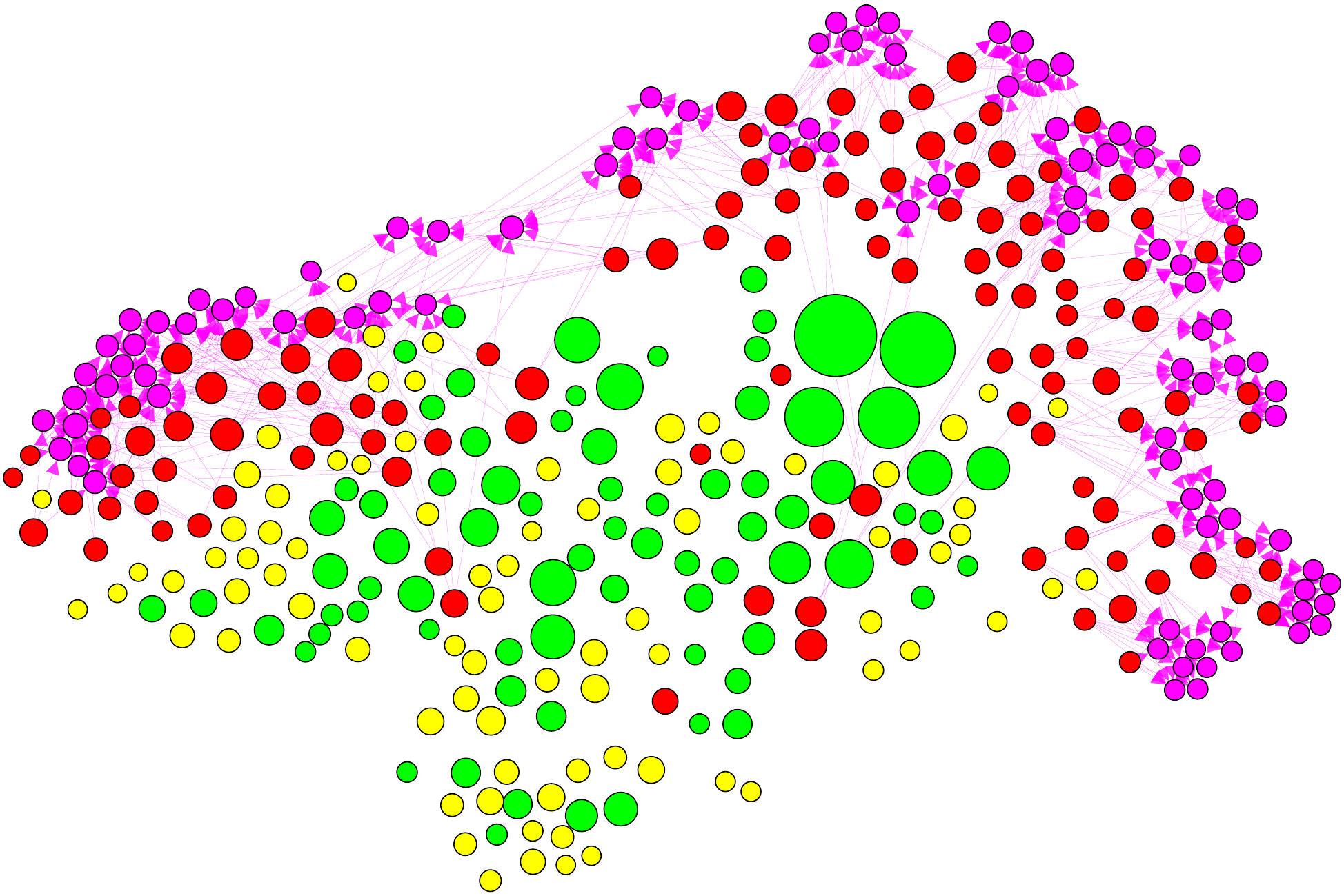} 
	  \caption{Neuromuscular layer ($15.5\%$)}
	  \label{fig:multilayer_network_mus}
	\end{subfigure}
    \caption[The \elegans\ multilayer network.]{\keyword{The \elegans\ multilayer network.} Nodes (illustrated as circles) represent 80 sensory neurons (yellow), 76 interneurons (green), 123 motor neurons (red), and 95 body wall muscles (magenta). A larger node size indicates a higher total node degree. In total, $3,538$ distinct directed connections exist between the nodes. The different connection types are illustrated as colored arrows. Panel \keyword{(a)} is an overlay of all other panels \keyword{(b)-(h)}. The placement algorithm for the neurons is \textsc{Force Atlas} available in the program \textsc{Gephi}.}
    \label{fig:multilayer_network}
\end{figure*}%
The mapped network consists of 279 neurons and 95 muscle cells (\Cref{fig:multilayer_network_all}). The neurons can be categorized as sensory neurons, interneurons, and motor neuron. Although the neurons can be multifunctional, only one type is highlighted in the figure (see also \Cref{tab:neurons_functions}). In total, there are $3,538$ distinct directed connections between the nodes formed by electrical, chemical, and neuromuscular synapses whereby the number of synapses is not taken into account (see also \Cref{tab:neuron_muscular_connectivity}). Since the chemical links are represented by corresponding transmitter types, the multilayer network is defined by all these modes of interactions 
which are depicted in \Cref{fig:multilayer_network_ach,fig:multilayer_network_glu,fig:multilayer_network_gaba,fig:multilayer_network_ma,fig:multilayer_network_pep,fig:multilayer_network_el,fig:multilayer_network_mus}:
Acetylcholine (ACh), glutamate (Glu), \textit{gamma}-aminobutyric acid (GABA), monoamine (MA), peptide, electrical, and neuromuscular connections, respectively. 
ACh covers $33.4\%$ of the $3,538$ connections in the network and forms the largest layer.
It is widely spread among all neurons and does not prefer a specific neuron type.
Glu uses $20.5\%$ of the connections. In comparison with the ACh layer, significantly fewer motor neurons are involved.
GABA represents the smallest network layer and is released on $3.8\%$ of the connections. Most of them are established by interneurons and motor neurons.
Thereafter, the MA transmitters follow with $5.8\%$.
These connections are independent of the neuron type, but most of the motor neurons are predominantly postsynaptic. In the group of MAs, dopamine accounts for about two thirds (cf. \Cref{fig:freq_Monoamines}).
The peptide layer utilizes $11.8\%$ of the connections in the network which exist primarily between interneurons and sensory neurons. 
About $31\%$ of the peptides are peptide transmitters and about $69\%$ are cotransmitters (cf. \Cref{fig:freq_Peptides,fig:freq_Trans}).
Electrical transmission accounts for $29.1\%$ of the connections and is therefore the second largest layer after ACh.
$15.5\%$ of the connections employ neuromuscular junctions that connect motor neurons and body wall muscles via ACh or GABA \cite{Rand.2007, Jorgensen.2005}.
Background information on the different layers is summarized in Appendix~\ref{appendix:layers}, and the data preparation process is detailed in Appendix~\ref{sec:Data description and data preparation}.

The transmitters utilized in the network allow for logistic regression analysis to predict the neurons involved in locomotion behavior as detailed in the next section.


\subsection{Logistic regression analysis}
\label{subsec:logistic}

\keyword{Binary model} - Logistic regression is a classification algorithm which can be used to predict a binary response variable $Y_i \in \{0, 1\}$ based on a set of independent explanatory variables $x_1,\dots,x_m$. Since the expectation value of the response variable $\mathbb{E}(Y_i)$ lies in the interval $[0,1]$, the binary model  
applies the sigmoid function as response function 
\begin{eqnarray}\label{eq:logistic_regression_model}
\sigma_i\left(z_i(\vecvar{c})\right)=\frac{1}{1+e^{-z_i (\vecvar{c})}},\quad i=1,2,...,n, 
\end{eqnarray}
where $i$ is the observation index, $z_i(\vecvar{c}) = c_0 + c{_1}x_{1i} + c{_2}x_{2i} + \dots + c{_m}x_{mi}$ is the linear combination of $m$ explanatory variables $x_{1i},x_{2i}, \dots, x_{mi}$ with $m+1$ parameters $\vecvar{c}=(c_0,c_1,c_2,$ $\dots,c_m)$. The outcome for observation $i$ can be interpreted as the probability that the response variable $Y_i$ is equal to one, $\sigma _i(z_i (\vecvar{c})) = \mathbb{E}(Y_i) = \mathbb{P}(Y_i = 1)$. The goal of logistic regression analysis is to find parameters that best fit empirical observed data in Eq.~\eqref{eq:logistic_regression_model}. This can be achieved using 
the maximum likelihood estimation \cite[Chapter 4]{Pruscha.2006}.
 
\keyword{Discriminative power} - The discriminative power $P$ measures the ability of a logistic regression model or potential explanatory variables to distinguish correctly between observations  ($0$ or $1$) of the response variable $y_i$. 
We define the power as 
\begin{eqnarray}\label{eq:power_calc}
 \begin{array}{ll}
 P = \dfrac{1}{N_{Y=1}N_{Y=0}}\sum\limits_{l=1}^{N_{Y=1}}\sum\limits_{m=1}^{N_{Y=0}}\Psi\left(x_{l,Y=1},x_{m,Y=0}\right),\\
\\
\Psi\left(x_{l,Y=1},x_{m,Y=0}\right)=\begin{cases}
    1, & \text{if $x_{l,Y=1}>x_{m,Y=0}$}\\
    0, & \text{if $x_{l,Y=1}=x_{m,Y=0}$}\\
    -1, & \text{if $x_{l,Y=1}<x_{m,Y=0}$}
  \end{cases}
 \end{array}
\end{eqnarray} 
where $x_{l,Y=1}$ and $x_{m,Y=0}$ are observations of the empirical distributions $X_{Y=1}$ and $X_{Y=0}$ with numbers of observations $N_{Y=1}$ and $N_{Y=0}$.\\
The power can be derived from the receiver operating characteristic (ROC) and the cumulative accuracy profile (CAP). Both ROC and CAP are important concepts to visualize the discriminative power or separation ability of a model. The information contained in a ROC or CAP curve can be aggregated into a single number, the area under the ROC curve (AUROC) or the accuracy ratio (AR). The AR can be interpreted as a simplified representation of AUROC since $\text{AR} = 2 \cdot \text{AUROC}-1$ and is also known as Gini coefficient or power statistics. 
Note that the Mann-Whitney statistics can be introduced as a quantity equivalent to AUROC \cite[Chapter 13]{Engelmann.2011}, which allows us to calculate the discriminative power as introduced in Eq.~\eqref{eq:power_calc}.
From a statistical point of view, the power $P$ can be interpreted as the probability to uncover a difference when there really is one. 
The advantage of defining the power using the Mann-Whitney statistics is that $95\%$ confidence intervals can be calculated which better account for the uncertainty associated with small sample sizes (see Appendix \ref{sec:Logistic regression analysis} for more details about logistic regression and discriminative power).


\subsection{Prediction of the locomotory subnetwork}
\label{subsec:locomotory_subnetwork}

Next, we apply the presented tools to predict neurons of the locomotory subnetwork. 

\begin{figure*}[!htbp] 
\centering
	\begin{minipage}{.49\textwidth}
\centering
\captionof{table}[Logistic regression datasets.]{\keyword{Logistic regression datasets.} The datasets represent the shortest paths between \keyword{(a)} the touch sensory neurons (ALML, ALMR, AVM, PLML, PLMR) and \keyword{(b)} all sensory neurons and the body wall muscles where the second neuron is an interneuron and the third and fourth (if existing) are both motor neurons. Logistic regression models are developed on \keyword{(a)} and tested on \keyword{(b)}.}
\label{tab:shortest_paths}
\centering
  \subcaption{Touch sensory neurons}
 	\centering
 	\small    
    \begin{tabular}{rrrr}
    \toprule
    \multicolumn{1}{l}{\textbf{Shortest path }} & \multicolumn{3}{c}{\textbf{Number of observations}} \\
    \multicolumn{1}{l}{\textbf{length}} & \multicolumn{1}{c}{\textbf{for $Y_i$ = 1}} & \multicolumn{1}{c}{\textbf{for $Y_i$ = 0}} & \multicolumn{1}{c}{\textbf{in total}} \\
    \midrule
    \multicolumn{1}{c}{3} & 1,691 & 540 & 2,231 \\
    \multicolumn{1}{c}{4} & 6,759 & 3,441 & 10,200 \\
    \midrule
    \textbf{Total} & \textbf{8,450} & \textbf{3,981} & \textbf{12,431} \\
    \end{tabular}
    \vspace{+2mm}
  \centering
  \subcaption{All sensory neurons}
 	\centering
 	\small     
    \begin{tabular}{rrrr}
    \toprule
    \multicolumn{1}{l}{\textbf{Shortest path }} & \multicolumn{3}{c}{\textbf{Number of observations}} \\
    \multicolumn{1}{l}{\textbf{length}} & \multicolumn{1}{c}{\textbf{for $Y_i$ = 1}} & \multicolumn{1}{c}{\textbf{for $Y_i$ = 0}} & \multicolumn{1}{c}{\textbf{in total}} \\
    \midrule
    \multicolumn{1}{c}{3} & 14,784 & 13,439 & 28,223 \\
    \multicolumn{1}{c}{4} & 58,128 & 70,537 & 128,665 \\
    \midrule
    \textbf{Total} & \textbf{72,912} & \textbf{83,976} & \textbf{156,888} \\
    \end{tabular}%
	\end{minipage}
	 \hfill
	\begin{minipage}{.49\textwidth}
	    \centering
	    \includegraphics[width=.95\linewidth]{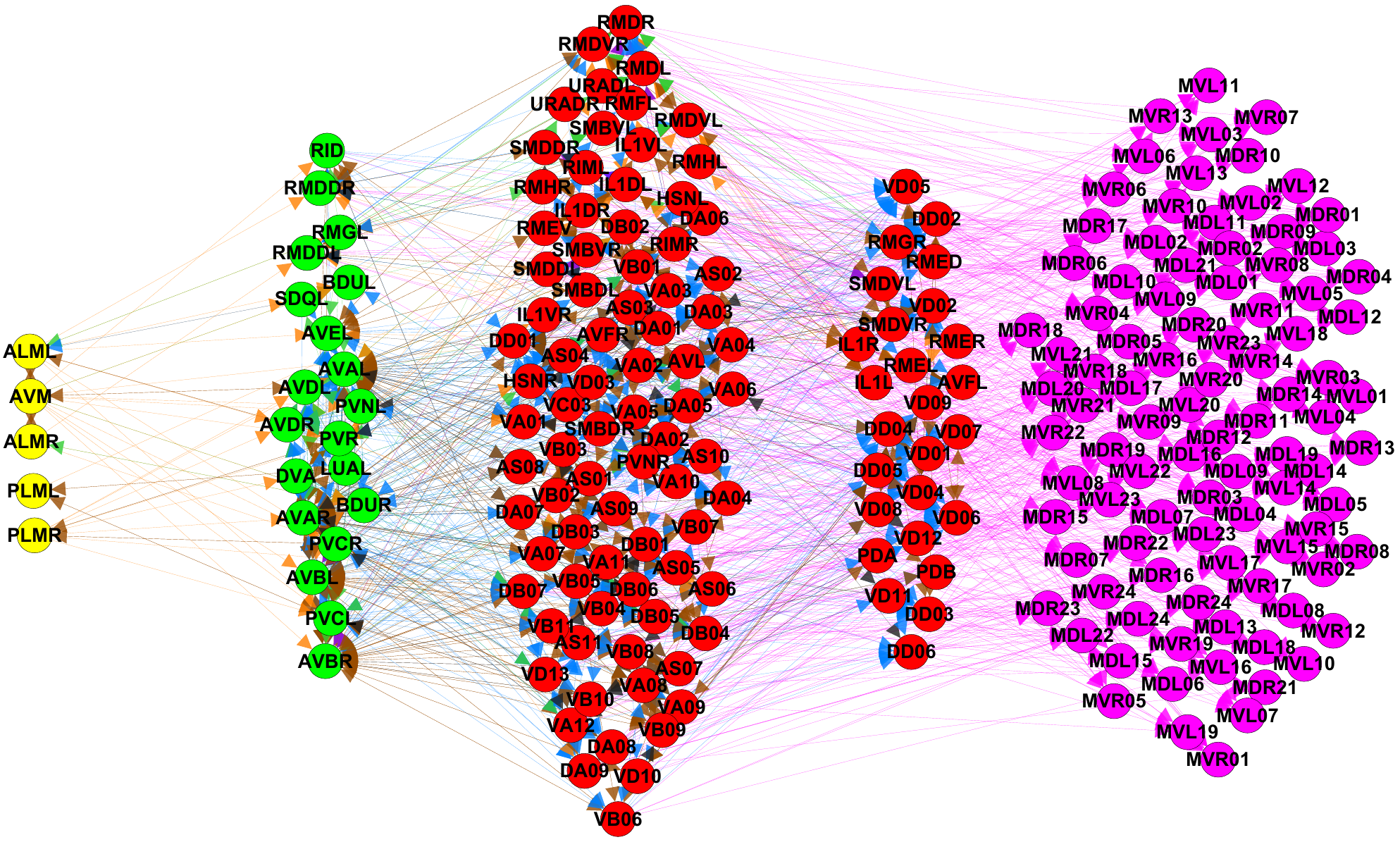}
        \captionof{figure}[The \elegans\ information flow from touch sensory neurons to muscle cells.]{\keyword{The \elegans\ information flow from touch sensory neurons to muscle cells.} The information flows along the shortest paths from the sensory neurons (ALML, ALMR, AVM, PLML, PLMR) to specific interneurons for processing and then to specific motor neurons to complete an action. The latter can be directly connected to the muscle cells, or they can previously activate a second layer of motor neurons. The color code is as in \Cref{fig:multilayer_network}.}
        \label{fig:shortest_paths}
    \end{minipage}
\end{figure*}%

\keyword{General approach} - In compiling the data basis for logistic regression, it must be ensured that the data are related to the locomotion behavior of the worm. We achieve this by considering the shortest paths between sensory neurons and muscle cells.
In the development of regression models, we restrict ourselves to 5 sensory neurons through which \elegans\ perceives gentle touches on the head and tail. The worm reacts on this by moving forward or backward. The shortest paths are illustrated in \Cref{fig:shortest_paths}. In general, the transfer of information in the network begins with sensory neurons and proceeds via interneurons to motor neurons. The latter in turn can activate muscle cells. The particularity of \elegans\ is that the stimulation and relaxation of muscle cells is effected by two types of motor neurons: excitatory and inhibitory. Moreover, the inhibitory cells are generally activated by excitatory cells which is included in the shortest paths. However, \Cref{fig:shortest_paths} has only an illustrative purpose and is not complete. Many of the motor neurons in the third group would also appear in the fourth group because the motor neurons themselves are interconnected. 
All locomotion-relevant interneurons and motor neurons are contained in the shortest paths.\\
The shortest paths are classified as paths going completely through the locomotory subnetwork and others.
This is done first on the level of the neurons (results are taken from \cite{Yan.2017}) and then on the level of the shortest paths: 
\begin{eqnarray*}
 \begin{array}{ll}
LS_p=\begin{cases}
    1, & \text{if neuron } p \text{ belongs to the class AVA (2), AVB (2),}\\
      & \text{AVD (2), PVC (2), AS (11), DA (9), DB (7),}\\
    & \text{DD (6), VA (12) , VB (11), VD (13), or PDB (1)}\\
    0, & \text{otherwise}
  \end{cases},
 \end{array}%
\end{eqnarray*} 
\begin{eqnarray*}
 \begin{array}{ll}
\quad y_i=\begin{cases}
    1, & \text{if the shortest path sp}_{i} \text{ between the sensory neurons}\\ 
    & \text{and the muscle cells only involves neurons with LS}_{p = 1}\\
    0, & \text{otherwise}\\
  \end{cases}.
 \end{array}%
\end{eqnarray*} 

\begin{figure*}[H]%
\begin{minipage}{.98\textwidth}
	\centering
	    \begin{subfigure}{.49\textwidth}
                \centering
                \includegraphics[width=0.99\textwidth]{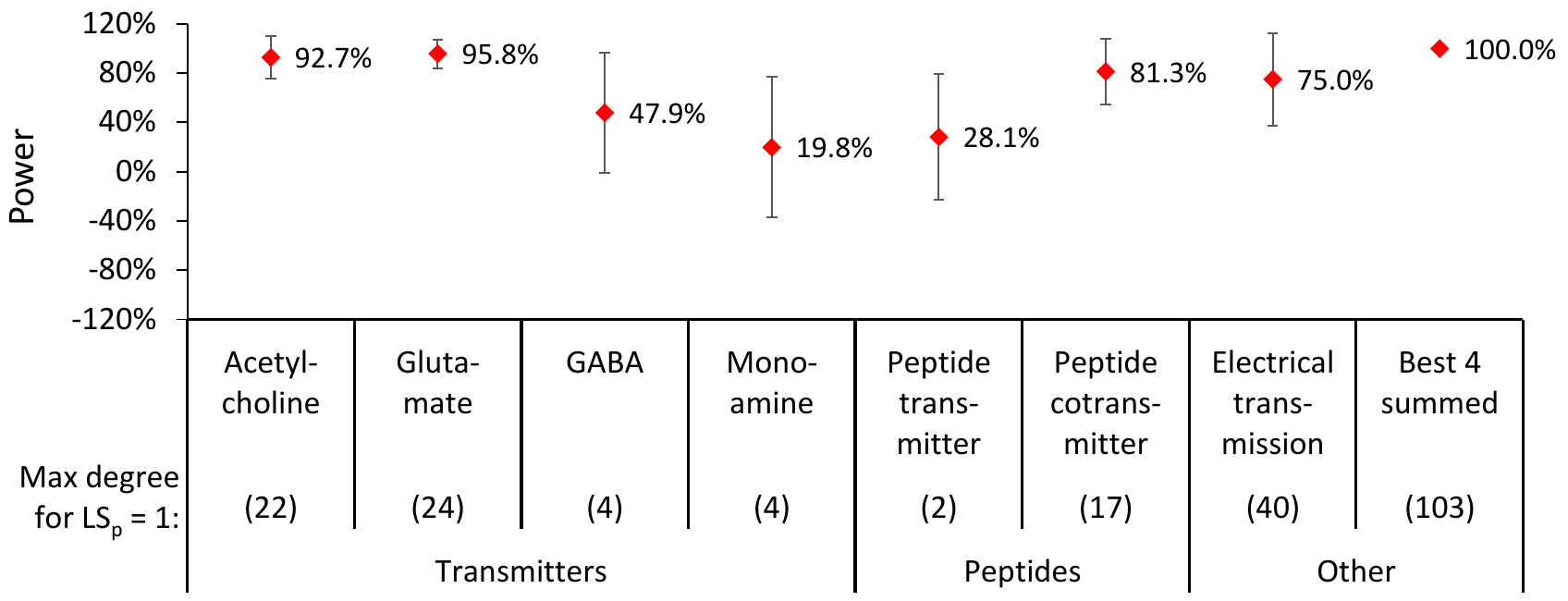}
                \caption{Interneurons}\label{fig:Power_IN_a}
        \end{subfigure}
        \hfill
        \begin{subfigure}{.49\textwidth}
                \centering
                \includegraphics[width=0.99\textwidth]{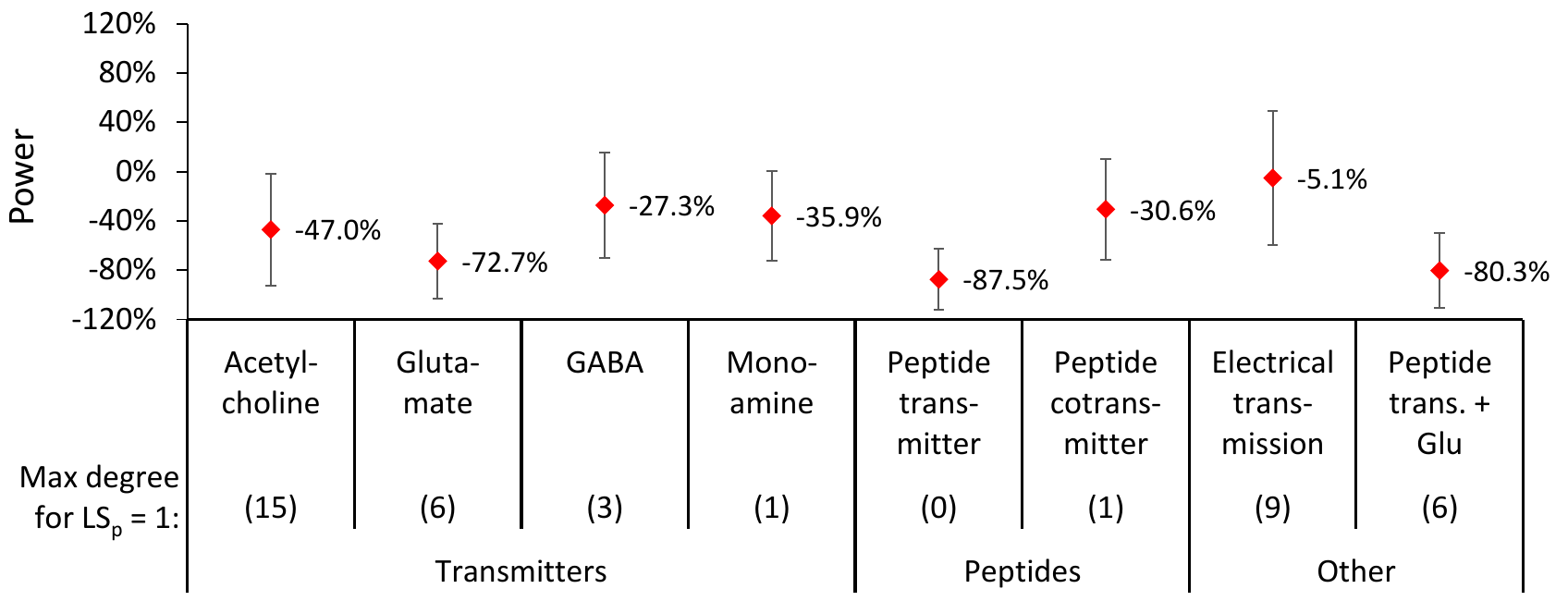}
                \caption{First layer motor neurons}\label{fig:Power_IN_b}
        \end{subfigure}
	\caption[Indegree power values.]{\keyword{Indegree power values.}}
    \label{fig:Power_IN}
	\end{minipage}
\end{figure*}%
The variable $LS_p$ contains the information of being part of the locomotory subnetwork for the neurons $neuron_p$ with index $p=1,2,\dots,279$, and $y_i$ holds the same information for the shortest paths $sp_i$ with index $i=1,2,\dots,n$.
While $LS_p$ plays a role in finding key factors for a logistic regression model in the univariate factor analysis, $y_i$ is used in the multivariate optimization where factors are combined and parameters are fitted. 
The numbers of the shortest paths are presented in \Cref{tab:shortest_paths}. Logistic regression models are developed on the dataset with $5$ touch sensory neurons. 
In total, there are $12,431$ pathways of which $8,450$ pass through the locomotory subnetwork.  
The validation of the developed models is based on the data set that includes all sensory neurons. 
In this case, $72,912$ of a total of $156,888$ pathways exploit the locomotory subnetwork.

\subsection{Prediction results}
\label{subsec:prediction_results}

After the data basis has been set up, the next step is to search for factors that have a high explanatory contribution regarding the locomotory subnetwork. 

\keyword{Univariate power analysis} - Since the question of connectivity is of central importance in a network, we calculate the power values \eqref{eq:power_calc} for different degree distributions regarding the different neuron types in the shortest paths (see exemplary \Cref{fig:shortest_paths}).
There are a total of $20$ interneurons, $8$ of them belong to the locomotory subnetwork $LS_p = 1$, and 12 do not $LS_p = 0$.
To analyze the power of these 20 neurons, the in- and outdegree within the multilayer network (see \Cref{fig:multilayer_network}) must be determined first.
In \Cref{fig:Power_IN_a}, the power $P$ is visualized for different transmitter types. For the peptides, it is also relevant whether they are released as classical transmitters or as cotransmitters (\Cref{fig:network_frequencies}). In the lower part of the figure, the highest observed values among the $8$ neurons of the locomotory subnetwork $LS_p = 1$ are also given for better understanding.\\
The indegree of the interneurons yields many factors with extremely good power. The values for ACh, Glu, electrical transmission, and peptide cotransmission are greater or equal to $75\%$ suggesting that most neurons of the locomotory subnetwork $LS_p = 1$ can be distinguished from other neurons $LS_p = 0$. The positive power values indicate that most neurons of the locomotory subnetwork should have higher indegree values than other neurons. 
Glu has the best power with $95.8\%$, and the highest observed indegree among the locomotory subnetwork neurons is $24$. The indegree of the electrical connections is also an excellent factor to separate the neurons. The power value of $75\%$ also signifies that there is a difference between the neurons but this time the uncertainty is a little higher which is shown by the broader confidence interval. On the other side, the power values of GABA, MA, and peptide transmission are not unsatisfactory. From probability theory point of view, however, it is very uncertain to state a difference. The confidence intervals indicate that the true power value could be lower than $5\%$ which means that the null hypothesis "There is no difference" should not be rejected (on a significance level of $5\%$). The reason for this uncertainty is the low sample size of $20$ neurons. Note that it is sufficient to postulate a difference as long as power values are greater than $60\%$.\\
In addition, the indegree values of the interneurons can be combined for different transmission types. The summation of the values for ACh, Glu, electrical transmission, and peptide cotransmission lead to a power of $100\%$ meaning that 
the locomotory subnetwork neurons can be perfectly separated from other neurons. 
The locomotory interneurons are generally characterized by a high in- and outdegree since they convey signals from sensory neurons to motor neurons. On the other hand, locomotion is the most important behavior of \elegans. The worms must be able to respond to environmental changes by changing their locomotion behavior. 
The interneurons of the locomotory subnetwork have many incoming connections concerning many different transmitters. This can already be seen in \Cref{fig:multilayer_network}. Note that the $8$ interneurons have the highest total degree including all transmission types (see \Cref{fig:Power_oudegree_IN} for outdegree power values).

Next, we consider the first layer of motor neurons (excitatory) in the shortest paths (see exemplary \Cref{fig:shortest_paths}). It is assumed that the $8$ interneurons of the locomotory subnetwork are known, and only those motor neurons are selected that are postsynaptic to them. This limits the number of motor neurons to $62$ of which $54$ belong to the locomotory subnetwork.
The power values are depicted in \Cref{fig:Power_IN_b}. There are two factors with negative power values lower than $-70\%$, Glu and peptide transmission. In this case, the minus sign indicates that the locomotory subnetwork motor neurons $LS_p = 1$ have more lower indegree values than other neurons $LS_p = 0$. The indegree for peptide transmission has the best power value with absolute $87.5\%$. The motor neurons in the locomotory subnetwork can be distinguished from most other neurons because they do not have peptide transmitters. The power cannot be further increased by simple combination of factors. All tested combinations result in a lower absolute power than for the peptide transmitters. For example, the combination with Glu decreases the power to absolute $80.3\%$ (see \Cref{fig:Power_outdegree_MO1} for outdegree power values). In addition, it can already be sufficient to consider only the existence of specific transmitters for the neurons. In this case, no degree distributions need to be calculated, and a better qualitative understanding of the neurons in the network can eventually be gained (see \Cref{fig:Power_qual_MO1} and Appendix~\ref{subsec:Complementary univariate power analyses for prediction of the locomotory subnetwork}). The in- and outdegree power values of the second layer of motor neurons (inhibitory) are provided in \Cref{fig:Power_MO2}. 
The results are comparable with those of the excitatory motor neurons, but they do not contribute to the multivariate optimization that we will investigate next.

\begin{figure*}[H] 
\centering
\begin{minipage}{.98\textwidth}
  \centering
  \captionof{table}[Logistic regression models to predict neurons involved in the locomotion behavior of \elegans.]{\keyword{Logistic regression models to predict neurons involved in the locomotion behavior of \elegans.} The table provides the estimated coefficients for the explanatory variables, the estimated constant, and the resulting negative two-fold log-likelihood for the regression models. The latter is given by the statistical program \textsc{SPSS}. Under the null hypothesis "The model has a perfect fit", the fit is the better the lower its value. "Best 4 summed" includes the indegree for ACh, Glu, electrical transmission, and peptide cotransmission.}
    \small
\begin{tabular}{ccccccc}
\toprule
    & \multicolumn{2}{c}{\textbf{Interneuron}} & \multicolumn{2}{c}{\textbf{First layer motor neuron}} &     &  \\
\multirow{2}[0]{*}{\textbf{Model}} & \multicolumn{1}{l}{\textbf{Indegree}} & \multicolumn{1}{l}{\textbf{Outdegree}} & \multicolumn{1}{l}{\textbf{Indegree}} & \multicolumn{1}{l}{\textbf{Quality factor}} & \multirow{2}[0]{*}{\textbf{Constant}} & 
\multirow{2}[0]{*}{\textbf{\specialcell{-2$\cdot$Log-\\Likelihood}}} \\
    & \multicolumn{1}{l}{\textbf{(Best 4 }} & \multicolumn{1}{l}{\textbf{(ACh +}} & \multicolumn{1}{l}{\textbf{(Peptide}} & \multicolumn{1}{l}{\textbf{(ACh -}} &     &  \\
    & \multicolumn{1}{l}{\textbf{summed)}} & \multicolumn{1}{l}{\textbf{ electrical)}} & \multicolumn{1}{l}{\textbf{transm.)}} & \multicolumn{1}{l}{\textbf{ NLP - PDF)}} &     &  \\
\midrule
1   & \multicolumn{2}{c}{0.164} & -25.3 &     & -10.3 & 3,478.8 \\
2   & 0.212 &     & -25.1 &     & -6.5 & 3,714.9 \\
3   & \multicolumn{2}{c}{0.157} &     & 9.7 & -19.5 & 3,908.7 \\
\bottomrule
\end{tabular}%
  \label{tab:Final_logistic_regression_models}%
\end{minipage}
	\begin{minipage}{.49\textwidth}
	\vspace{+4mm}
	    \centering
    \includegraphics[width=0.85\textwidth]{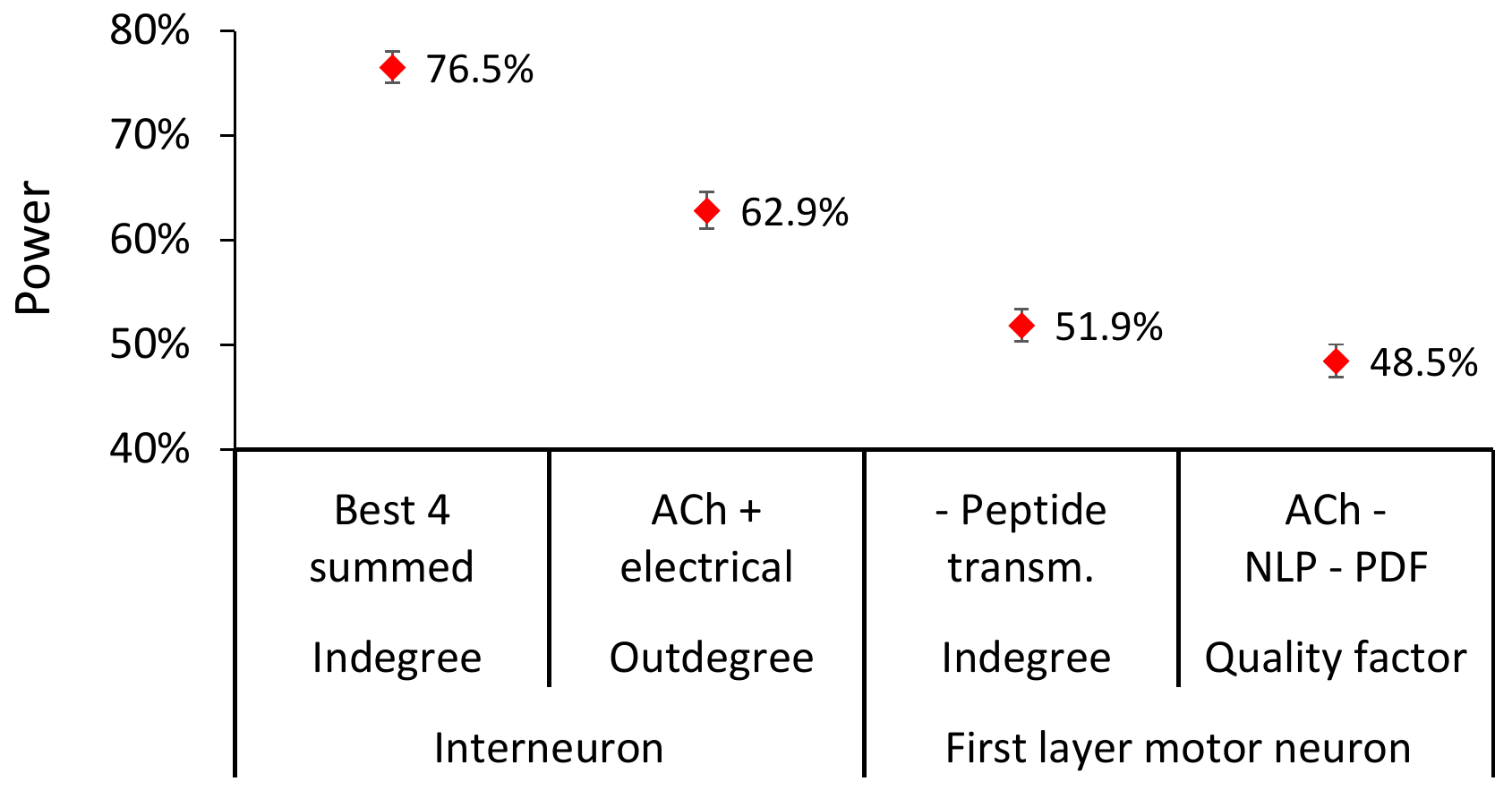}
    \caption[Individual power values for the chosen regression factors on the development dataset.]{\keyword{Individual power values for the chosen \\ regression factors on the development dataset.}}
    \label{fig:Power_Factors}
	\end{minipage}
	\begin{minipage}{.49\textwidth}
	\vspace{+7mm}
	\centering
    \includegraphics[width=0.85\textwidth]{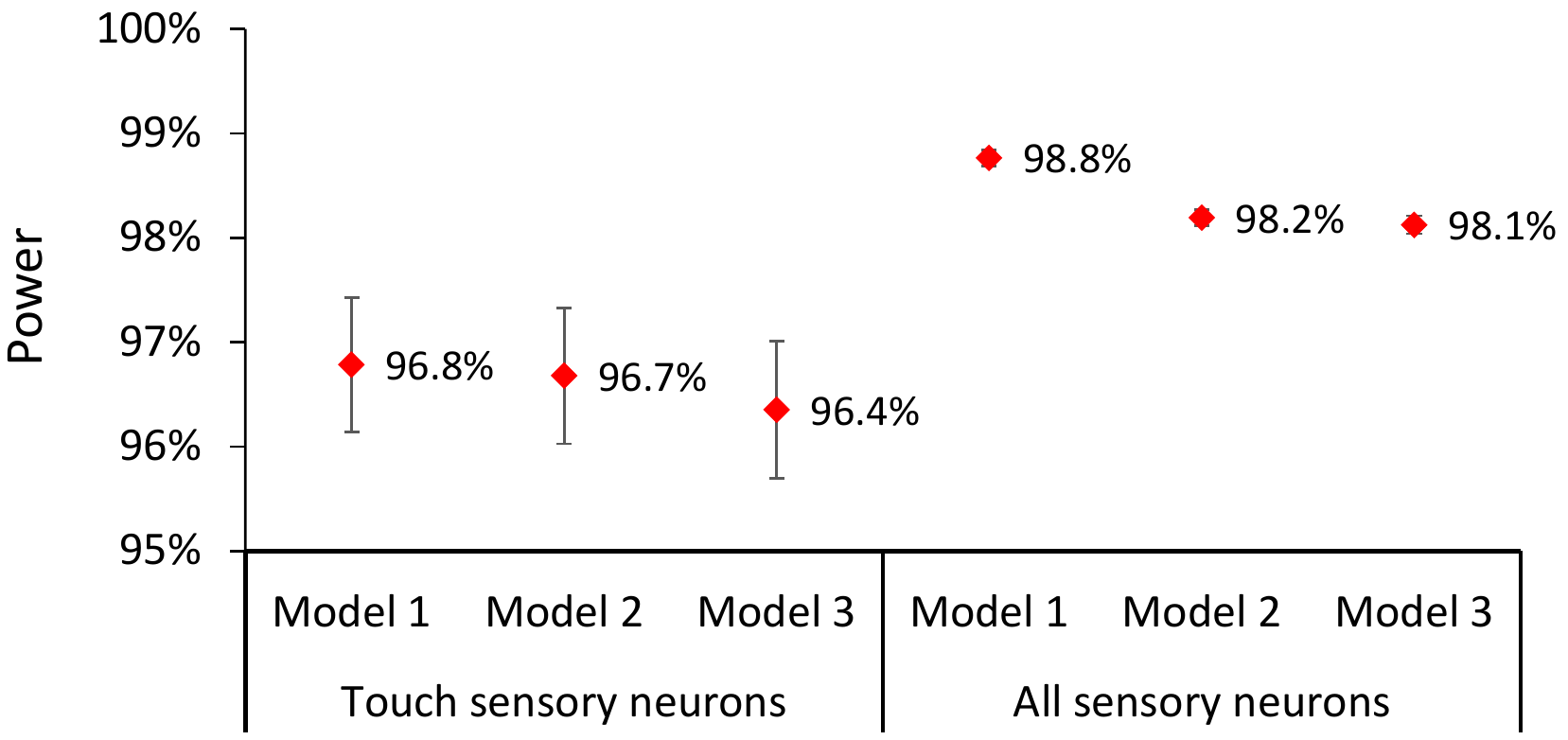}
    \caption[Power values of the logistic regression models.]{\keyword{Power values of the logistic regression models.}}
    \label{fig:Power_Model}
    \end{minipage}
\end{figure*}

\keyword{Multivariate optimization} - In the multivariate analysis, the factors discussed in the previous section are combined with logistic regression \eqref{eq:logistic_regression_model} to further increase their power \eqref{eq:power_calc}. Therefore, we now look at the shortest paths which put the locomotory interneurons and locomotory motor neurons in relation to each other.
Finally, three models are available in the shortlist. The considered factors and parameters including the resulting negative two-fold log-likeli-hood are provided in \Cref{tab:Final_logistic_regression_models}.\\
All three models include only two factors.
The models 1 and 3 consider the combined in- and outdegree of the interneurons as the first factor. The best factor from the indegree analysis is summed up with the best factor from the outdegree analysis (for details see \Cref{fig:Power_IN_a,fig:Power_oudegree_IN}). 
The inclusion of the outdegree is not absolutely necessary because the indegree power in the univariate analysis is already $100\%$, but it makes the prediction of the logistic regression models a little more robust.
On the other hand, model 2 considers only the indegree of the interneurons as the first factor.
Models 1 and 2 utilize as the second factor the peptide transmitter indegree of the first layer motor neurons. Model 3 considers a qualitative factor which aims only at the existence of transmitters or peptides. In this case, the existence of the ACh transmitter, along with the existence of NLP and PDF peptides is binary coded (1 or 0) and added up (for details \Cref{fig:Power_qual_MO1}). 
The peptide transmitter indegree is the only factor with a negative coefficient because motor neurons of the locomotory subnetwork do not have such connections which is indicated by negative power values in the univariate analysis.
For the other factors applies that locomotory subnetwork neurons can be recognized by having more higher values than other neurons. 
The negative two-fold log-likelihood indicates that model 1 should have the highest power followed by model 2 and model 3.
Considering the shortest paths, the power values of the selected factors are compared in \Cref{fig:Power_Factors}. 
The interneuron indegree only achieves a power value of $76.5\%$ meaning that the information concerning the interneurons is no longer unique. For example, many paths can lead through a particular interneuron of the locomotory subnetwork, but in dependence of the postsynaptic motor neurons, each path is classified as path through the locomotory subnetwork $y_i=1$ or not $y_i=0$. Therefore, the same interneuron can be part of pathways marked as 1 as well as part of pathways marked as 0 which makes the information redundant. 
Therefore, the power cannot be $100\%$ as it is the case in the univariate analysis where a distinct group of neurons is considered. The idea behind this procedure is as follows: If the neurons of the distinct groups can be separated by certain factors, then these factors should also be able to distinguish the shortest paths in combination. The other factors in \Cref{fig:Power_Factors} have a lower power than the interneuron idegree, but their separation ability is still more than acceptable. The quality factor of the first layer motor neurons has the lowest value with $P=48.5\%$.
Finally, \Cref{fig:Power_Model} shows the resulting power for the fitted regression models on the development (left) and the validation dataset (right).
The models already show an outstanding power between $96\%$ and $97\%$ for the touch sensory neurons. 
If the models are applied on the shortest paths with all sensory neurons, comparably good values between $98\%$ and $99\%$ are obtained.
Here, we have $90$ neurons with sensory function (instead of 5), and $93$ interneurons are postsynaptic to them (instead of 20) but only $8$ belong to the locomotory subnetwork $LS_p=1$ as assumed. And the interneurons in turn are presynaptic to 106 motor neurons (instead of 62) where $54$ belong to the locomotory subnetwork.
The high power values indicate that these shortest can also be well separated by the logistic regression models. One interesting question is what predictions can be made on this basis, which we will investigate later.\\
It remains to be mentioned that none of the regression models employs factors for the second layer motor neurons. The reason is that the power on the development dataset could not be significantly increased by using these factors which is probably due to high correlation effects between the first layer and second layer motor neurons. The good thing is that the wanted inhibitory motor neurons in the second layer are almost all just postsynaptic to the wanted excitatory motor neurons in the first layer. Therefore, it is sufficiant to predict the interneurons and the excitatory motor neurons.
However, the power can be slightly increased  by taking more factors into account, but the power gain does not justify such complications. In the following, the prediction ability for the models 1 and 2 is examined in more detail.
\begin{figure*}[H] 
\centering
  \captionof{table}[Model performance on the development dataset (touch sensory neurons).]{\keyword{Model performance on the development dataset (touch sensory neurons).}}
  \small
\begin{tabular}{rrrcrrc}
\toprule
\multicolumn{1}{l}{\multirow{3}[2]{*}{\textbf{\specialcell{Locomotory\\subnetwork}}}} & \multicolumn{3}{c}{\textbf{Prediction Model 1}} & \multicolumn{3}{c}{\textbf{Prediction Model 2}} \\
    & \multicolumn{1}{c}{\multirow{2}[1]{*}{\textbf{correct}}} & \multicolumn{1}{c}{\multirow{2}[1]{*}{\textbf{incorrect}}} & \textbf{correct} & \multicolumn{1}{c}{\multirow{2}[1]{*}{\textbf{correct}}} & \multicolumn{1}{c}{\multirow{2}[1]{*}{\textbf{incorrect}}} & \textbf{correct} \\
    &     &     & \textbf{in \%} &     &     & \textbf{in \%} \\
\midrule
\multicolumn{1}{l}{yes} & 8,450 & 0   & 100.0 & 8,450 & 0   & 100.0 \\
\multicolumn{1}{l}{no} & 3,876 & 105 & 97.4 & 3,594 & 387 & 90.3 \\
\midrule
\textbf{Total} & \textbf{12,326} & \textbf{105} & \textbf{99.2} & \textbf{12,044} & \textbf{387} & \textbf{96.9} \\
\bottomrule
\end{tabular}%
  \label{tab:performance_training}%
\end{figure*}%

\keyword{Prediction accuracy} - The output of the logistic regression models on the shortest paths is classified by the threshold $0.5$. Values greater than or equal to the threshold indicate paths through the locomotory subnetwork, and values smaller than the threshold indicate other paths. In practice, the threshold should be adjusted so that the predicted proportion of paths through the locomotory subnetwork corresponds to the actual observed proportion. Since models 1 and 2 have an outstanding discriminative power, this is not necessary. \Cref{tab:performance_training} provides the performance of the models for the touch sensory neurons. The results for all sensory neurons can be found in \Cref{tab:performance_test}.

The prediction of model 1 is very close to a perfect result. In total, the model correctly predicts $99.2\%$ of all shortest paths, and none of the paths through the locomotory subnetwork $y_i=1$ are misclassified. 
On the other hand, some paths that do not completely pass through the locomotory subnetwork $y_i=0$ are incorrectly classified as paths through the locomotory subnetwork. However, the proportion of incorrectly classified paths is only $2.6\%$.
The prediction accuracy of model 2 is slightly lower with a total performance of $96.9\%$ because there is a larger misclassification of paths $9.7\%$ that do not belong to the locomotory subnetwork.
The reason for this is that model 2 does not take into account the outdegree of the interneurons. Nevertheless, both models deliver an excellent performance suggesting that they should be appropriate for use on the larger data set with all sensory neurons.

\keyword{Predictions with all sensory neurons} - We now consider the shortest paths which are classified as paths through the locomotory subnetwork by the models 1 and 2 with threshold $0.5$ and analyze the underlying neurons.
Both model predictions include all neurons of the locomotory subnetwork $LS_p=1$.
Besides, model 1 (2) predicts $8$ ($29$) further neurons which are presented in \Cref{tab:Additional_locomotory_subnet_neurons}. Note that for all of these neurons except for three it can be assumed on the basis of literature research and connectivity analysis that they are involved in locomotion behavior of \elegans. For example, the SMD and RMD motor neurons are included. These are connected with muscles in the head and neck and contribute to multiple navigation behaviors. 
The locomotory subnetwork can be divided in different circuits. 
While the starting neurons $LS_p = 1$ belong to a circuit that mainly initiates forward and backward locomotion (main motor program) of \elegans\ \cite[Section I]{Driscoll.1997}, the SMD and RMD neurons can be included in the circuit for navigation \cite{Gray.2005}.
Model 2 is capable of capturing both circuits since its first factor is little less strict. Without considering the outdegree, a few more interneurons are allowed which are presynaptic to additional potential motor neurons. In total, the results can be evaluated as very good.
In the next step, the underlying dynamics will be examined.


\section{Simulation of dynamics}
\label{sec:simulation}

\begin{figure*}[htbp] 
\centering
	    \centering
    \includegraphics[width=.85\textwidth]{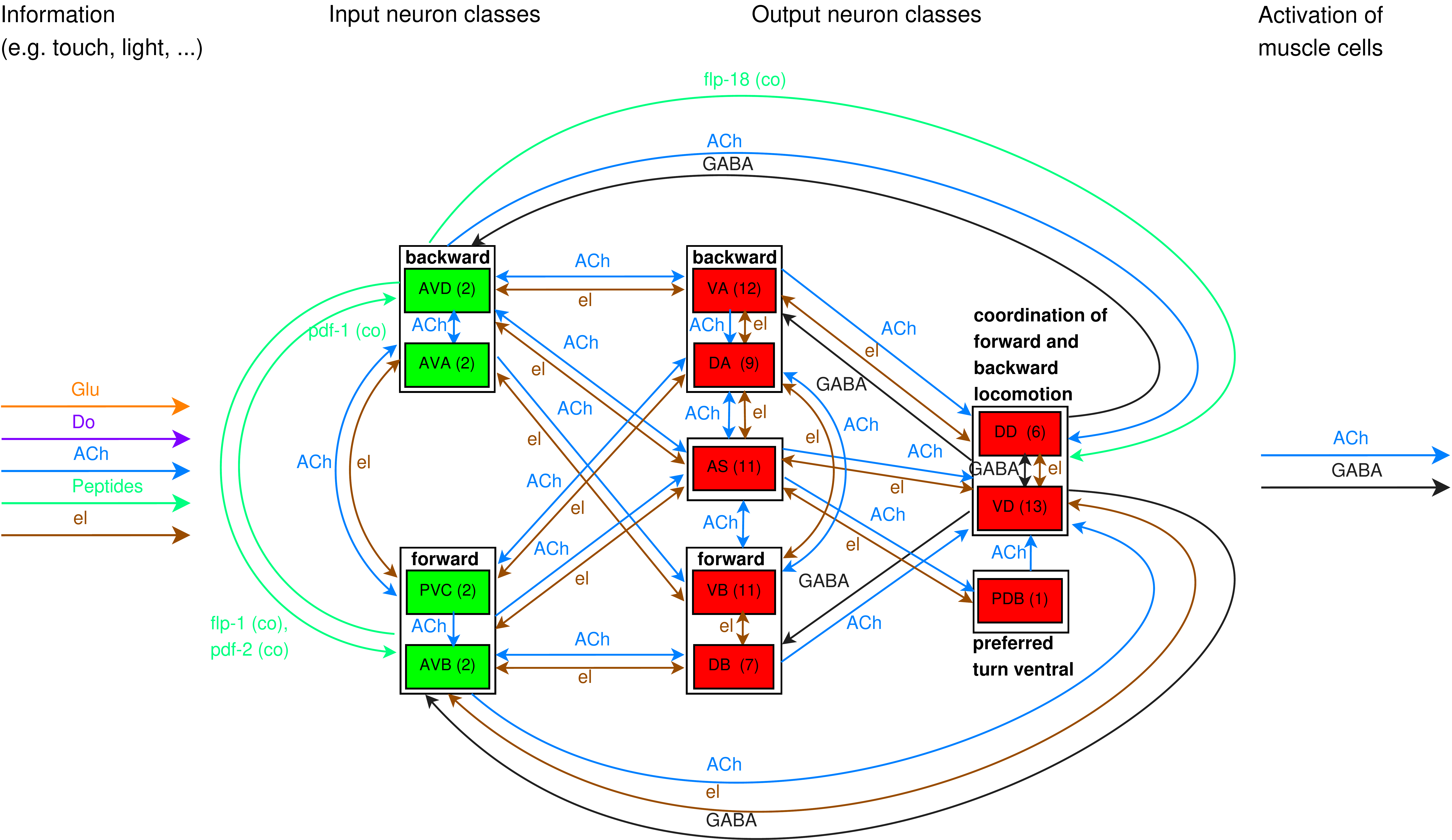}
    \caption[The \elegans\ main motor program for forward and backward locomotion (locomotory circuitry).]{\keyword{The \elegans\ main motor program for forward and backward locomotion (locomotory circuitry).} The command interneurons (green) initiate the forward and backward locomotion by primarily activating the first layer motor neurons. These stimulate muscle contraction via the transmitter ACh. In addition, they activate the second layer motor neurons, the DD and VD neurons, which relax muscle contraction via the transmitter GABA.}
    \label{fig:locomotory_circuitry}
\end{figure*}%

In this section, we investigate the patterns of neuromuscular activity that occur in response to a touch stimulus on the tail.
Since \elegans\ can only react with forward locomotion, we restrict ourselves to the main motor program that initiates forward and backward locomotion to certain stimuli.
Therefore, the results $LS_p=1$ by \cite{Yan.2017} including the 5 touch sensory neurons serve as neuronal basis.

Figure \ref{fig:locomotory_circuitry} depicts the utilized circuit for forward and backward locomotion. 
The neurons are summarized in classes, such as AS and PDB, and components of classes, such as the forward locomotion component of interneurons.
If the worm experiences a gentle touch on its tail, the sensory neurons PLML, PLMR will register this and provide input to the forward interneurons PVC via Glu and electrical transmission but also to the backward interneurons AVD and AVA only via Glu. 
The forward component primarily activates the B-type motor neurons VB and DB via ACh and electrical transmission.
Subsequently, the VB (DB) neurons stimulate ventral (dorsal) muscle cells via ACh and at the same time the second layer motor neurons DD (VD) via ACh and electrical transmission which are connected to muscle cells on the opposite side.
The D-type motor neurons DD (VD) have an inhibitory effect and relax muscle cells via GABA so that the simultaneous contraction of ventral and dorsal muscles is prevented. 
Therefore, these neurons enable the coordination of the movement of the worm. The explanation for the backward locomotion is analog \cite[Section I]{Driscoll.1997} .
The AS neurons are not included in many models
and studies of locomotion of \elegans.
These are only connected to dorsal muscle cells and therefore support the DA and DB neurons to cause dorsal muscle contraction.
The PDB neuron probably causes a ventral bias when performing large wavelength body bends that occur during turning \cite{Yan.2017}.
It was included for the sake of completeness but should not play a decisive role for the initiation of forward locomotion.
Furthermore, some peptides acting as ACh cotransmitters can be identified in \Cref{fig:locomotory_circuitry}. These are neglected in the following. 
Note that the layered structure shown in the figure is also in good agreement with a community analysis of dynamical correlations \cite{Pournaki.2019}.

During locomotion, the worms create rhythmic body undulations.
This raises the important question of how these can be generated.


\subsection{Modeling neuromuscular activity}
\label{subsec:dynamics}

The neuron dynamics can be described in terms of the three-dimensional Hindmarsh-Rose system \cite{Hizanidis.2016}
\begin{subequations}\label{eq:neuron_dynamics}
\begin{flalign}
\begin{split} \label{eq:neuron_dynamics2}
\dot{p}_{i}(t) ={}&q_{i}(t) - ap_{i}(t)^3 + bp_{i}(t)^2 - n_{i}(t) + I_{\text{ext}, i}\\
&-g_{\text{Glu}}\left(p_{i}(t)-V_{\text{exc}}\right)\sum\limits_{k=1}^{N}{C}_{\text{Glu}, ki}S\left(p_{k}(t)\right)\\
&-g_{\text{ACh}}\left(p_{i}(t)-V_{\text{exc}}\right)\sum\limits_{k=1}^{N}{C}_{\text{ACh}, ki}S\left(p_{k}(t)\right)\\
&-g_{\text{GABA}}\left(p_{i}(t)-V_{\text{inh}}\right)\sum\limits_{k=1}^{N}{C}_{\text{GABA}, ki}S\left(p_{k}(t)\right)\\
&+ g_{\text{el}}\sum\limits_{k=1}^{N}{L}_{ik}p_{k}(t)
\end{split}\\
\begin{split} \label{eq:neuron_dynamics3}
\dot{q}_{i}(t) ={}&c - dp_{i}(t)^2 - q_{i}(t)
\end{split}\\
\begin{split} \label{eq:neuron_dynamics4}
\dot{n}_{i}(t) ={}&r\left[s\left(p_{i}(t)-p_0\right)-n_{i}(t)\right]
\end{split} 
\end{flalign}
\end{subequations}
where $p_i(t)$, $i=1,\dots,N$, denotes the membrane potential of the $i$-th neuron at time $t$, $q_i(t)$ represents the
fast current, either Na$^+$ or K$^+$, and $n_i(t)$ the slow current, for example, Ca$^{2+}$. 
The parameter $r$ is the timescale separation between fast and slow variables.

The $N\times N$ coupling matrices $\mathbf{C}_{\text{Glu}}$, $\mathbf{C}_{\text{ACh}}$ and $\mathbf{C}_{\text{GABA}}$ contain the number of chemical synapses 
between pairs of neurons for the transmitters Glu, ACh, and GABA, respectively. 
The function $S$ reflects the operating principle of the nonlinear chemical coupling. The membrane potential $p_k(t)$ is transformed by the sigmoid function 
\begin{equation}
   S(p)=\frac{1}{1+e^{-\lambda(p-\theta)}},\nonumber
\end{equation}
which acts as a continuous mechanism for the activation and deactivation of chemical synapses.
The connectivity of the electrical synapses is described in terms of the Laplacian matrix $\mathbf{L} = \mathbf{D} - \mathbf{G}$ where the coupling matrix $\mathbf{D}$ and the diagonal degree matrix $\mathbf{G}$ comprise the number of electrical synapses between each neuron pair and  the total number of electrical synapses per neuron, respectively. 
The reversal potentials for excitatory and inhibitory chemical synapses are denoted by $V_\text{exc}$ and $V_\text{inh}$, respectively.

Throughout this paper, the system parameters are set as follows: $a = 1$, $b = 3$, $c = 1$, $d = 5$, $r = 0.005$, $s = 4$, $p_{0} = -1.6$, $V_\text{exc} = 2$, $V_\text{inh} = -1.5$, $I_{\text{ext}, i} = 9$ for PLM neurons else $0$, $\lambda = 10$, and $\theta = -0.25$.
The electrical synapses are considered to have the strongest coupling. The coupling of the GABA transmitter is assumed to be slightly stronger than that of Glu and ACh. Accordingly, for the coupling strengths, we use the values: $g_{\text{el}} = 0.2$, $g_\text{Glu} = 0.1$, $g_\text{ACh} = 0.1$, and $g_\text{GABA} = 0.15$. 

The muscle dynamics can be modeled as leaky integrators \cite{Boyle.2012,Izquierdo.2018} 
\begin{flalign}\label{eq:muscle_dynamics}
\hspace{2mm}
\begin{split}
\dot{A}_{l}(t) ={}&\frac{1}{\eta}\biggl[\sum\limits_{s=1}^{N}\Bigl(g_{\text{ACh}}{E}_{\text{ACh}, sl}p_{s}(t)\Bigr.\biggr.\\
&\; \biggl.\Bigl.- g_{\text{GABA}}{E}_{\text{GABA},sl}p_{s}(t)\Bigr)-A_{l}(t)+const\biggr] 
\end{split} &&
\end{flalign}%
with a characteristic time scale of $\eta=100$~ms.
The activation state of the muscle cells at time $t$ is represented by the dimensionless variable $A_l(t)$, $l=1,\dots,M$, where $M$ denotes the number of muscle cells. 
The $N \times M$ coupling matrices $\mathbf{E}_\text{ACh}$ and $\mathbf{E}_\text{GABA}$ contain the number of neuromuscular synapses
between the neuron-muscle pairs 
for the transmitters ACh and GABA, respectively.
The coupling strengths $g_\text{ACh}$ and $g_\text{GABA}$ are as specified above.
The cosmetic parameter $const = 1$ ensures that the muscle activivation is more positive than negative. 


\subsection{Modeling forward locomotion}
\label{subsec:Modeling}

The thrust for the forward locomotion of \elegans\ is provided by 95 body wall muscles located in 4 quadrants along the midline \cite{Altun.2009}. 
For simplicity, we limit our mathematical description to the dorsal and ventral quadrant on the right-hand side and assume that the muscles are arranged in rows and numbered $1,2,\dots,24$ from head to tail, with each dorsal muscle having a ventral partner. This takes into account that the worm crawls on its side and generates sinusoidal undulations. Then, the locomotion of the worm can be described as harmonic wave. Since the undulations spread from the head to posterior, we model the elongation (orthogonal to the direction of motion) of each muscle pair $m=1,2,\dots,24$ consisting of a dorsal and a ventral muscle as harmonic wave 
\begin{flalign}\label{eq:body_dynamics}
\hspace{2mm}
\begin{split}
f_m\left(x,\Delta t_{r_m},t\right) &{}= B\sin\left[ \gamma\left(x,\Delta t_{r_m},t\right)\right]\\
& =B\sin\biggl[\frac{2\pi}{\lambda}x - \overline{\omega}\left(\Delta t_{r_m}\right)t + \phi_m \biggr.\\ 
& \; \biggl.+\sum\limits_{s_m=1}^{r_m \geq 1}%
\left[\overline{\omega}(\Delta t_{s_m}) - \overline{\omega}(\Delta t_{s_m-1})\right]
t_{s_m}\biggr]  
\end{split} &&
\end{flalign} 
%
where $x$ is the position of muscle cells, $t$ is the time, $B$ is the amplitude, $\lambda$ the wavelength, $\overline{\omega}$ the mean angular frequency between extreme body bends, $\phi_m$, 
the phase constant, and $\Delta\overline{\omega}$ the change of the mean angular frequency.
The coordinates are chosen such that $x=m$ for all muscle pairs $m=1,\dots,24$ (see \Cref{{fig:schematics}}).
The index $r_m = 0,1,2,\dots,N_m$ indicates extreme values in the time series of subtracted muscle activity $A^\text{ventral}_m(t) - A^\text{dorsal}_m(t)$ (cf. \Cref{fig:feedback_control}). 

\begin{figure}
    \centering
	\includegraphics[width=.95\linewidth]{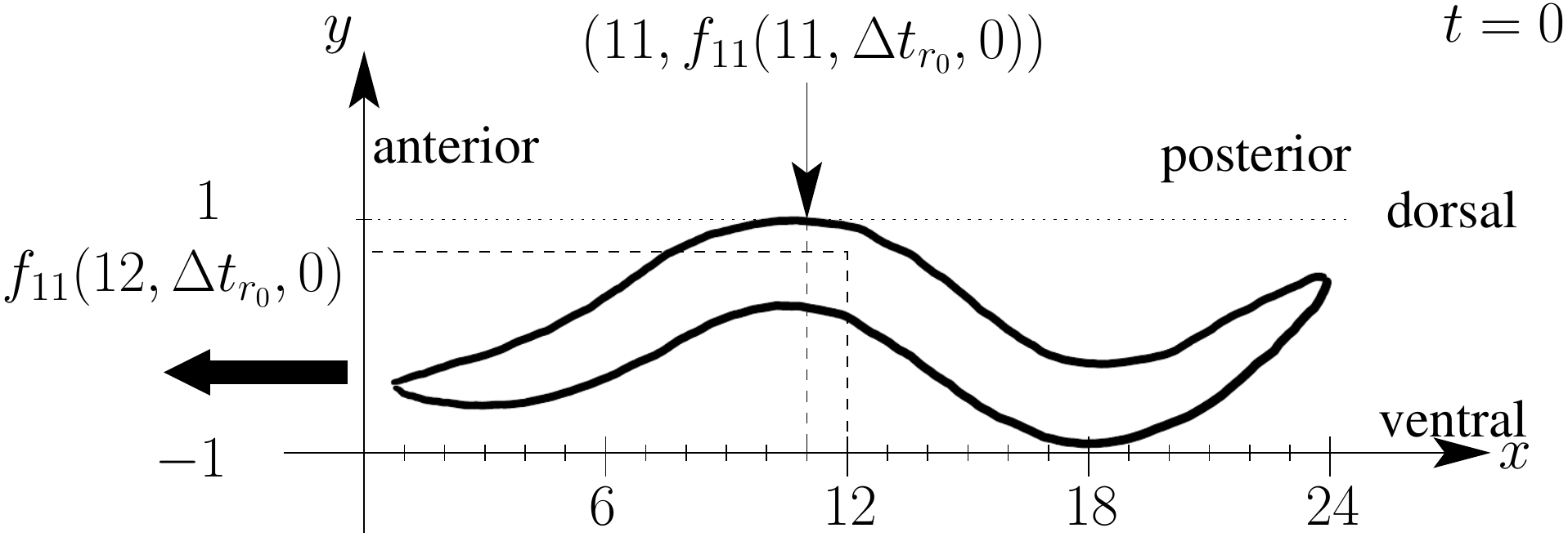}
	\caption[Spatial embedding of \elegans.]{\keyword{Schematic diagram of coordinates.} At time $t=0$, the muscle pairs are located at positions $x=m$, $m=1,\dots,24$. The worm moves forward in negative $x$-direction. The harmonic wave $f_m$ [cf. Eq.~\eqref{eq:body_dynamics}] corresponds to the undulation in $y$-direction of the worm based on the location of the $m$-th muscle pair. Note that at time $t = 0$ the function $f_m$ has either a minimum or a maximum at position $x = m$ which is adjusted by the phase constant $\phi_m$ (cf. \Cref{tab:extracted_extrema}), as exemplified for $f_{11}$.}
\label{fig:schematics}
\end{figure}%

The first term $2\pi x/ \lambda$ in the sine function defines the position of the harmonic wave $f_m$ at time $t = 0$ without phase shift. The wavelength is estimated to be $\lambda=18$ which represents three fourths of the length of the worm.
The second term describes the propagation of the harmonic wave in positive $x$-direction so that the worm moves forward in negative $x$-direction.
The mean angular frequency results from the temporal difference between two 
subsequent minimum and maximum values $r_m$ and $r_m+1$ (or vice versa)
leading to $\overline{\omega}(\Delta t_{r_m})=\pi / \Delta t_{r_m}$ with $\Delta t_{r_m} = t_{r_m+1}-t_{r_m}$ and $t_{r_m} \leq t<t_{r_m+1}$.
The third term represents the initial phase $\phi_m$ (see \Cref{tab:extracted_extrema}).
With every subsequent extreme value, phase jumps occur because the mean angular frequency changes. 
These jumps are compensated by the fourth term. 
The corresponding change of the mean angular frequency 
$\overline{\omega}(\Delta t_{s_m}) - \overline{\omega}(\Delta t_{s_m-1})$ multiplied by time $t_{s_m}$ results in the required phase correction, which is cumulative as expressed by the summation.
Finally, the amplitude of the harmonic waves is set to $B=1$. The forward locomotion of the worm is described by the body wave which results as an average value over all harmonic waves. 


\subsection{Simulation results}
\label{subsec:results_sim}

In order to integrate the dynamical equations \eqref{eq:neuron_dynamics} and \eqref{eq:muscle_dynamics}, the statistical software \textsc{R} 
is utilized. 
The simulation time is $25$ time units with timesteps of $0.005$. 
The wave model \eqref{eq:body_dynamics} is applied as a downstream process on the simulated time series for muscular activity
with timesteps of $0.05$.

\keyword{Neuronal activity} - 
Many neurons in \elegans, especially those involved in locomotion, do not fire classical action potentials \cite{Wen.2018}. This, however, does not exclude signal processing in \elegans\ \cite{Goodman.1998}.
Therefore, the time series represent the behavior of the membrane potential below action potential threshold, and there is a huge operating range where different patterns (regenerative events) can occur \cite{Lockery.2009}.

\begin{figure}%
	\centering
	\begin{subfigure}{.49\linewidth}
        \centering
        \includegraphics[width=.99\linewidth]{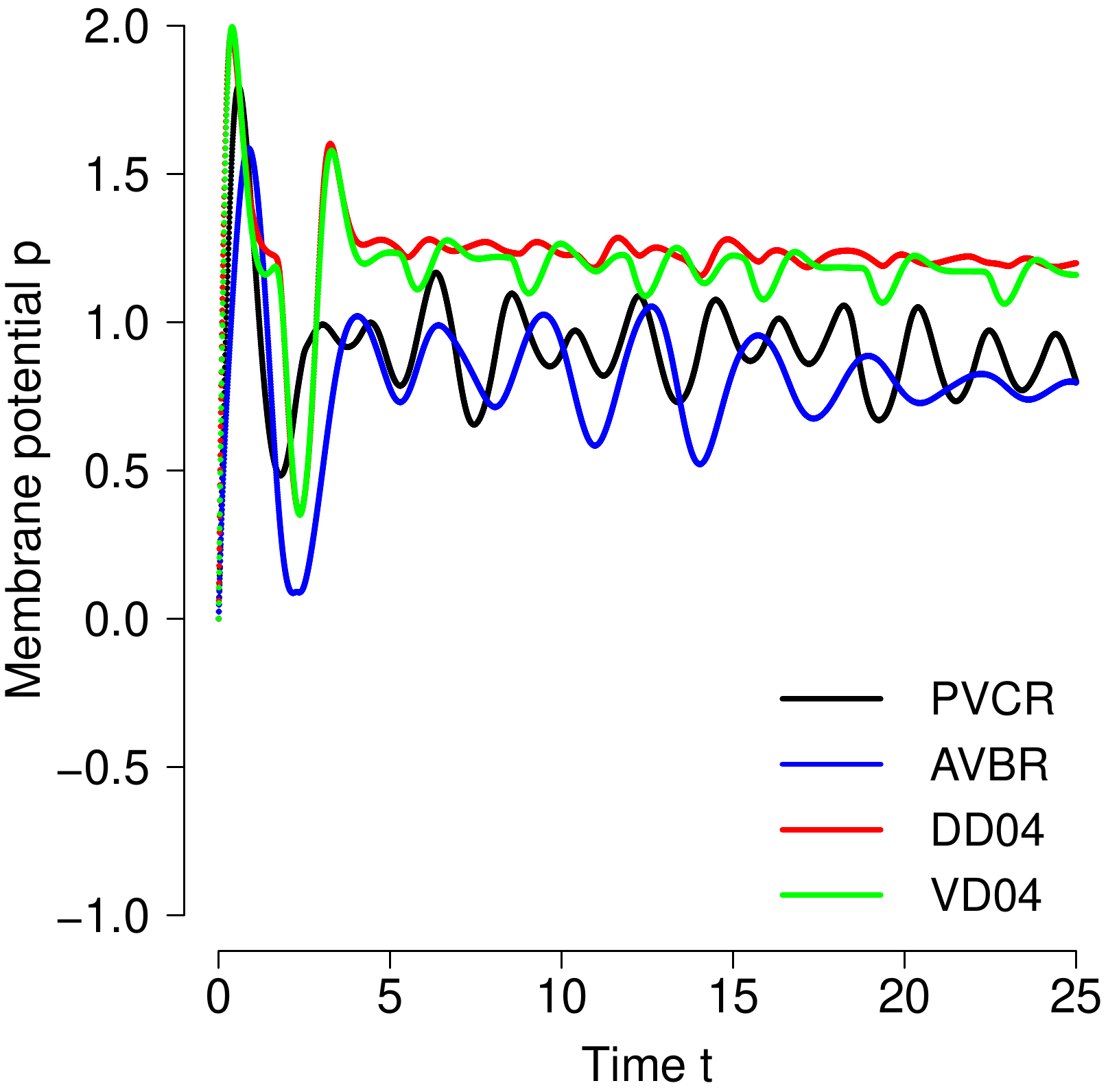}  
        \caption{Isopotential neurons}
        \end{subfigure}
    	\begin{subfigure}{.49\linewidth}
        \centering
        \includegraphics[width=.99\linewidth]{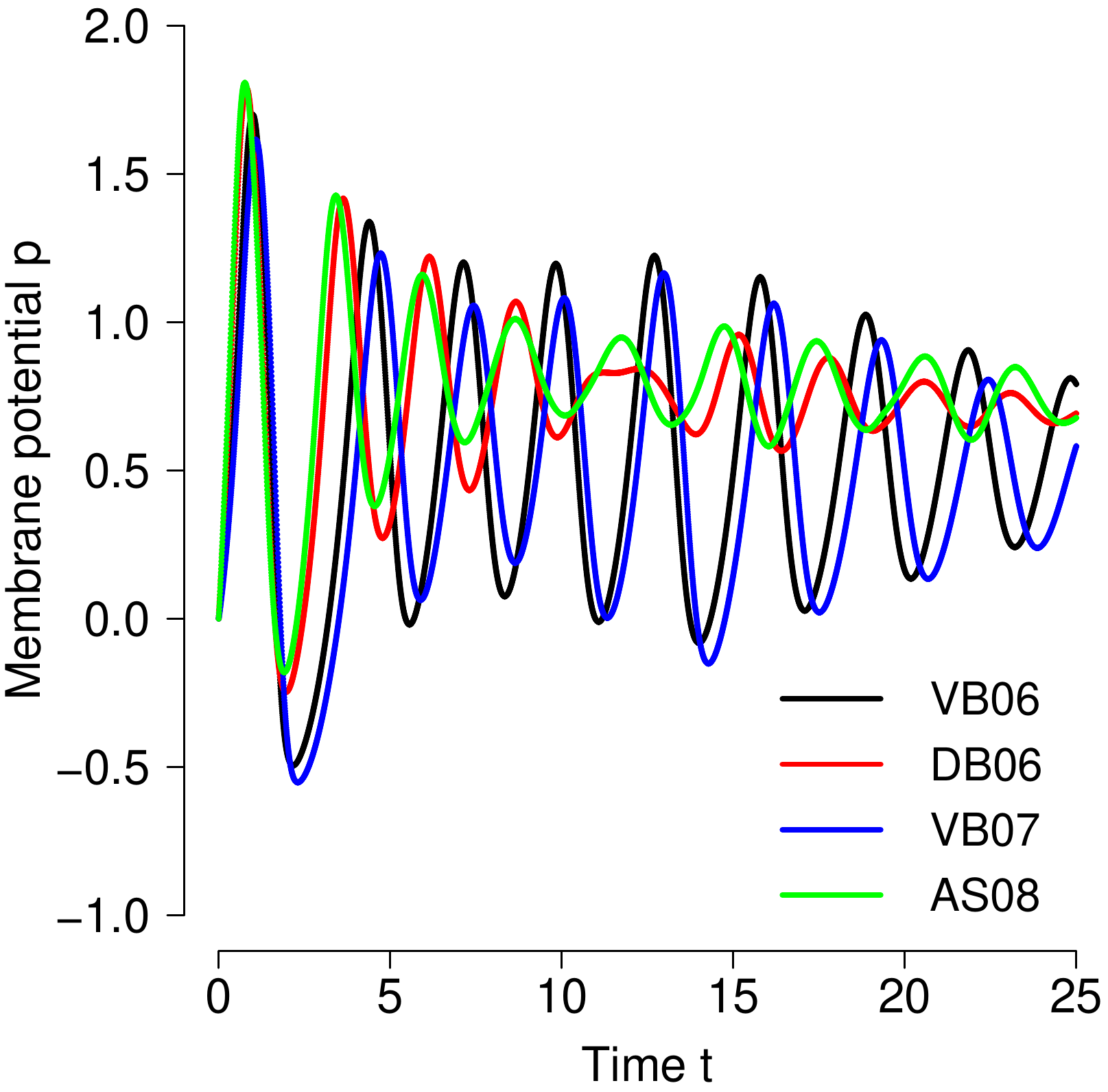}  
        \caption{Oscillatory motor neurons}
        \end{subfigure}
        \caption[Neuronal activity patterns.]{\keyword{Neuronal activity patterns.} Two types of activity patterns can be identified among the neurons of the locomotory circuitry: \keyword{(a)} isopotentials and \keyword{(b)} local oscillations. 
        }
        \label{fig:neuron_timeseries}
\end{figure}

The neurons of the locomotory circuitry are essentially either isopotentials or local oscillators  (see \Cref{fig:neuron_timeseries}). 
The neurons PVCR and AVAR are representative for the most interneurons and show a rather constant activity (very small oscillations) after considering a transient time of about 5 time units.
Ablation experiments proved that these are active and non-oscillating during forward locomotion \cite{Gjorgjieva.2014}.

The neurons DB06, VB06, VB07, and AS08 are representative for most first layer motor neurons (cf. \Cref{fig:locomotory_circuitry})
and exhibit harmonic oscillations with non-constant amplitude over time. The latter is not subject of this study.
Comparing different neurons within a class, their oscillations appear to be slightly shifted to each other.
In this context, the neurons DB06, VB06, and VB07 are shown in \Cref{fig:neuron_timeseries} because they reproduce the experimentally observed activity of comparable neurons
\cite[Fig.~5]{Gao.2018}.
The simulated membrane potential of DB06 is in anti-phase to those of VB06 and VB07, whereas the membrane potential of VB07 precedes that of VB06.
Therefore, the 
motor neurons function as local oscillators to drive body bending and generate forward and backward movements.
Moreover, the oscillations are regulated by the interneurons \cite{Wen.2018}.
The same must also apply to AS neurons because they exhibit a similar oscillatory behavior, as indicated by AS08 in \Cref{fig:neuron_timeseries}.

The neurons DD04 and VD04 represent the behavior of all DD and most VD neurons and show a very constant activity over time.
These neurons share many electrical synapses with each other which could be a reason for this behavior because they must be able to quickly transmit signals from excitatory motor neurons to opposite muscle cells in order to coordinate locomotion.
Similar to the shifts of membrane potential oscillations for excitatory motor neurons, different levels of membrane potential can be detected within the classes DD and VD or among interneurons.

In summary, the main motor program of \elegans\ is a neural circuit with intrinsic oscillatory activities and therefore fulfils the prerequisites of a central pattern generator (CPG) \cite{Fouad.2018,Gjorgjieva.2014}.
The simulation adequately reflects this.
Therefore, the results indicate that the coupling of neurons within the circuit (\Cref{fig:locomotory_circuitry}) is responsible for the oscillatory behavior. As a consequence, the CPG is an intrinsic property of the multilayer network of \elegans\ which generates the sinus rhythm for locomotion by the interaction of different neuron types. 

\begin{figure*}[H]
	\begin{subfigure}{.32\textwidth}
	  \centering
	  \includegraphics[page=18,width=.95\linewidth]{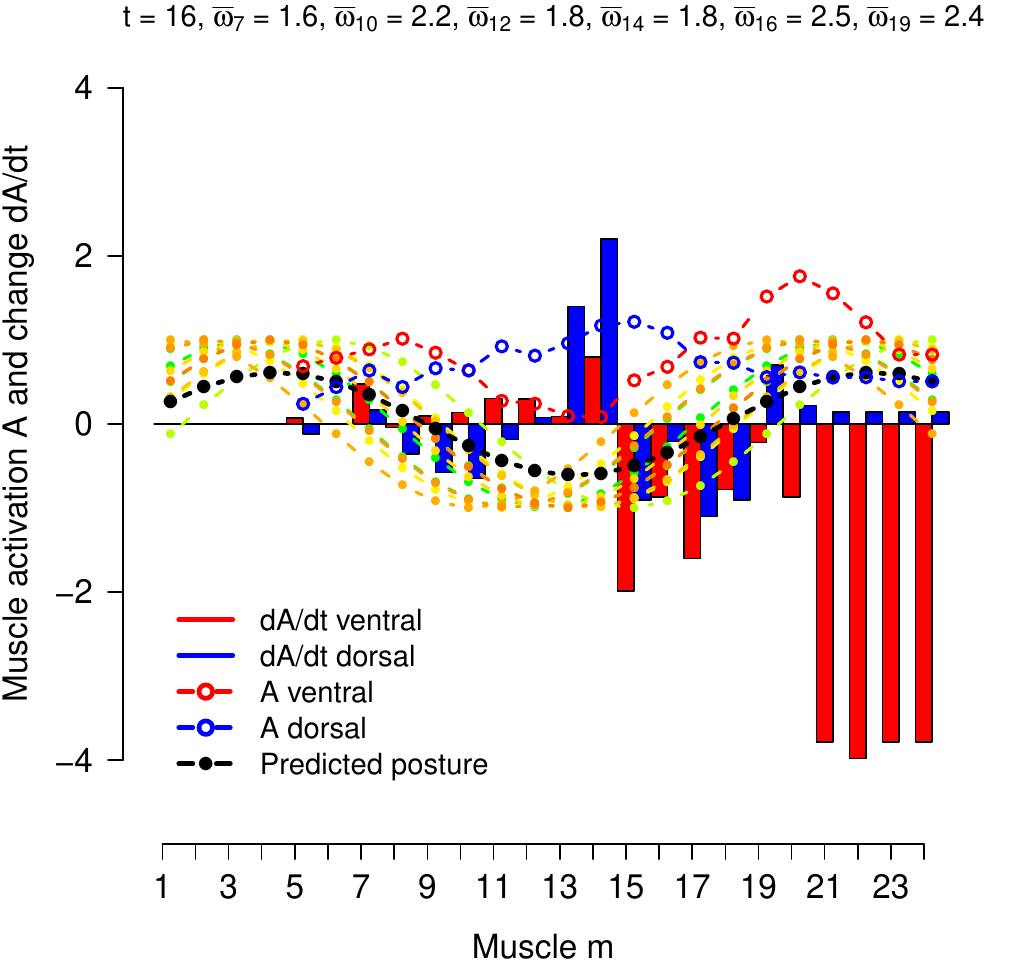}
	  \caption{Time $t = 16$}
	\end{subfigure}
	\begin{subfigure}{.32\textwidth}
	  \centering
	  \includegraphics[page=19,width=.95\linewidth]{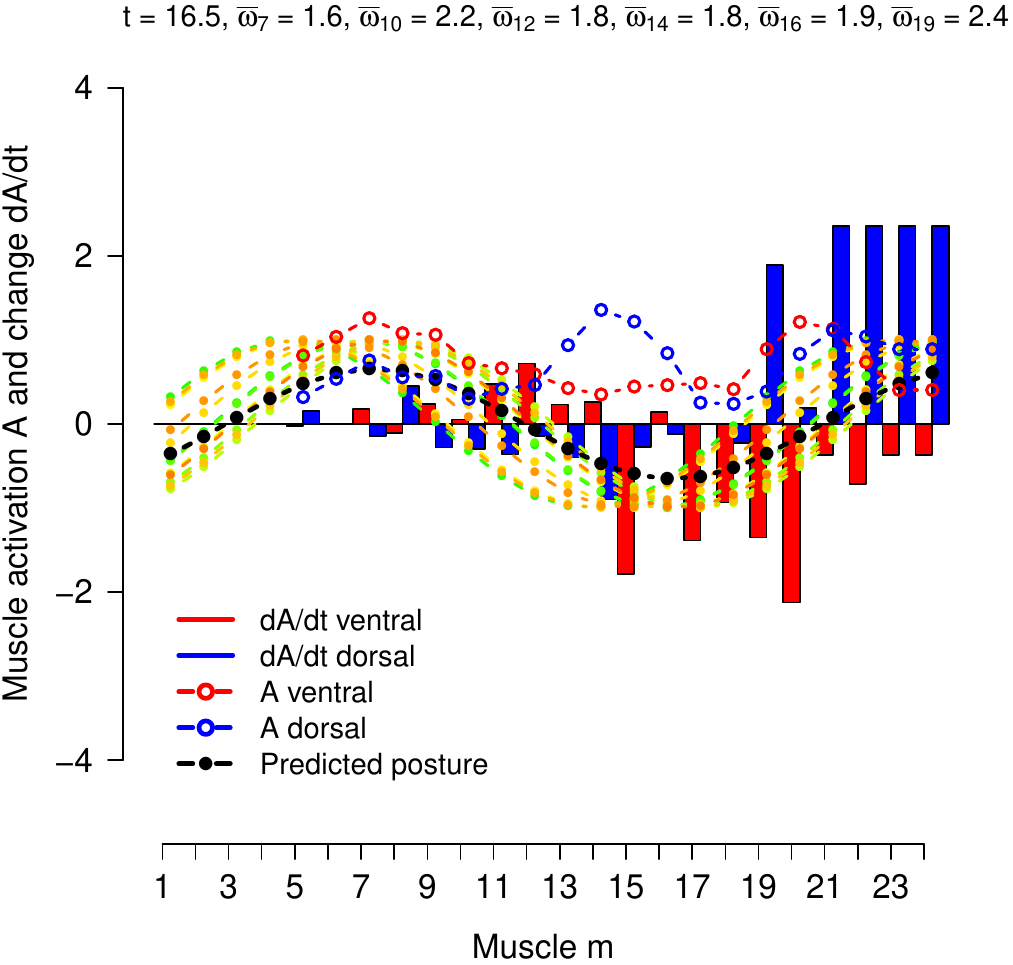}  
	  \caption{Time $t = 16.5$}
	\end{subfigure}
	\begin{subfigure}{.32\textwidth}
	  \centering
	  \includegraphics[page=20,width=.95\linewidth]{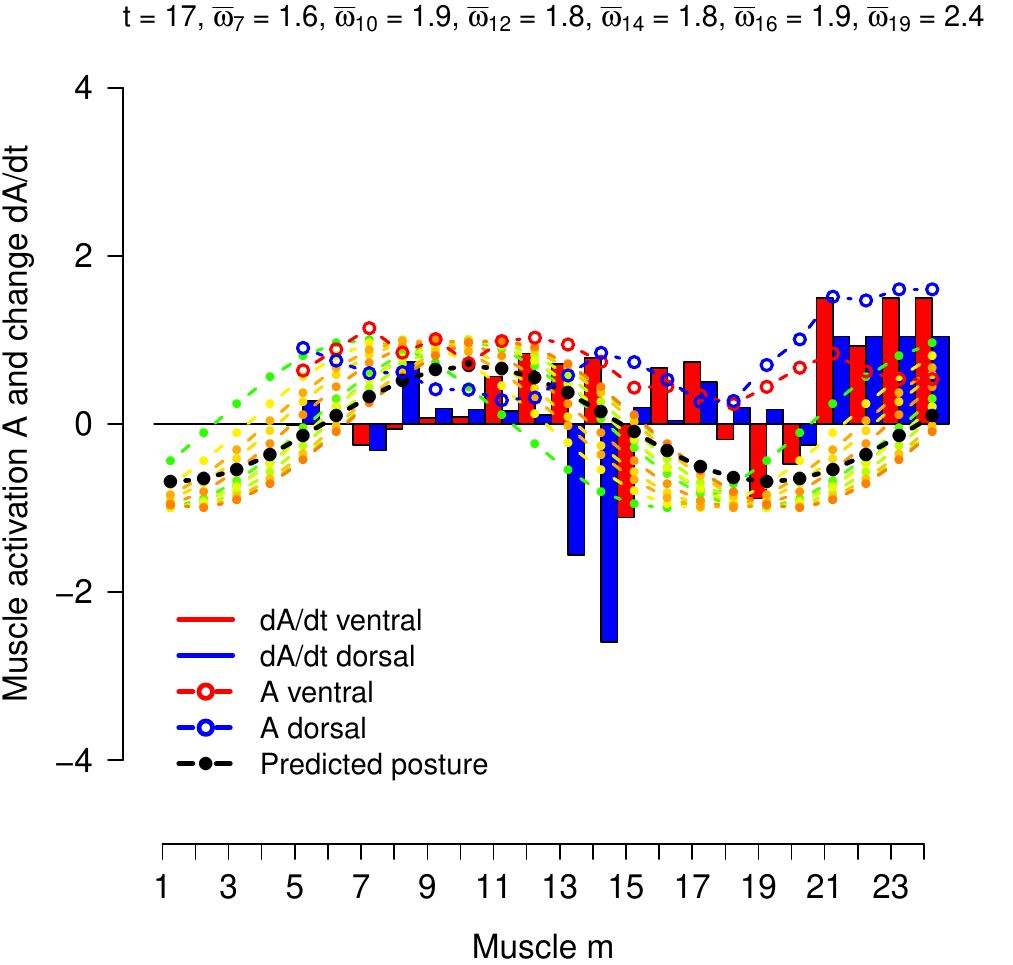}
	  \caption{Time $t = 17$}
	\end{subfigure}
	\newline
	\begin{subfigure}{.32\textwidth}
	  \centering
	  \includegraphics[page=21,width=.95\linewidth]{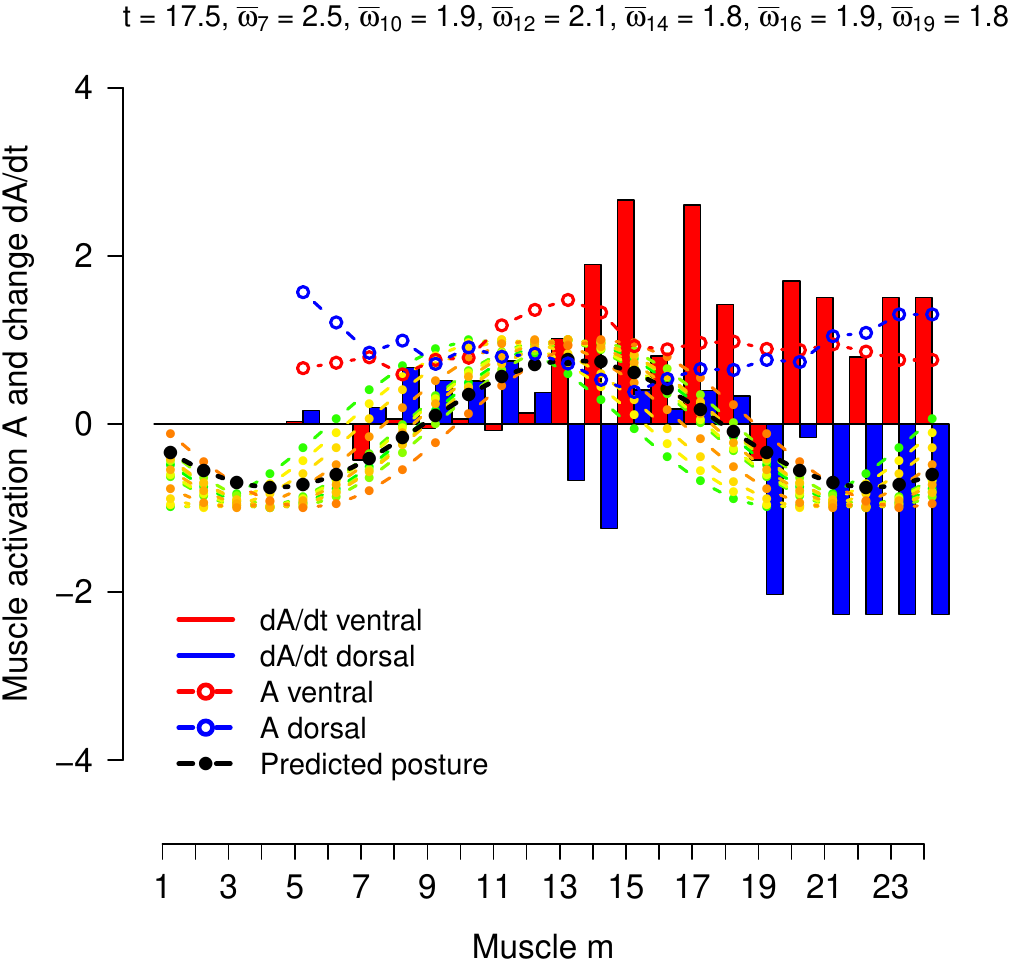}
	  \caption{Time $t = 17.5$}
	\end{subfigure}
	\begin{subfigure}{.32\textwidth}
	  \centering
	  \includegraphics[page=22,width=.95\linewidth]{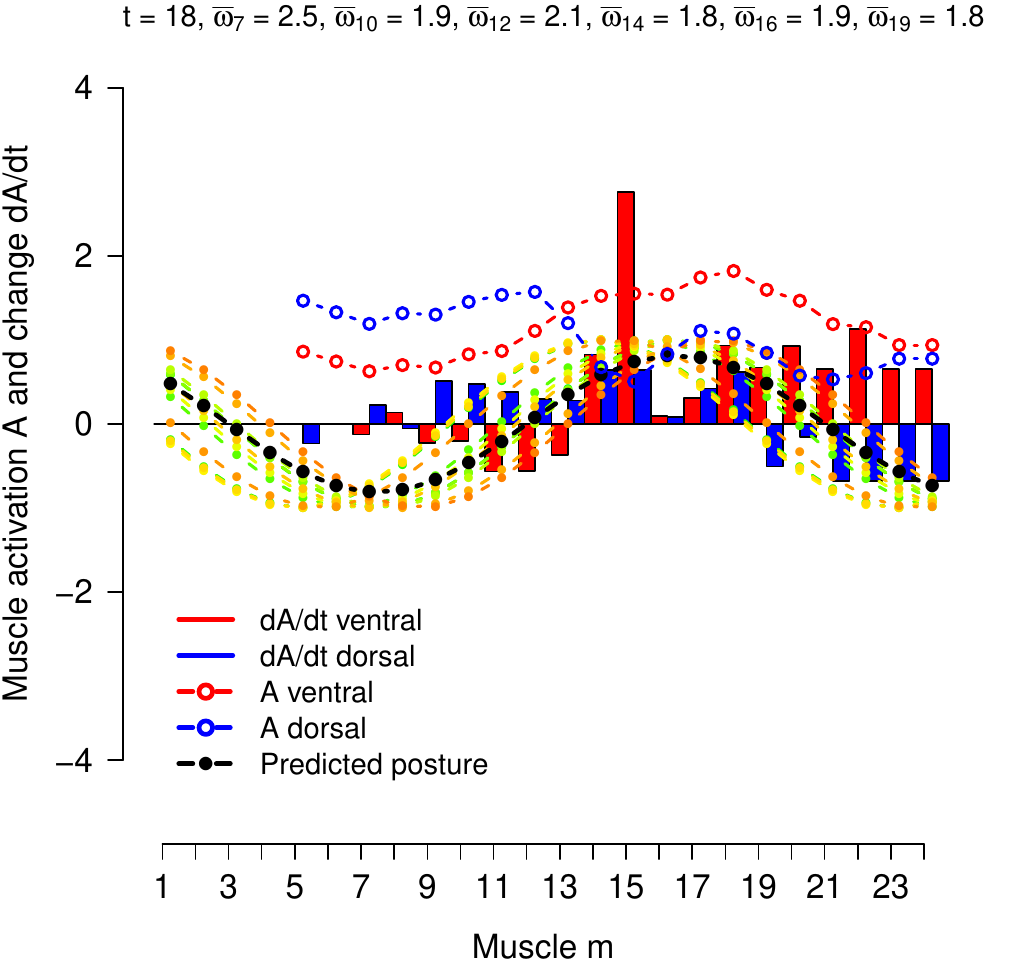}
	  \caption{Time $t = 18$}
	\end{subfigure}
	\begin{subfigure}{.32\textwidth}
	  \centering
	  \includegraphics[page=23,width=.95\linewidth]{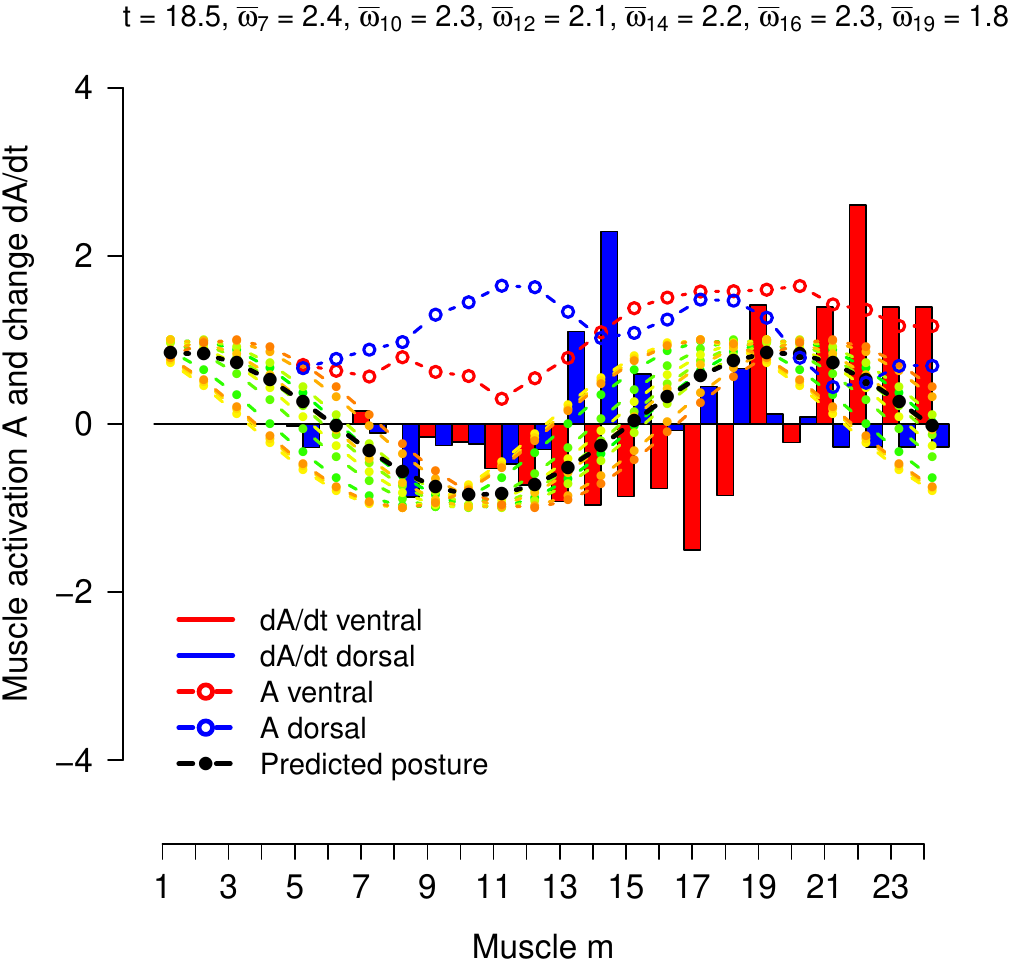}
	  \caption{Time $t = 18.5$}
	\end{subfigure}
	\caption[Forward locomotion of \elegans.]{\keyword{Forward locomotion of \elegans\ with time-delayed feedback control.} The posture of \elegans\ can be described as harmonic wave (black curve) and results from ventral (red curve) and dorsal (blue curve) muscle activity which is simulated with Eq.~\eqref{eq:neuron_dynamics} and \eqref{eq:control_dynamics} (parameters of feedback control are provided in \Cref{tab:time_delays,tab:feedback_strengths}) and plotted over the muscle index $m$.
	Based on this, the individual waves $f_m(x)$, $m=5,7,8,\cdots,24$, $x \in [1,24]$ are calculated [cf. Eq.~\eqref{eq:body_dynamics}] and averaged. Besides the black wave, 12 muscular waves with index $m = 5,7,9-14,16,17,19,20$ are shown in green, orange, and yellowish color, which can be interpreted as its fluctuations.
	Over time, the black body wave slowly propagates to the right meaning that the worm moves to the left.
	In addition to the muscle activation, its temporal change is indicated by bars.
	On top of the diagrams, chosen mean angular frequencies of the harmonic wave model are given. The closer these are together, the better the coordinated locomotion of \elegans.
	Note that the muscle activity is scaled and smoothed for illustrative purposes, but the calculation of the harmonic waves is performed before and is independent of this.}
\label{fig:forward_locomotion}
\end{figure*}%

\keyword{Muscular activity} - The neuronal harmonic oscillations generated in the locomotory circuitry are transferred to body wall muscles for locomotion. For a coordinated locomotion of \elegans, an anti-phase muscular activity is required between dorsal and ventral muscle cells. However, this is not the case for most muscle pairs in our simulation.
The harmonic waves \eqref{eq:body_dynamics} -- adjusted on the basis of muscular activity -- diverge significantly. Later, we will see that the problem can be largely solved by using time-delayed feedback control. 

Figure \ref{fig:forward_locomotion} captures the dynamics of the muscles at different times. 
The dorsal (ventral) muscle activation $A$ is represented by the dotted blue (red) curve and its temporal change $\dot{A}$ by the blue (red) bars. 
Since the locomotory circuitry (\Cref{fig:locomotory_circuitry}) mainly covers the forward and backward locomotion of \elegans, some of the muscles in the head and neck are missing. This concerns the dorsal and ventral muscles $1-4$ and $6$.  For the latter, the mean activation of the muscles $5$ and $7$ is calculated.
Note that the displayed muscle activation $A$ does not reflect the actual simulated values.
For better visibility, the time series are first normalized to values between $0$ and $2$ in the time interval $[8,20]$ and then smoothed with a simple second order moving average in order to avoid a serrated appearance.
However, the modelling of forward locomotion is independent of this and done beforehand.
The black curve of the body wave results from the average over all $19$ muscular waves $f_m(x)$, $m=5,7,8,\cdots,24$, $x \in [1,24]$, calculated with Eq.~\eqref{eq:body_dynamics} and constitutes the final posture of the worm.
In addition to the averaged body wave, $12$ individual muscular waves are shown in the graphics in green, orange, and yellowish color, which can be interpreted as its fluctuation. 
For some of them, the mean angular frequency is given at the top of the diagrams indicating a coordinated locomotion if they are close together.
The body wave propagates to the right in time implying that the worm moves forward to the left.


\section{Synchronicity of motion}
\label{sec:synchronocity}

\begin{figure}[H] 
\centering
	    \centering
	    \begin{subfigure}{.235\textwidth}
            \centering
            \includegraphics[width=.99\linewidth]{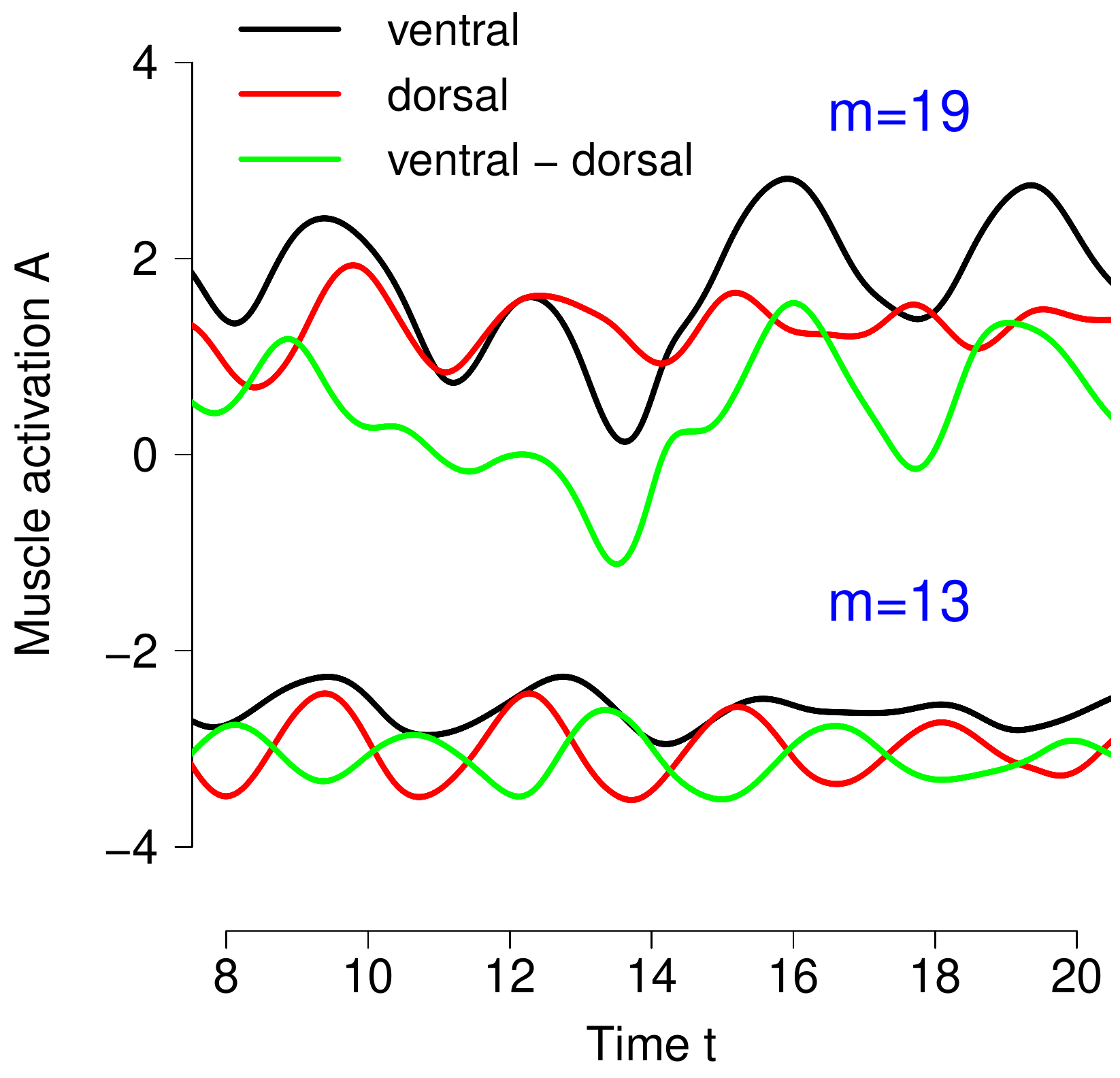}  
            \caption{Uncontrolled}
        \end{subfigure}
        \begin{subfigure}{.235\textwidth}
            \centering
            \includegraphics[width=.99\linewidth]{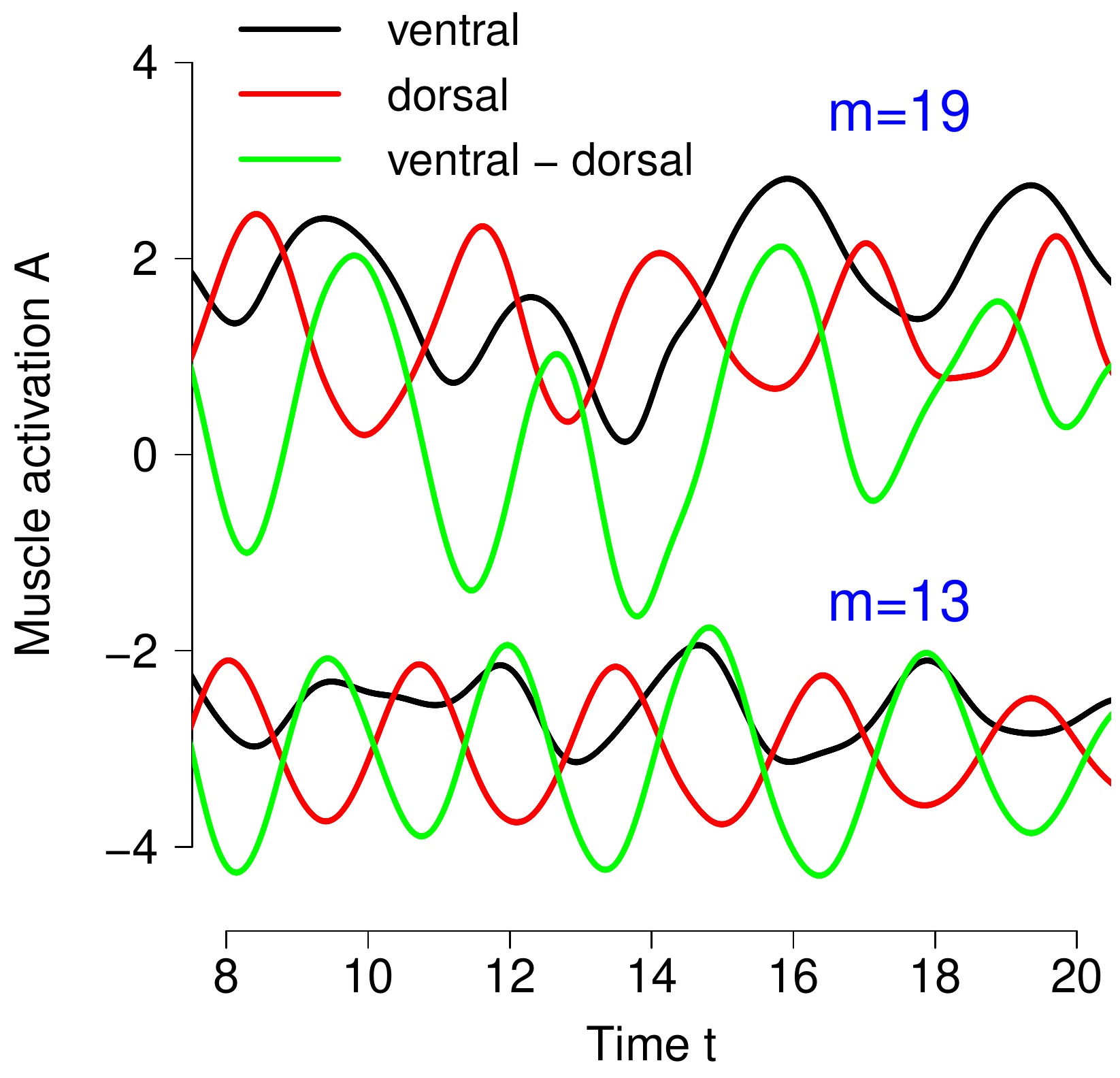}  
            \caption{With feedback control}
        \end{subfigure}
        \caption[Effect of time-delayed feedback control on muscular activity.]{\keyword{Effect of time-delayed feedback control on muscular activity for two muscle pairs.} Panel (a) and (b) correspond to the uncontrolled [simulated with Eq.~\eqref{eq:neuron_dynamics} and \eqref{eq:muscle_dynamics}] and controlled case [simulated with Eq.~\eqref{eq:neuron_dynamics} and \eqref{eq:control_dynamics} - additional parameters are provided in \Cref{tab:time_delays,tab:feedback_strengths}]. The black (red) curve refer to the ventral (dorsal) muscle and the green curve shows their difference. The time series for the muscle pair $m = 13$ do not reflect the actual muscle activation but are shifted downwards from the original level.}
        \label{fig:feedback_control}
\end{figure}

Based on the simulations of muscular activity \eqref{eq:muscle_dynamics}, the harmonic waves \eqref{eq:body_dynamics} are not well synchronized. This corresponds to an uncoordinated locomotion of \elegans.
In order to enhance the synchronicity of harmonic waves, we consider a feedback control scheme.
Subsequently, synchronicity is quantified over relevant simulation time using the Kuramoto order parameter as introduced in later Section~\ref{subsec:kuramoto_order_parameter}.
This parameter then forms the reference for further simulations in which certain neurons of the locomotory circuitry are silenced.
As a consequence, conclusions can be drawn about the importance of neurons for a coordinated locomotion of \elegans. 

\subsection{Time-delayed feedback control} 
\label{subsec:feedback_control}

In \Cref{fig:neuron_timeseries}, neuronal oscillations are shifted to each other. This can be interpreted as the existence of different time delays between the neurons.
The oscillations should be transferred to muscle cells via neuromuscular synapses in such a way that there is an anti-phase activation between ventral and dorsal muscles. 
Only then, a smooth and coordinated locomotion of \elegans\ is possible.
However, the neuromuscular connections are not perfect but partly based on estimates. This concerns in particular the given numbers of synapses \cite{wormatlas.2020}.
Therefore, anti-phase behavior is not observed and a coordinated locomotion of the worm is not given.
However, the coordination of locomotion can be improved with time-delayed feedback control 
which is a general powerful control method in nonlinear systems \cite{Pyragas.1992,Scholl.2007,Scholl.2016}. Such delayed feedback mechanisms are often present in neuronal systems due to intrinsic propagation and processing delays.

For this purpose, the muscle dynamics is remodeled as 
\begin{flalign}\label{eq:control_dynamics}
\hspace{1mm}
\begin{split}
\dot{A}_{l}(t) ={}&H\left(A_l(t),p_s(t)\right)\\
& -\frac{K_l}{\eta}\biggl[g_{\text{ACh}}\sum\limits_{s=1}^{N}{E}_{\text{ACh}, sl}\bigl(p_{s}(t) - p_{s}(t-\tau_s)\bigr)\biggr.\\
& \; \biggl. - g_{\text{GABA}}\sum\limits_{s=1}^{N}{E}_{\text{GABA}, sl}\bigl(p_{s}(t) - p_{s}(t-\tau_s)\bigr)\biggr]
\end{split} &&
\end{flalign}
where function $H$ corresponds to the right side of Eq.~\eqref{eq:muscle_dynamics}.
The second term represents the control force that entails two new components: (i) the time delay $\tau_{s}$ for the $N$ neurons and (ii) the feedback strength $K_l$ for the $M$ muscles.
One peculiarity is that the control itself is delayed. It takes place after the integration of the dynamical system and is based on the comparison of synchronization between different harmonic waves.

With feedback control, the muscle activation $A$ is simulated via Eq.~\eqref{eq:control_dynamics}. From the resulting time series, the harmonic waves are computed using Eq.~\eqref{eq:body_dynamics}. The latter are then gradually calibrated to a chosen set of reference waves (cf. \Cref{fig:forward_locomotion}). 
For each harmonic wave to be calibrated, all combinations of time delay and feedback strength resulting from the intervals $[0.25,5]$ and $[0.25,3]$ in steps of $0.25$ are simulated.
Subsequently, the propagation of the time-delayed waves is successively compared with the reference waves in the time interval $[8,20]$.
In total, $35$ motor neurons and $51$ muscle cells are affected by the control force. The estimated parameters are provided in \Cref{tab:time_delays,tab:feedback_strengths}. 

Figure~\ref{fig:feedback_control} illustrates the effect of time-delayed feedback control for the activation of muscles with index $m = 13$ and $m = 19$.
The dorsal and ventral muscle activity exhibit most of the time a rather in-phase behavior. By activating the control, the oscillations of the dorsal and ventral muscles are more regular and almost in anti-phase.

\begin{figure}[H] 
	\centering
	\begin{subfigure}{.21\textwidth}
            \centering
            \includegraphics[width=.99\linewidth,height=4cm]{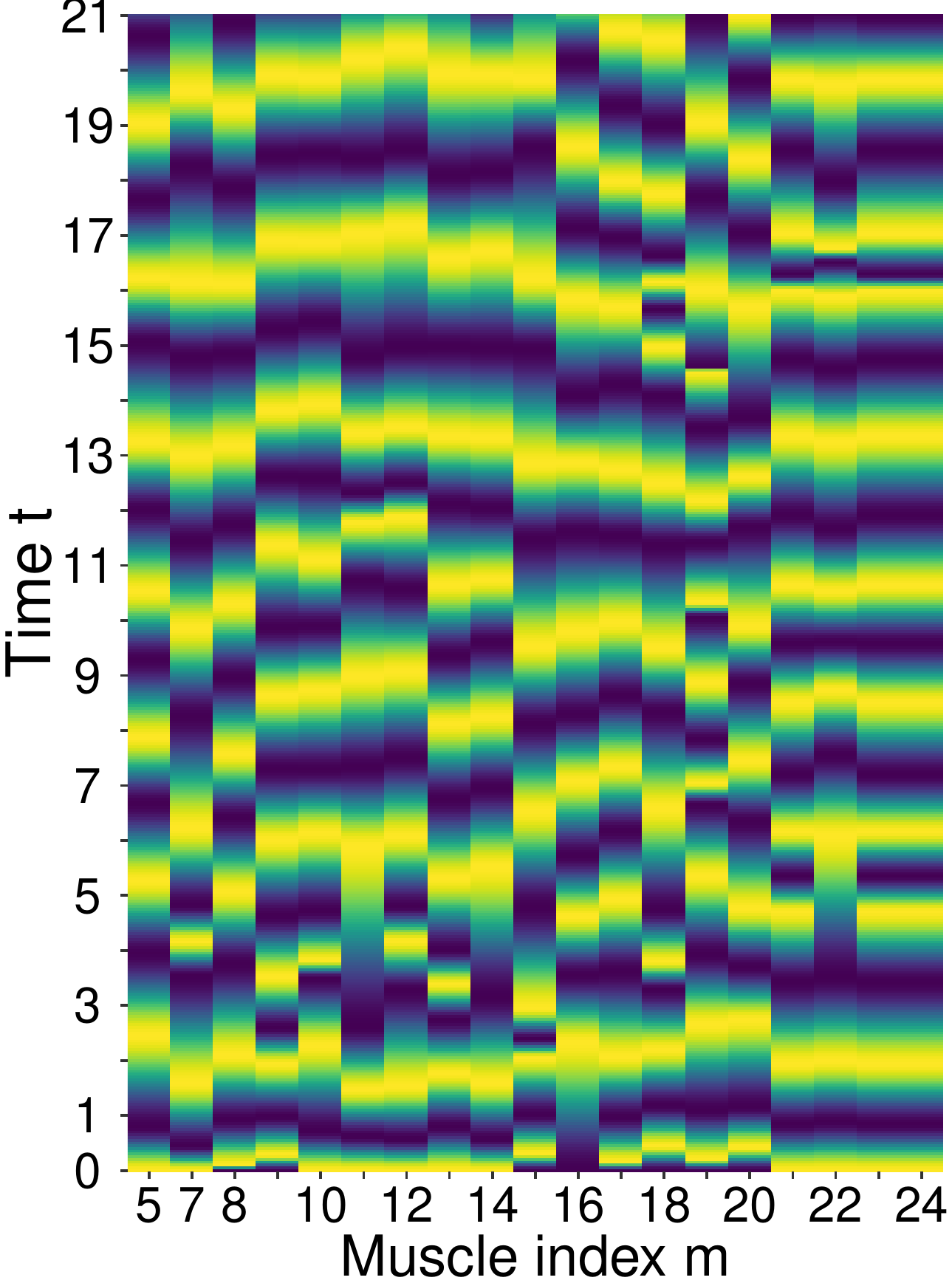}  
            \caption{Uncontrolled}
        \end{subfigure}
        \begin{subfigure}{.26\textwidth}
            \centering
            \includegraphics[width=.99\linewidth,height=4cm]{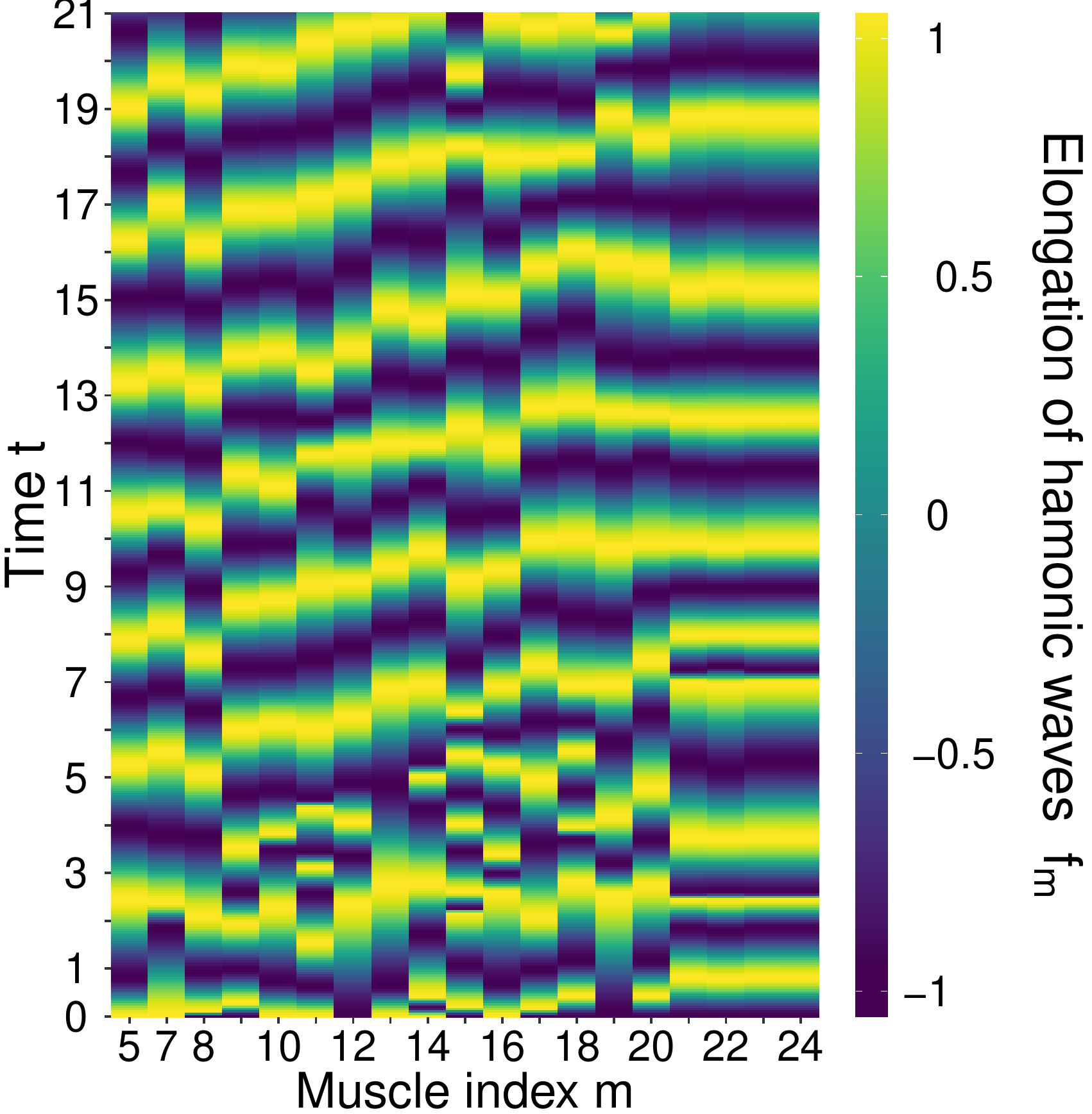}
            \caption{With feedback control}
        \end{subfigure}
	\caption[Synchronization of harmonic waves with time-delayed feedback control.]{\keyword{Synchronization of harmonic waves with time-delayed feedback control.} The elongation $f_m(x=m)$ is plotted over time $t$ for each muscular wave with index $m=5,7,8,\cdots,24$ in the absence [simulated with Eq.~\eqref{eq:neuron_dynamics} and \eqref{eq:muscle_dynamics}] and presence [simulated with Eq.~\eqref{eq:neuron_dynamics} and \eqref{eq:control_dynamics} - additional parameters are provided in \Cref{tab:time_delays,tab:feedback_strengths}] of time-delayed feedback control in panels (a) and (b), respectively. Without feedback control, no pattern of locomotion can be detected. With feedback control, the amplitude generated in the anterior body propagates almost linearly in time to posterior muscles which makes it a phase-shifted synchronicity. The more apparent the linear behavior, the better is the coordination of forward locomotion.
	For the muscles in the tail, the waves spread rather constantly. This is due to the fact that the same muscular connections are assumed for them \cite{wormatlas.2020}. }
	\label{fig:sync_harmonic_waves}
\end{figure}

In addition, all relevant 19 harmonic waves are visualized in \Cref{fig:sync_harmonic_waves} in the absence and presence of feedback control.
The harmonic wave function $f_m(x=m)$ is plotted for each ventral and dorsal muscle pair with index $m=5, 7, 8, $ $\cdots, 24$ up to time $t = 21$.
Without feedback control, no synchronization pattern can be identified.
With feedback control, the amplitude generated in the anterior body indicated by yellow or dark blue color propagates almost linearly in time to posterior muscles in the considered time interval $[8,20]$. 
Since the anterior waves show similar behavior with a slight time lag, this is a phase-shifted synchronization.
A coordinated locomotion behavior of \elegans\ becomes more probable, the more clearly the linearity can be seen in the synchronization plot, resembling a travelling wave.
This is not the case for the muscles in the tail of the worm by design of the network. Actual high-power electron microscopes are not able to cover the neuromuscular connections in this region \cite{Cook.2019}. Therefore, the utilized data for neuromuscular connectivity only refer to estimated connections between motor neurons and muscle cells which are all the same in the tail \cite{wormatlas.2020}. 
Figure~\ref{fig:forward_locomotion} also illustrates this very clearly. The temporal changes of muscle activation $\dot{A}$ indicated as the corresponding bars display nearly identical behavior for the last muscles.


\subsection{Kuramoto order parameter} 
\label{subsec:kuramoto_order_parameter}

In order to measure the synchronization between harmonic waves, we define the Kuramoto order parameter $R$ as 
\begin{align}\label{eq:order_parameter}
R(t) &= \dfrac{1}{W}\left|\sum\limits_{m=5,7-24}\exp\left[i\gamma\left(x=0,\Delta t_{r_m},t\right)\right]\right| \;. 
\end{align} 
The order parameter represents the phase coherence of $W=19$ waves. As explained earlier, the muscle pairs with index $m=1-4,6$ are not considered. The phase $\gamma(x,\Delta t_{r_m},t)$ is equal to the argument of the sine function in Eq.~\eqref{eq:body_dynamics}. Since the phase differences between the muscular waves are the same for each position on the x-axis, the position is fixed and can be set to any desired value. Here, we use $x = 0$ without loss of generality. 
$R=1$ corresponds to full in-phase sychronization and will be denoted by $100\%$, and $R=0$ corresponds to complete desynchronization.
\begin{figure*}[htbp] 
\centering
\begin{minipage}{.98\textwidth}
	\centering
	    \begin{minipage}{.49\textwidth}
                \centering
	            \includegraphics[width=0.6\linewidth]{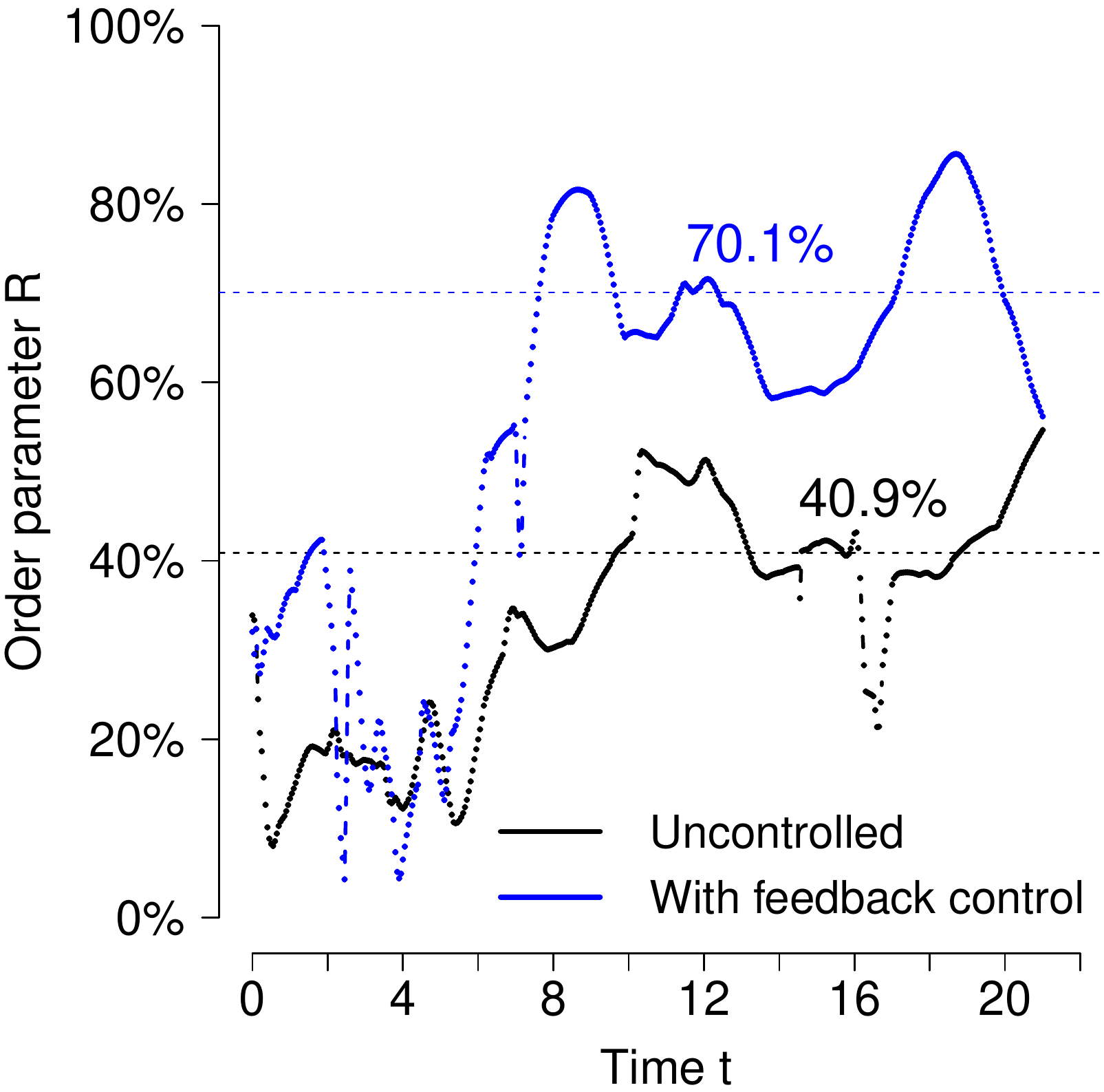}    
	        \captionof{figure}[Synchronicity of muscular waves during forward locomotion of \elegans.]{\keyword{Synchronicity of muscular waves during forward locomotion of \elegans.} The black and blue curve correspond to the Kuramoto order parameter given by Eq.~\eqref{eq:order_parameter} without [simulated with Eq.~\eqref{eq:neuron_dynamics} and \eqref{eq:muscle_dynamics}] and with feedback control [simulated with Eq.~\eqref{eq:neuron_dynamics} and \eqref{eq:control_dynamics} - additional parameters are provided in \Cref{tab:time_delays,tab:feedback_strengths}]. The dotted lines marks the average value in the time interval $[8,20]$.}
	        \label{fig:order_parameter}
        \end{minipage}
        \hfill
        \begin{minipage}{.49\textwidth}
                \centering
                \captionof{table}[Most significant motor neurons for a coordinated locomotion of \elegans.]{\keyword{Most significant motor neurons for a coordinated locomotion of \elegans\ identified by silencing the neuronal activity of singles, doubles, and triples (see \Cref{fig:Deletion_VB_DD,fig:Deletion_IN_AS,fig:Deletion_DA_DB,fig:Deletion_VA_VD}).}} 
                \small
                \begin{tabular}{ll}
                \toprule
                \textbf{Class} & \textbf{Neurons} \\
                \midrule
                DA (9) & DA01-DA05, DA07 \\
                DB (7) & DB01-DB04, DB06, DB07 \\
                DD (6) & DD01, DD03-DD06 \\
                VA (12) & VA04, VA08, VA10, VA12 \\
                VB (11) & VB02, VB06-VB09, VB11 \\
                VD (13) & VD02, VD09, VD10, VD12, VD13 \\
                \bottomrule
                \end{tabular}%
                \label{tab:Deletion_summary}%
        \end{minipage}
	\end{minipage}
\begin{minipage}{.98\textwidth}
\vspace{+3mm}
	\centering
	  	\begin{subfigure}{0.49\textwidth}
	  \centering
	  \includegraphics[width=1.0\linewidth]{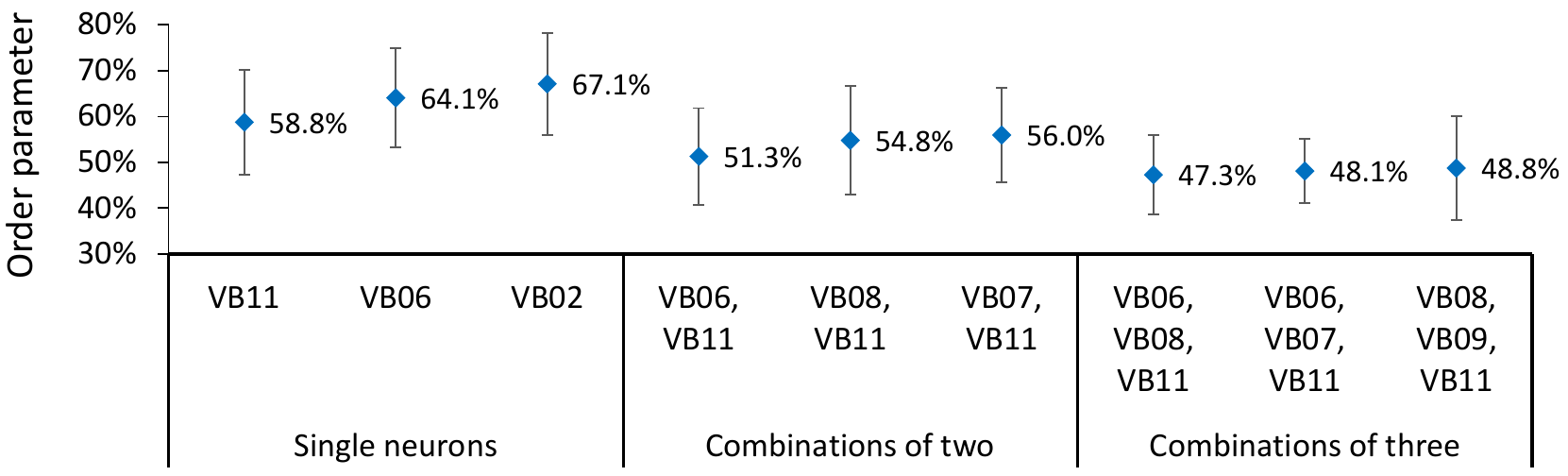}    
	  \caption{Silencing activity of VB neurons}
	\end{subfigure}
	\begin{subfigure}{0.49\textwidth}
	  \centering
	  \includegraphics[width=1.0\linewidth]{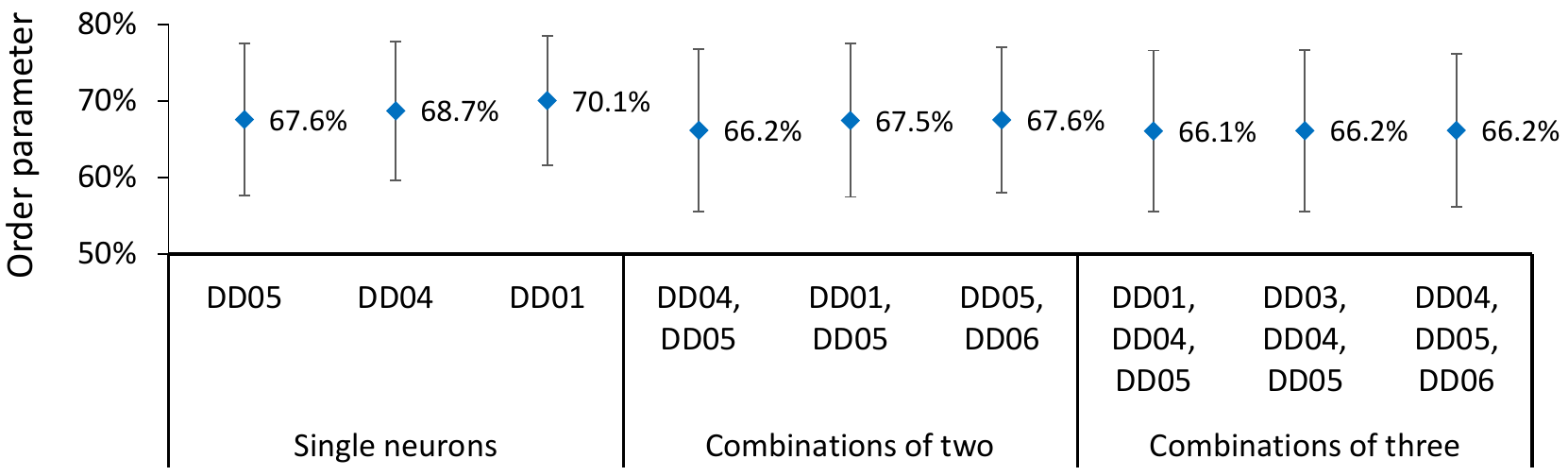}    
	  \caption{Silencing activity of DD neurons}
	\end{subfigure}
	\caption[Most significant VB and DD motor neurons for synchronicity of muscular waves.]{\keyword{Most significant VB and DD motor neurons for synchronicity of muscular waves.} Without silencing of neuronal activity, the reference value of the time-averaged order parameter is $70.1\%$.}
    \label{fig:Deletion_VB_DD}
	\end{minipage}
\end{figure*}%

Based on the simulations of the harmonic waves, the order parameter (\ref{eq:order_parameter}) is calculated in time steps of $0.05$ and considered in the time interval $[8,20]$.
The results are graphically displayed in \Cref{fig:order_parameter}.
The synchronicity of the $19$ muscular waves expressed in the order parameter fluctuates over time around its average value.
With time-delayed feedback control, the time-averaged order parameter is about $70.1\%$. 
This corresponds to an improvement of $29.2\%$ over the uncontrolled system.
Even if the time-averaged order parameter of $40.9\%$ for the uncontrolled system may indicate that there is some partial synchronization between the muscular waves, it is still at a level that does not facilitate a coordinated forward locomotion of \elegans\ (see \Cref{fig:sync_harmonic_waves}). 
The time-averaged order parameter of the controlled system serves as reference for simulations where neuronal activity is silenced.


\subsection{Silencing of neuronal activity}
\label{sec:Synchronicity_in_absence_of_neurons}

The starting point is the nearly synchronous propagation of muscular waves in the presence of feedback control.
If the activity of neurons is silenced, the synchronicity of locomotion of \elegans\ is disturbed.
This disturbance is considered in the time interval $[8,20]$ and can be quantified by the time-averaged order parameter $\left\langle R(t) \right\rangle_t$. 
Note that the time interval $[0,8]$ displays transient effects and is therefore disregarded.
The neurons that cause the strongest decline of the time-averaged order parameter are the most significant.
All neurons within relevant classes are successively silenced individually, in combinations of two, and combinations of three. 
Our silencing strategy is to hold the membrane potential at the constant value zero throughout the simulation.
For the different selections, the top three results with the lowest time-averaged order parameter $\pm$ its standard deviation are considered.  

Figure \ref{fig:Deletion_VB_DD} displays the results for the classes VB and DD. 
While silenced activity of VB neurons reduces the order parameter most, it causes the smallest changes for DD neurons.
If neurons within the VB class are silenced individually, VB11, VB06, and VB02 have the strongest impact on the coordinated locomotion of \elegans. 
In combination with these, the neurons VB07-VB09 become also relevant. The combined silencing of VB06, VB08, and VB11 leads to a significant decrease of the order parameter by more than $10\%$ compared to the individual silencing.
Therefore, the locomotion of the worm is highly uncoordinated.
Within the DD class, the individual silencing of DD05, DD04, and DD01 shows a small to minor impact on the order parameter which can only be marginally enhanced by the combined silencing of neurons.
For example, silencing DD05 individually leads to an order parameter of $67.6\%$, but silencing DD05 together with DD01 and DD04 only results in a reduction by $1.5\%$.
Although the effect may seem very weak, the individual silencing of neuronal activity regarding DD04 or DD05 is already sufficient to impair the \elegans\ locomotion \cite{Yan.2017}.
In this case, the combined silencing of neuronal activity hardly provides additional information.
The results for the other classes can be found in \Cref{fig:Deletion_IN_AS,fig:Deletion_DA_DB,fig:Deletion_VA_VD}.
\Cref{tab:Deletion_summary} summarizes the identified neurons in the top three single, double, and triple selection.
 

\section{Discussion \& conclusions}
\label{sec:discussion}

We have modeled the somatic nervous system of \textit{C. elegans} as a multilayer network whose nodes comprise both neurons and muscle cells.
The different layers are defined by different types of interactions between the nodes and include important neurotransmitters and neuropeptides.
This physiological approach allows for a better understanding of the network.
On the one hand, it enables to conduct a logistic regression analysis in order to predict neurons involved in locomotion behavior.
On the other hand, it allows one to study the dynamics of complex circuits like the locomotory circuitry which initializes forward and backward locomotion of \elegans. 

Logistic regression analysis can be operationalized in the multilayer network by considering the shortest paths between touch sensory neurons and body wall muscles.
The starting point is the identification of key factors for the neurons in the shortest paths. This is based on the measurements of the discriminative power.
Subsequently, logistic regression models can be built on them.
As a result, we have explored various models with only two factors that predict the shortest paths with excellent accuracy.
While the best model correctly predicts $99.2\%$ of the shortest paths on the development dataset, the second best model achieves about $96.9\%$.
Both models correctly classify all the shortest paths that only consist of neurons involved in locomotion behavior. A misclassification only occurs in pathways that also involve other neurons.
In contrast to the development dataset, the shortest paths on the test dataset are not limited to 5 sensory neurons but consider all sensory neurons in the network.
Based on the test dataset, our model predicts $29$ additional neurons involved in locomotion behavior and therefore extends the initial selection. 
For $26$ of them, it can be confirmed by literature research and connectivity analysis that these are indeed involved in locomotion behavior.
For this reason, the models are well suited to predict the locomotory subnetwork.
Moreover, the achieved results suggest that this method can also be used to predict other subnetworks.
Since different subnetworks are characterized by their structure along with the underlying neuron and transmitter types, there is a good chance to identify them using logistic regression.
This applies in particular for \elegans\ since all neurons and their connectivity are already known. Even if the information about neurons and connectivity is incomplete, logistic regression analysis is not necessarily useless as long as there is a tendency in the data.
The procedure proposed in this study can applied to identify circuits in the brain that control the movement of certain parts of the body since the information processing is the same in humans and \elegans.
However, the application of this method is limited by the fact that sufficient information about the neurons must already be gathered. In conclusion, logistic regression analysis can be seen as a reasonable complement to cell ablation experiments.

Concerning the dynamics, we have shown that the forward locomotion of \elegans\ can be described using a harmonic wave model.
The particularity of our model is that harmonic waves are adjusted for each pair of muscles consisting of a ventral and a dorsal muscle.
This seems to be reasonable since the worms generate harmonic oscillations that spread from the head to posterior.
The average value over the harmonic waves then represents the body posture of \elegans\ while moving.
The neuronal basis is the locomotory circuitry which generates the rhythmicity of locomotion based on the three-dimensional Hindmarsh-Rose system.
Our simulations suggest that the generated sinus rhythm is network-based and results from couplings within the locomotory circuitry and interactions between different neuron types.
Therefore, the circuit can be referred to as central pattern generator. 
Since it initiates both forward and backward locomotion, oscillatory behavior can be observed in all classes of the first layer motor neurons.
In the simulations, these oscillations are very robust against input of touch sensory neurons. With respect to a touch stimulus on the head or tail, only tiny phase shifts of them can be observed. This raises the question whether these tiny changes can already explain triggering forward and backward locomotion.
Importantly, the expected patterns of neuronal activity are correctly reflected by the Hindmarsh-Rose system although it is basically a model for describing action potential behavior.
Since no external current is applied to the neurons of the locomotory circuitry, the simulated dynamics shows pure coupling effects. Note that a current is only considered for the sensory neurons PLML/R which perceive gentle touches on the tail.
The set parameter value causes a fast oscillatory behavior rather than spiking behavior.   
However, this does not play a decisive role as they can only marginally influence the behavior of the other neurons due to the chosen coupling strengths.
A higher current was applied because it had a slightly positive impact on the alignment of the harmonic waves.
As a potential outlook, it would be worthwhile to examine the locomotory circuitry in more detail in order to understand exactly how the coupling must be designed to generate rhythmic oscillations. 
This could also play an important role in the brain. 
For example, the brainstem contains centers that control the rhythm for heartbeat and respiration.
Since both are coupled together, the heartbeat can synchronize with the respiratory rhythm \cite{Perry.2019}.
On the other side, there are neuromechanical models in the literature that can realistically reproduce the forward locomotion of \elegans\ physical body 
\cite{Boyle.2012,Izquierdo.2018}.
However, these include only small parts of the locomotory circuitry which are largely approximated and do not allow for an unrestricted study of the underlying neuronal dynamics.
Together with other studies, those models suggest that proprioception is closely associated with the generation of body rhythms without excluding the coexistence with central pattern generators \cite{Fouad.2018,Fouad.2018b,Gjorgjieva.2014}.
Although the concept of central pattern generator is not well articulated, our findings additionally support the latter and fill the gap.

The generated neuronal harmonic oscillations are subsequently transferred to body wall muscles via neuromuscular synapses.
Regarding the muscular activity, many problems occur that prevent a plausible description of locomotion behavior.
The biggest problem is that the simulated activation of opposite dorsal and ventral muscles is usually not at a comparable level and does not show an anti-phase behavior. In addition, there is only little correlation between activities of the different muscle pairs along the midline of the worm.
One reason for this is probably that the locomotory circuitry may still have missing parts. Logistic regression analysis has revealed that 26 additional neurons could be involved in locomotion behavior.
This is especially true for the SMD and RMD motor neurons which drive dorsoventral undulations in the head and neck of \elegans\ and propagate them posteriorly through stretch-receptor feedback.
Therefore, proprioceptive mechanisms also contribute to the coordination of locomotion \cite{Izquierdo.2018}. 
Moreover, extrasynaptic neurotransmission between head motor neurons may also play an important role in ensuring optimal efficiency of forward locomotion \cite{Shen.2016,KagawaNagamura.2018}.
Apart from that, the SMD and RMD motor neurons are involved in multiple navigation behaviors and can therefore be assigned to a circuit for navigation \cite{Gray.2005}. We have excluded navigation in our analyses because we wanted to examine first whether rhythmicity is generated in locomotory circuitry. Nevertheless, it cannot be excluded that a smooth coordinated locomotion of \elegans\ at least depends on both circuits. The next step would be to analyze the dynamics if they are coupled together.
Another important reason is that the neuromuscular connectivity data is far from perfect. The neuron to muscle connections in the tail of the worm are unknown, so the same connections are assumed for the latter muscles.
Furthermore, the underlying number of synapses is estimated as an average value for many connections \cite{wormatlas.2020}. 
However, this does not properly account for the asymmetric structure of the motor neurons in the locomotory circuitry \cite{Tolstenkov.2018}.
Besides, we have assumed that dorsal and ventral muscles lie in a row so that each dorsal muscle faces a ventral muscle.
In fact, the muscle pairs behind the neck of the worm are not directly opposite to each other but slightly staggered along the main body axis \cite{Altun.2009}.
As a consequence, the simulated muscle dynamics does not seem plausible across all muscles. For this reason, we have tried to improve the quality of the time series with scaling and smoothing efforts for visualization (\Cref{fig:forward_locomotion}).

In terms of the harmonic wave model on top of muscular activity, the aforementioned factors prevent a synchronous propagation of adjusted harmonic waves.
In order to enhance synchronicity, we make use of time-delayed feedback control which is effective in the time interval $[8,20]$.
The overall synchronicity can be quantified with the time-averaged order parameter.
As a result, we were able to increase the synchronicity by $29.2\%$ to $70.1\%$.
Although the initial value of $40.9\%$ for the uncontrolled system may indicate some synchronizations between muscular harmonic waves, these do not contribute to a coordinated locomotion behavior of \elegans. 
The effect of time-delayed feedback control can clearly be seen in the synchronization plots (\Cref{fig:sync_harmonic_waves}).
With the exception of muscle pairs in the tail of the worm, all adjusted harmonic waves propagate nearly synchronized but phase-shifted in time. 
This means that undulations created in the anterior body spread linearly in time towards the tail which describes a coordinated forward locomotion. Without feedback control, such a pattern cannot be detected.
Therefore, the harmonic wave model in combination with time-delayed feedback control solves many problems faced by insufficient quality of connectivity data or considering only parts of the locomotory subnetwork.
The approach also seems to be reasonable since the investigated neuronal activity supports the existence of different time delays for motor neurons.
However, the downside of the model is that the parameterization in terms of complex circuits can prove to be very difficult.

Finally, the modelling of forward locomotion enables us to perform synchronicity analyses from which conclusions can be drawn about the significance of neurons within the locomotory circuitry. 
The synchronization of muscular waves expressed in the time-averaged order parameter can be determined for different simulations where certain neurons are silenced.
This in turn interferes with the coordination of forward locomotion of \elegans. 
The stronger the disturbance, the more important are the silenced neurons.
In all motor neuron classes, we have silenced neurons individually, in combinations of two, and combinations of three by holding their membrane potential at the constant value zero throughout the simulation. 
Note that this strategy still includes coupling effects with other neurons, but it should not weaken the relevance of the analysis. 
Since the neurons of the locomotory circuitry are essentially isopotential with the ability to produce regenerative responses \cite{Lockery.2009}, the constant value of zero may be due to a very high membrane resistance.
A similar effect could be achieved using optogenetics where neuronal activity is manipulated with light in order to study the impact on behavior \cite{Deisseroth.2015,Husson.2013}.
For specifying the most significant neurons, we consider the top three results for each selection and highlight all neurons included.
Drawing a hard line regarding the order parameter does not capture important neurons properly because the measure can take on different levels within the different classes.
In general, silenced VB and VA motor neurons have the strongest influence on the coordination of locomotion. These generate oscillatory behavior for ventral muscles.
The DB and DA motor neurons do the same for dorsal muscles, but they share the role with the AS motor neurons which may be one reason that the effect is slightly lower.
In comparison, silencing the activity of DD and VD motor neurons leads to the smallest decrease of the order parameter which changes only marginally among the different selections singles, doubles, and triples.
This could be due to the fact that those neurons generally exhibit a rather constant behavior (\Cref{fig:neuron_timeseries}). 
Nevertheless, the individual silencing of their neuronal activity can already be sufficient to impair the locomotion of \elegans\ \cite{Yan.2017}.
Moreover, we have examined the interneurons in the locomotory circuitry in the same way. As a result, all interneuron classes are important for a coordinated locomotion behavior. 
Beyond that, similar statements can be made by applying network control principles to the \elegans\ connectivity data \cite{Yan.2017}.
Since our investigations are not only based on neuronal and muscular connectivity but also include the underlying dynamics of neurons and muscle cells, our findings should provide a good indication of significant neurons in the locomotory circuitry.
However, it must be pointed out that these are reserved for the implemented parametrizations of the utilized models.
The better the harmonic waves can be synchronized in advance, the more meaningful such results will be.

In summary, we have shown how a specific circuit of locomotion can be identified in the multilayer network of \elegans\ using logistic regression.
Furthermore, we have introduced various dynamical models and physical methods in order to understand the underlying patterns of neuronal and muscular activity during forward locomotion.
This provides a good basis for explaining further circuits and understanding how these can be coupled with each other.

\section*{Acknowledgements}
This work was supported by Deutsche Forschungsgemeinschaft (DFG) in the framework of Collaborative Research Center 910. PH acknowledges further support by DFG under grant number under grant no. HO4695/3-1. JR acknowledges support by the German Academic Exchange Service (DAAD) and by the National Agency for Research and Development (ANID): Scholarship Program DAAD/BECAS Chile, 2016 (57221134).



\pagebreak


\clearpage
\appendix
\renewcommand\thefigure{\thesection.\arabic{figure}}  
\setcounter{figure}{0}   
\renewcommand\thetable{\thesection.\arabic{table}}  
\setcounter{table}{0}   
\renewcommand\theequation{\thesection.\arabic{equation}}  
\setcounter{equation}{0}   

\section{Data description and data preparation}
\label{sec:Data description and data preparation}

\subsection{Neuronal and muscular connectivity}
\label{subsec:Neuronal connectivity}

\keyword{Neuronal connectivity} - The nervous system of the adult \elegans\ hermaphrodite contains $302$ neurons and can be divided into two nearly completely isolated systems: a large somatic nervous system ($282$ neurons) and a small pharyngeal nervous system ($20$ neurons).
The wiring diagram of the somatic nervous system can be found on \cite{wormatlas.2020} and was provided by
\cite{Varshney.2011}.

In the dataset, the numbers of electrical and chemical synapses are specified for each neuron pair. The electrical connections are labeled "EJ". For chemical synapses, the type of synapse must be taken into account:
\begin{itemize}
\item \keyword{Monoadic}: Send ("S") -- Neuron 1 is presynaptic to Neuron 2.
\item \keyword{Polyadic}: Send poly ("Sp") -- Neuron 1 is presynaptic to more than one postsynaptic partner. Neuron 2 is just one of these postsynaptic neurons.
\end{itemize}
Nearly two thirds of the synapses are polyadic, but not all polyadic synapses have been faithfully marked as such.
In this study, the type of synapse is not of interest. To get rid of the synapse type, the number of synapses must be aggregated (summed) over identical neuron pairs in the selection "Send" plus "Send poly".
In total, the wiring diagram consists of $2,194$ unidirectional chemical connections and $514$ bidirectional electrical connections. 
Since the nervous system is modeled as a directed network, the number of electrical connections doubles to $1,028$. The numbers of chemical and electrical synapses (gap junctions) are $6,394$ and $1,774$ in total (\Cref{tab:neuron_muscular_connectivity}). Note that the connections are made by 279 neurons. The neurons CANL, CANR, and VC06 are excluded since they do not have connections with other neurons.

\begin{table}[h]
  \centering
  \caption[Data basis connectivity data.]{\keyword{Data basis connectivity data.} The somatic nervous system of \elegans\ is modeled as a directed network whose nodes represent neurons and body wall muscles connected by chemical synapses, gap junctions, and neuromuscular junctions. The multilayer network in \Cref{fig:multilayer_network} has a total of $3,538$ distinct connections because chemical and electrical connections overlap.}
 	\small
    \begin{tabular}{rrr}
    \toprule
    \multicolumn{1}{p{6.5em}}{\textbf{Connection}} & \multicolumn{1}{l}{\textbf{Number of}} & \multicolumn{1}{l}{\textbf{Number of}} \\
    \multicolumn{1}{p{6.5em}}{\textbf{type}} & \multicolumn{1}{l}{\textbf{connections}} & \multicolumn{1}{l}{\textbf{synapses}} \\
    \midrule
    \multicolumn{1}{l}{chemical} & 2,194 & 6,394 \\
    \multicolumn{1}{l}{electrical} & 1,028 & 1,774 \\
    \multicolumn{1}{l}{muscular} & 548   & 1,791 \\
    \midrule
    \textbf{Total} & \textbf{3,770} & \textbf{9,959} \\
    \bottomrule
    \end{tabular}%
  \label{tab:neuron_muscular_connectivity}%
\end{table}

\keyword{Neuromuscular connectivity} - If the worm is cut from above (dorsally) along the midline and splayed out laterally, four muscle quadrants can be identified: from left to right these are the quadrants \emph{Dorsal Left}, \emph{Ventral Left}, \emph{Ventral Right}, and \emph{Dorsal Right}. Each quadrant contains $24$ muscle cells with the exception of the ventral left quadrant which contains $23$ cells. This results in a total of $95$ muscle cells.
Within each quadrant, the muscles lie in double rows and are numbered from head to tail. 
The first $16$ muscles across the four quadrants belong to the head, the next $16$ to the neck, and the rest of them to the body. While the muscles lie side by side in the head, they are slightly staggered along the main body axis in the remaining regions (\cite{Altun.2009}, \cite[Section III]{Moerman.1997}). 
Neuron to muscle connections are also available on \cite{wormatlas.2020} and are based on the works of \cite{White.1986,Dixon.2005,Varshney.2011}.
It is assumed that body wall muscles are activated by motor neurons. In total, $552$ motor neurons make connections to body wall muscles. The connections of the neurons CEPVL, CEPVR, ADEL, and AVKR are not considered because they do not function as motor neurons. 
The total number of neuromuscular synapses is $1,791$ (\Cref{tab:neuron_muscular_connectivity}).
Note that a large proportion of neuromuscular synapses is estimated only and that the neuron to muscle connections in the tail of the worm are based only on assumptions which are the same for all muscles \cite{wormatlas.2020}.

\subsection{Neuron functions}
\label{subsec:Neuron functions}

\begin{table}[htbp]
  \centering
  \caption[Neuron functions in \elegans.]{\keyword{Neuron functions in \elegans.} The somatic nervous system contains $282$ neurons in total which can be can be classified into sensory neurons, interneurons, and motor neurons.}
  	\small
    \begin{tabular}{ccrr}
    \toprule
    \multicolumn{1}{l}{\textbf{Neurons that}} & \multicolumn{1}{l}{\multirow{2}[2]{*}{\textbf{Number}}} & \multicolumn{1}{l}{\textbf{also}} & \multicolumn{1}{l}{\multirow{2}[2]{*}{\textbf{Number}}} \\
    \multicolumn{1}{l}{\textbf{function as}} &       & \multicolumn{1}{l}{\textbf{function as}} &  \\
    \midrule
    \multicolumn{1}{l}{\multirow{3}[2]{*}{Sensory neurons}} & \multicolumn{1}{r}{\multirow{3}[2]{*}{82}} & \multicolumn{1}{l}{-} & 73 \\
          &       & \multicolumn{1}{l}{Interneurons} & 3 \\
          &       & \multicolumn{1}{l}{Motor neurons} & 6 \\
    \midrule
    \multicolumn{1}{l}{\multirow{3}[2]{*}{Interneurons}} & \multicolumn{1}{r}{\multirow{3}[2]{*}{89}} & \multicolumn{1}{l}{-} & 78 \\
          &       & \multicolumn{1}{l}{Motor neurons} & 7 \\
          &       & \multicolumn{1}{l}{Sensory neurons} & 4 \\
    \midrule
    \multicolumn{1}{l}{\multirow{3}[2]{*}{Motor neurons}} & \multicolumn{1}{r}{\multirow{3}[2]{*}{111}} & \multicolumn{1}{l}{-} & 83 \\
          &       & \multicolumn{1}{l}{Interneurons} & 24 \\
          &       & \multicolumn{1}{l}{Sensory neurons} & 4 \\
    \midrule
          &       & \textbf{Total} & \textbf{282} \\
    \bottomrule
    \end{tabular}%
  \label{tab:neurons_functions}%
\end{table}%

In \elegans, three main types of neurons can be distinguished: sensory neurons, interneurons, and motor neurons. 
This information can be extracted from \cite{Oshio.2003} but is also available on \cite{wormatlas.2020}.
In general, neurons can have multiple functions and combine the properties of all three main types.
The dataset contains up to two functions whereby the first specified function is assumed to be the main function.
There are a total of $82$ sensory neurons, $89$ interneurons, and $111$ motor neurons in the somatic nervous system of \elegans. Six of the sensory neurons and seven of the interneurons also operate as motor neurons resulting in a total of $124$ neurons with the function of a motor neuron (\Cref{tab:neurons_functions}). 
We changed the function of the neurons PVDL and PVDR from interneuron to sensory neuron. For the neurons RIML and RIMR, the function as interneuron is additionally assigned to the function as motor neuron \cite{wormatlas.2020}.

\subsection{Neurotransmitter and neuroreceptor data}
\label{subsec:Neurotransmitter and neuroreceptor nata}

By considering actual transmitter and receptor data collected for the individual neurons, the chemical layer in the network can be divided into several sublayers. \Cref{tab:trans_recep} provides all transmitters and receptors utilized in this study.
%
\begin{table*}[htbp]
  \centering
\begin{threeparttable}
  \caption[Neurotransmitter and neuroreceptor data.]{\keyword{Neurotransmitter and neuroreceptor data.} For each transmitter type, the corresponding receptors are specified.}
  	\small
    \begin{tabular}{lllll}
    \toprule
    \textbf{Class} & \textbf{Family} & \textbf{Type} & \textbf{Matching receptors} & \textbf{Source} \\
    \midrule
        &     &     & dop-1, dop-2, dop-3, &  \\
        &     & Dopamine & dop-4, dop-5, dop-6, &  \\
        &     &     & lgc-53 &  \\
\cmidrule{3-4}        & Mono- & Octopamine & octr-1, ser-3, ser-6 & Bentley et al. \\
\cmidrule{3-4}        & amines & \multirow{2}[2]{*}{Serotonin} & mod-1, ser-1, ser-4, & (2016) \\
        &     &     & ser-5, ser-7 &  \\
\cmidrule{3-4}        &     & \multirow{2}[2]{*}{Tyramine} & lgc-55, ser-2, tyra-2, &  \\
        &     &     & tyra-3 &  \\
\cmidrule{2-5}        &     &     & acc-1, acc-2, acc-4, &  \\
        &     &     & acr-12, acr-14, acr-15, &  \\
        &     &     & acr-16, acr-18, acr-2, &  \\
    \multirow{2}[0]{*}{Transmitters} &     & Acetyl- & acr-23, acr-5, deg-3, &  \\
        &     & choline & des-2, gar-1, gar-2, &  \\
        &     &     & gar-3, lev-8, lgc-12, &  \\
        &     &     & lgc-27, lgc-46, unc-29, &  \\
        & Classical &     & unc-63 & Jorge Ruiz \\
\cmidrule{3-4}        & transmitters & $gamma$- & exp-1, gab-1, gbb-1, & (2017) \\
        &     & Amino- & ggr-1, ggr-2, lgc-35, &  \\
        &     & butyric acid & lgc-37, lgc-38 &  \\
\cmidrule{3-4}        &     &     & avr-15, glc-3, glr-1, &  \\
        &     &     & glr-2, glr-3, glr-4, &  \\
        &     & Glutamate & glr-5, glr-6, glr-8, &  \\
        &     &     & mgl-1, mgl-3, nmr-1, &  \\
        &     &     & nmr-2 &  \\
    \midrule
        &     & flp-1 & npr-11, npr-4 &  \\
        &     & flp-4 & npr-4 &  \\
        &     & flp-5 & npr-11 &  \\
        &     & flp-10 & egl-6 &  \\
        &     & flp-13 & frpr-4 &  \\
        & \multirow{2}[1]{*}{FLPs\tnote{*}} & flp-15 & npr-3 &  \\
        &     & flp-17 & egl-6 &  \\
\cmidrule{3-4}        &     & \multirow{2}[2]{*}{flp-18} & npr-1, npr-11, npr-4, &  \\
    Peptides &     &     & npr-5 & Bentley et al. \\
\cmidrule{3-4}        &     & \multirow{2}[2]{*}{flp-21} & npr-1, npr-11, npr-2, & (2016) \\
        &     &     & npr-5 &  \\
\cmidrule{3-4}        &     & flp-24 & npr-17 &  \\
\cmidrule{2-4}        & \multirow{2}[2]{*}{NLPs\tnote{*}} & nlp-1 & npr-11 &  \\
        &     & nlp-12 & ckr-2 &  \\
\cmidrule{2-4}        & NTCs\tnote{*} & ntc-1 & ntr-1 &  \\
\cmidrule{2-4}        & \multirow{2}[2]{*}{PDFs\tnote{*}} & pdf-1 & pdfr-1 &  \\
        &     & pdf-2 & pdfr-1 &  \\
    \bottomrule
    \end{tabular}%
  \label{tab:trans_recep}%
\begin{tablenotes}\footnotesize
\item[*] FLP -- FMRFamide-like peptide
\item[\ ] NLP -- Neuropeptide-like protein
\item[\ ] NTC -- Nematocin (Oxytocin/Vasopressin-related peptide)
\item[\ ] PDF -- Pigment-dispersing factor
\end{tablenotes}
\end{threeparttable}
\end{table*}%
The data for the MAs can be found by \citet{Bentley.2016}. Hypothetical DA receptors, such as \emph{dop-5} and \emph{dop-6}, are also taken into account to increase the number of connections in the network. 
In total, there are $20$ neurons with MA transmitters and $232$ neurons with MA receptors in network. 
The data for the classical transmitters ACh, Glu, and GABA were collected by ourselves.
There are a total of $218$ neurons with classical transmitters and $210$ neurons with matching receptors.
Another important class of neurochemicals are neuropeptides which are also provided by \cite{Bentley.2016}.
These can act either as neurotransmitters or neuromodulators.
In total, there are $164$ neurons with neuropeptides and $185$ neurons with corresponding receptors.
In this study, we use the transmitter and receptor data including the subset of peptides to map the chemical connections with respect to the underlying transmitter types. 

Note that for MAs and neuropeptides large extrasynaptic signalling networks exist in \elegans.
Extrasynaptic connections are generated by the diffusion of neurotransmitters and neuropeptides to adjacent (extracellular) synapses. Such connections occur mainly outside the synaptic connectome and are referred to as wireless connections. It is well established that they play an important role for brain function \cite{Bentley.2016}.
The inclusion of extrasynaptic neurotransmission would add many more layers to the network which is not done.

\subsection{Mapping of chemical connections}
\label{subsec:Mapping of Transmitter Types for Chemical Connections}

Only $1,529$ connections (about $70\%$) of the $2,194$ chemical connections can be covered with the utilized transmitter and receptor data. 
For the remaining $665$ connections (about $30\%$), either the information about the transmitter or the receptor is missing, or the transmitters and receptors are not compatible. To have a more complete view of the network of \elegans, the missing transmitter types are estimated for these connections. Furthermore, it is assumed that the estimated transmitters are valid for all neurons of the same class. The classes can be found on \cite{wormatlas.2020}. 

%

If the transmitter information for one neuron is missing, the least common set of transmitters is searched for all postsynaptic receptors in order to establish at least one connection with all unassigned postsynaptic neurons. If several options remain afterwards, not all transmitter types are assigned but preferred ones are chosen. Sometimes a decision can be made with the help of WormAtlas \cite{wormatlas.2020} by studying the transmitter and receptor information for the individual neurons. If still no result is found, the transmitter with the highest probability is utilized (compare exemplary with \Cref{fig:network_frequencies}).
Two examples are given below:
\begin{itemize}
\item \keyword{Example 1}: Neuron A has no transmitter information, neuron X has the ACh receptor \emph{acc-1} as well as the glutamate (in the following Glu) receptor \emph{glr-1}, neuron Y has the dopamine (in the following DA) receptor \emph{dop-1}, and neuron Z has the ACh receptor \emph{acc-2}. The least common set of transmitters to connect to the neurons X, Y, and Z is given by the transmitters ACh and DA which are assigned to Neuron A. 
\item \keyword{Example 2}: If neuron Z from the first example has also a Glu receptor like \emph{glr-2}, no unique decision can be made about the transmitters ACh and Glu.  If no solution can be found with the help of WormAtlas 
either, frequency distributions are considered. In this case, the transmitter ACh has the higher probability of occurrence and is assigned to Neuron A in combination with the DA transmitter.
\end{itemize}

The same procedure is used when the receptor information is missing for specific neurons. This time, the least common set of transmitters among all presynaptic neurons is desired. For the different transmitters found, no specific receptors can be specified. It can be indicated that there must be at least one receptor for certain types of transmitters. 
Finally, the last approach is also applied in the case that the information for transmitter and receptor does not correspond. 
In general, neurons have significantly fewer transmitters than receptors. In \Cref{tab:trans_recep}, it can be seen that there is a huge variety for the latter. For certain neurons, many receptors may not play a role in establishing synaptic connections since they can only couple with transmitters with are not present. These may have use in extrasynaptic connections, but this is not considered. In the case of synaptic transmission, it is therefore more likely that the existence of receptors can be indicated for particular transmitters than vice versa. That is why transmitters are considered and receptors are assigned. The general approach for estimating transmitter types does not claim to be the best solution, but it can be done without much time and the results can be plausibilised.
The findings are provided in Tables \ref{tab:Pred_Trans} and \ref{tab:Pred_Recep}.

\begin{figure}[!htbp]
\centering
\begin{subfigure}{.235\textwidth}
            \centering
            \includegraphics[width=.93\linewidth]{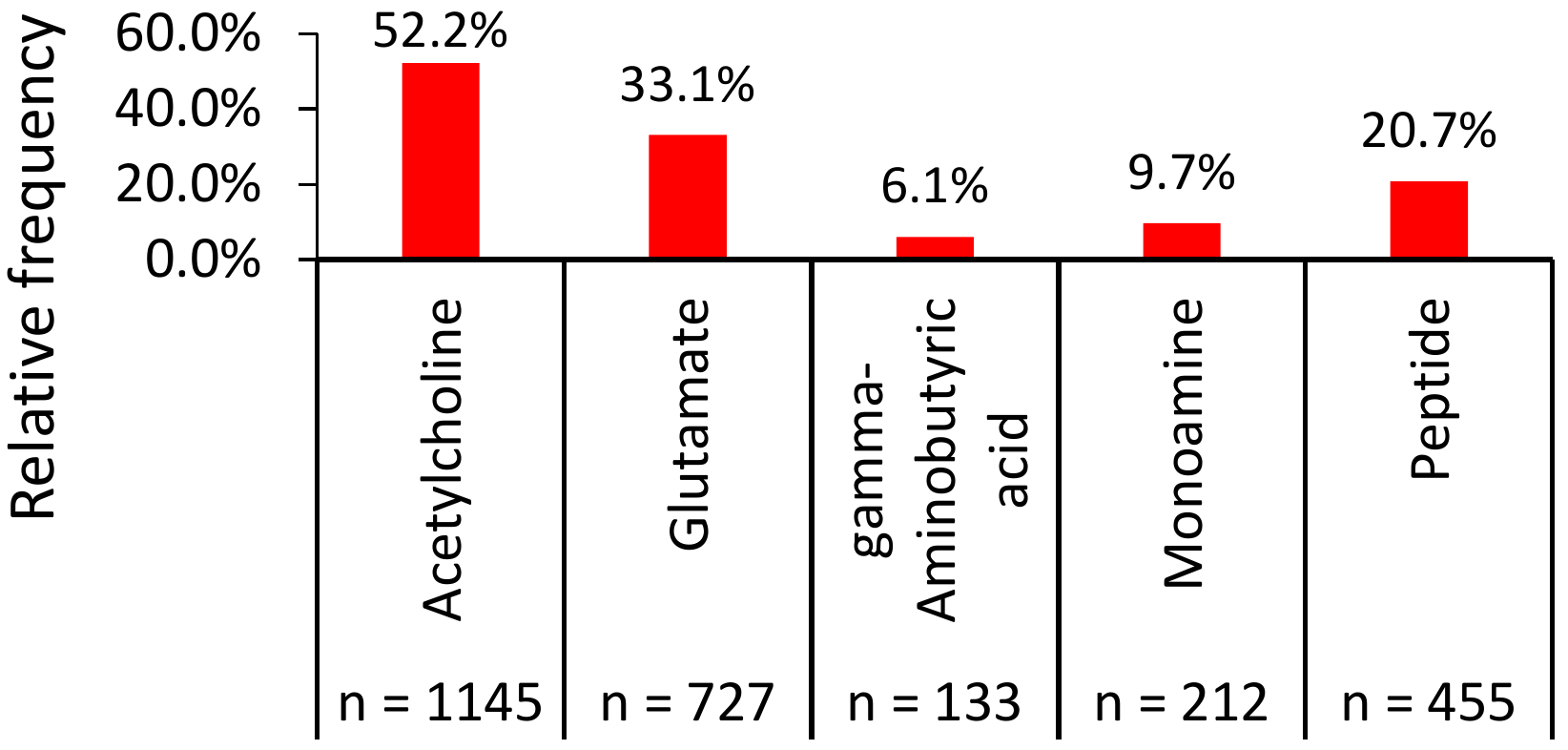}  
            \caption{Neurochemicals}
            \label{fig:freq_Neurochemicals}
        \end{subfigure}
        \begin{subfigure}{.235\textwidth}
            \centering
            \includegraphics[width=.93\linewidth]{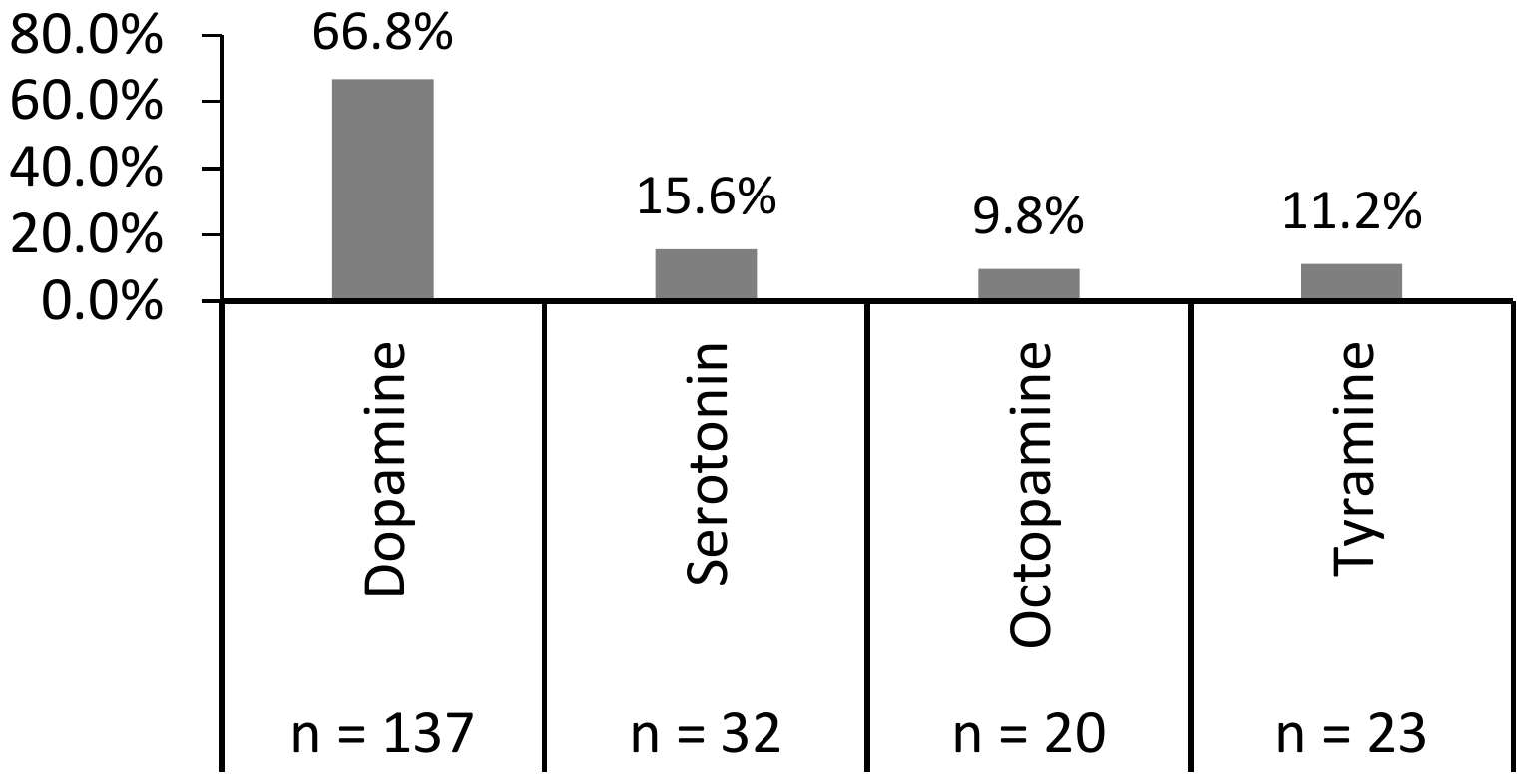}  
            \caption{Monoamines}
            \label{fig:freq_Monoamines}
        \end{subfigure}
        \newline
        \begin{subfigure}{.235\textwidth}
            \centering
            \includegraphics[width=.93\linewidth]{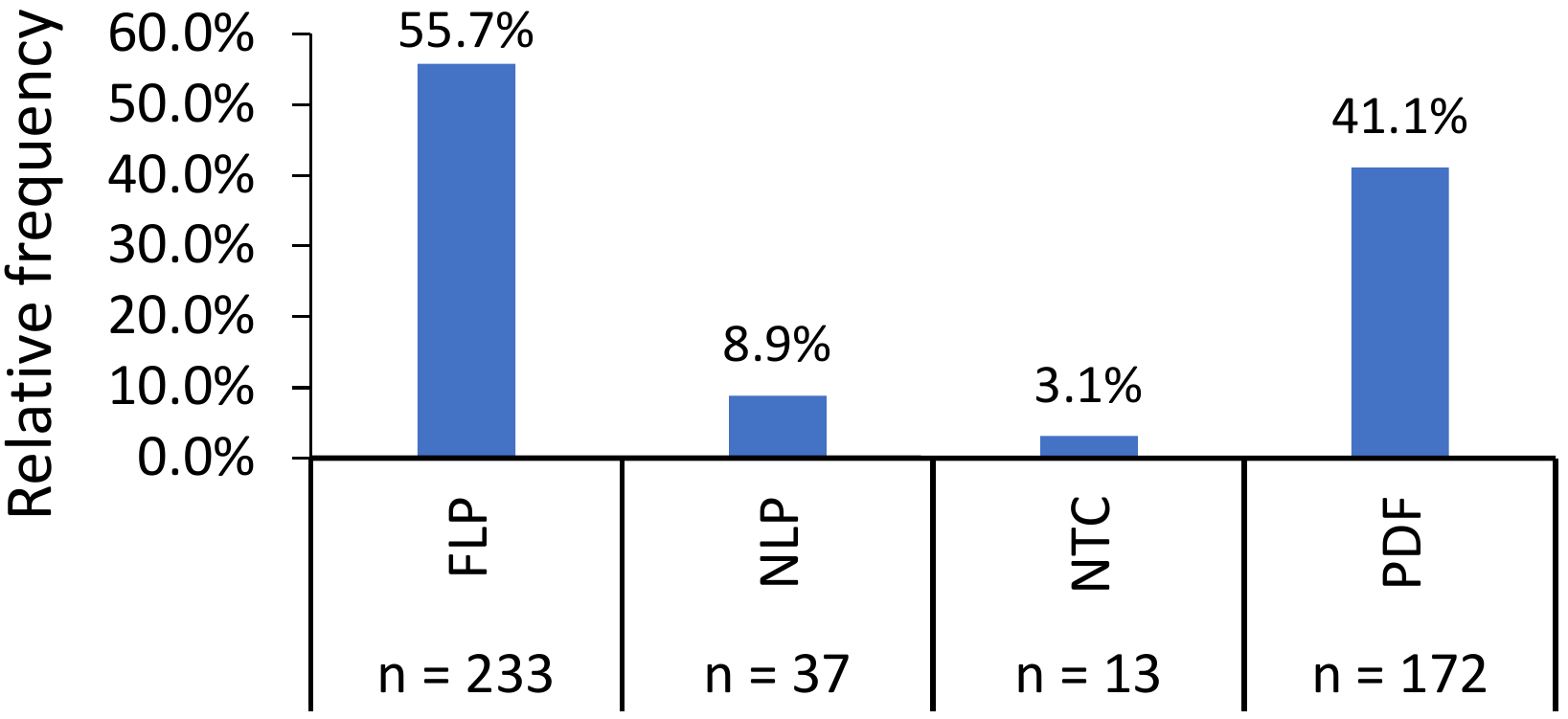}  
            \caption{Peptides}
            \label{fig:freq_Peptides}
        \end{subfigure}
        \begin{subfigure}{.235\textwidth}
            \centering
            \includegraphics[width=.93\linewidth]{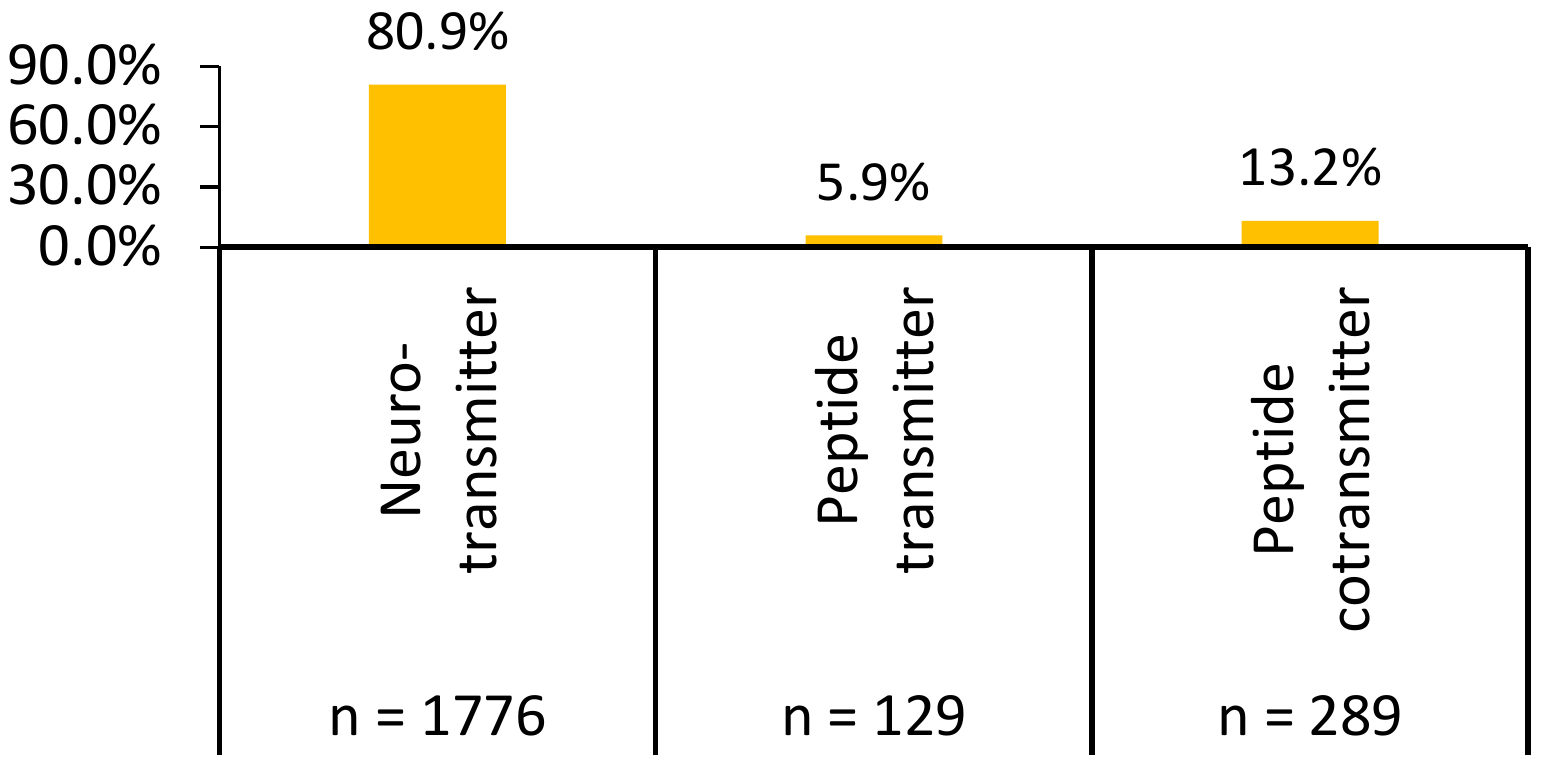}  
            \caption{Transmission types}
            \label{fig:freq_Trans}
        \end{subfigure}
    \caption[Frequency distributions for the chemical connections.]{\keyword{Frequency distributions for the chemical connections.} \keyword{(a)} Relative frequency of different neurochemicals on a total of $2,194$ chemical connections (overlap is $21.8\%$). \keyword{(b)} Relative frequency of the monoamine transmitters on a total of $205$ monoamine connections (overlap is $3.4\%$). \keyword{(c)} Relative frequency of the peptide families on a total of $418$ peptide connections (overlap is $8.9\%$). \keyword{(d)} Relative frequency of neurotransmission in general (only neurotransmitters, only peptide transmitters, and neurotransmitters with peptide cotransmitters).}
    \label{fig:network_frequencies}
\end{figure}

\begin{table*}[H]
  \centering
  \caption[Estimated transmitters for chemical connections.]{\keyword{Estimated transmitters for chemical connections.}}
  	\small
    \begin{tabular}{ll|ll}
    \toprule
    \textbf{Neuron} & \textbf{Predicted} & \textbf{Neuron} & \textbf{Predicted} \\
    \textbf{class} & \textbf{transmitters} & \textbf{class} & \textbf{transmitters} \\
    \midrule
    \multirow{2}[1]{*}{AVH (2)} & ACh, flp-21, flp-24, & PVM (1) & dopamine, GABA, Glu \\
        & Glu, ntc-1 & PVW (2) & Glu, ntc-1, serotonin \\
    AVJ (2) & ACh, Glu & RID (1) & ACh \\
    \multirow{2}[1]{*}{AWA (2)} & dopamine, flp-21, Glu, & RIP (2) & ACh \\
        & tyramine &     &  \\
    \bottomrule
    \end{tabular}%
  \label{tab:Pred_Trans}%
\vspace{+3mm}
  \centering
  \caption[Estimated transmitter receptors for chemical connections.]{\keyword{Estimated transmitter receptors for chemical connections.}}
	\small
    \begin{tabular}{ll|ll}
    \toprule
    \textbf{Neuron} & \textbf{Predicted receptors} & \textbf{Neuron} & \textbf{Predicted receptors} \\
    \textbf{class} & \textbf{(unspecified) for} & \textbf{class} & \textbf{(unspecified) for} \\
    \midrule
    \multirow{2}[1]{*}{ADA (2)} & Ach, Glu, octopamine, & OLL (2) & Ach, GABA, Glu \\
        & pdf-1 & OLQ (4) & dopamine \\
    ADE (2) & Glu, nlp-1 & PDA (1) & Ach \\
    ADF (2) & ACh & PDB (1) & ACh \\
    ADL (2) & Ach, dopamine & PDE (2) & Ach, flp-1, GABA, Glu \\
    AFD (2) & Ach, Glu & PHC (2) & GABA, Glu \\
    AIA (2) & Ach & PLN (2) & ACh \\
    AIB (2) & Ach, flp-21, pdf-1 & PVC (2) & GABA \\
    AIM (2) & Ach, Glu, pdf-1 & PVM (1) & dopamine, flp-1 \\
    AIN (2) & Glu & \multirow{2}[0]{*}{PVN (2)} & ACh, flp-10, GABA, \\
    AIZ (2) & Glu &     & Glu \\
    ALA (1) & dopamine, Glu & PVP (2) & Ach, GABA, Glu \\
    ALN (2) & Glu, pdf-1 & PVR (1) & ACh, dopamine, Glu \\
    AQR (1) & ACh & PVT (1) & Glu \\
    AS (11) & Glu & PVW (2) & ACh, GABA \\
    ASE (2) & Ach & RIB (2) & flp-21, GABA \\
    ASG (2) & Ach, Glu & RIC (2) & Ach \\
    ASH (2) & Ach & RID (1) & Glu, pdf-1 \\
    ASJ (2) & ACh, GABA, Glu & RIF (2) & Ach \\
    ASK (2) & Ach, GABA, Glu & RIG (2) & dopamine, pdf-1 \\
    AUA (2) & Ach & RIH (1) & Glu, pdf-1 \\
    \multirow{2}[0]{*}{AVA (2)} & dopamine, nlp-1, & \multirow{2}[0]{*}{RIM (2)} & flp-1, GABA, \\
        & octopamine &     & octopamine \\
    AVB (2) & flp-1 & RIP (2) & ACh, dopamine, Glu \\
    AVD (2) & Ach & RIR (1) & Ach, pdf-1 \\
    AVE (2) & dopamine, flp-1, GABA & RIS (1) & ACh \\
    AVF (2) & GABA, Glu & \multirow{2}[0]{*}{RIV (2)} & ACh, dopamine, Glu, \\
    AVG (1) & GABA &     & octopamine \\
    AVH (2) & Ach, nlp-1 & RMD (6) & flp-1, GABA, pdf-1 \\
    \multirow{2}[0]{*}{AVJ (2)} & ACh, dopamine, GABA, & \multirow{2}[0]{*}{RMF (2)} & ACh, flp-1, Glu, \\
        & nlp-1, pdf-1 &     & octopamine, pdf-1 \\
    \multirow{2}[0]{*}{AVK (2)} & ACh, dopamine, GABA, & RMG (2) & Ach \\
        & octopamine, pdf-1 & \multirow{2}[0]{*}{RMH (2)} & Ach, dopamine, Glu, \\
    AVL (1) & Ach, GABA, Glu, pdf-1 &     & pdf-1 \\
    AVM (1) & flp-1 & SAA (4) & dopamine, flp-1, nlp-1 \\
    AWA (2) & Ach, Glu & SAB (3) & Ach, GABA \\
    AWB (2) & ACh & SDQ (2) & Glu \\
    AWC (2) & Ach, flp-21 & SIA (4) & flp-1, Glu \\
    BAG (2) & Ach & SIB (4) & pdf-1 \\
    BDU (2) & ACh & \multirow{2}[0]{*}{SMB (4)} & dopamine, flp-1, Glu, \\
    CEP (4) & GABA, Glu, pdf-1 &     & octopamine, pdf-1 \\
    DA (9) & Glu & SMD (4) & flp-1, octopamine \\
    DD (6) & Glu & URA (4) & Ach, nlp-1 \\
    DVA (1) & flp-1 & URB (2) & Ach, dopamine \\
    DVC (1) & GABA, pdf-1 & URX (2) & Glu \\
    IL1 (6) & dopamine, Glu & URY (4) & ACh, GABA \\
    IL2 (6) & Glu & VA (12) & Glu \\
    LUA (2) & Ach & VD (13) & GABA, Glu \\
    \bottomrule
    \end{tabular}%
  \label{tab:Pred_Recep}%
\end{table*}%

Using the estimated transmitters and receptors, all chemical connections in the network are mapped with at least one type of transmitter. The results of the mapping are illustrated in \Cref{fig:network_frequencies}. 
The chemical layer consists of $2,194$ neuron connections. About $52\%$ of the connections are covered by the transmitter ACh which represents the largest sublayer. Then follows Glu with about $33\%$ and the group of the MAs with almost $10\%$. In the group of the MAs, the DA transmitter has the largest proportion with about $67\%$. The DA sublayer is slightly larger than the GABA sublayer which occupies about $6\%$ of the chemical connections. Although the peptides have a larger proportion at around $21\%$, only about $31\%$ of them act as peptide transmitters. The sublayer of peptide transmitters is comparable to the sublayer of GABA.
The remaining $69\%$ of the peptides are cotransmitters which always occur together with neurotransmitters. Among all pepdites, the FLP and PDF family are the most frequently encountered with about $56\%$ and $41\%$.


\newpage
\null\newpage
\null\newpage
\section{Functions of layers in the network}
\label{appendix:layers}
This appendix provides details on the different layers depicted in \Cref{fig:multilayer_network_ach,fig:multilayer_network_glu,fig:multilayer_network_gaba,fig:multilayer_network_ma,fig:multilayer_network_pep,fig:multilayer_network_el,fig:multilayer_network_mus} that give rise to the overall network shown in \Cref{fig:multilayer_network_all}.

\keyword{Acetylcholine layer} - ACh covers 
$33.4\%$ of the $3,538$ connections in the network and forms the largest layer. It is widely spread among all neurons and does not prefer a specific neuron type. In humans, ACh acts at skeletal neuromuscular junctions, at neuromuscular junctions between the vagus nerve and cardiac muscle fibers, and at a variety of locations within the central nervous system. While the function of ACh at neuromuscular junctions is well known, its role in the central nervous system is not well understood \cite[Chapter 6]{Purves.2018}.
In \elegans, ACh is involved in many behaviors like locomotion \cite{Winnier.1999,Hallam.2000}, egg-laying \cite{Bany.2003}, feeding \cite{Raizen.1995}, and defecation \cite{Thomas.1990}.

\keyword{Glutamate layer} - Glu uses $20.5\%$ of the connections in the network and is therefore the third largest layer. In comparison with the ACh layer, significantly fewer motor neurons are involved. In the human brain, Glu is the most important transmitter for normal brain function. It is estimated that more than half of all brain synapses release this substance \cite[Chapter 6]{Purves.2018}. In \elegans, it contributes to foraging behavior \cite{Hills.2004}, long-term memory \cite{Rose.2003}, and spontaneous switches from forward to backward movement referred to as reversals \cite{Zheng.1999, Brockie.2001}. 

\keyword{\textit{gamma}-Aminobutyric acid layer} - GABA represents the smallest layer in the network and is released on $3.8\%$ of the connections. Most of them are established by interneurons and motor neurons. About one third of the synapses in the brain use GABA as their neurotransmitter which is most frequently found in interneurons of local circuits. In contrast to ACh and Glu, GABA has an inhibitory effect \cite[Chapter 6]{Purves.2018}. In \elegans, it can also act as excitatory transmitter which depends on the neuroreceptor. As an inhibitory transmitter, GABA regulates the head movements while foraging \cite{White.1986} or relaxes muscle cells during locomotion \cite{McIntire.1993}.

\keyword{Monoamine layer} - MA transmitters have a proportion of $5.8\%$ of the network connections. These are independent of the neuron type, but most of the motor neurons are predominantly postsynaptic. In the group of MAs, dopamine accounts for about two thirds (\Cref{fig:network_frequencies}). In humans, MAs regulate many brain functions and are also present in the peripheral nervous system. They are entangled in a wide range of behaviors which range from central homeostatic functions to cognitive phenomena such as attention. The transmitters play an important role in the brain because defects in the MA function can lead to psychiatric disorders. The dopamine transmitter, for instance, is necessary for the coordination of body movements. A degeneration of particular dopaminergic neurons can lead to characteristic motor dysfunction as it is the case with parkinson's disease \cite[Chapter 6]{Purves.2018}. In \elegans, the MAs affect a variety of behaviors including egg-laying, pharyngeal pumping, locomotion, and learning. A good overview is given by \cite{Chase.2007}. The dopamine transmitter, for example, is responsible for the modulation of locomotion behavior and for learning.
The modulation of locomotion behavior enables the worms to react on environmental changes \cite{Sawin.2000} and to search efficiently for new food sources \cite{Hills.2004}. Learning allows the worms to change their behavior based on previous experience. For example, the animals react to a non-localized mechanical stimulus such as plate tapping by either moving backwards or increasing their forward locomotion rate. Repeated tapping on the plate causes the worms to become habituated to the stimulus, and they exhibit a reduced frequency of reversals \cite{Rose.2001}. 

\keyword{Peptide layer} - Peptides are utilized in $11.8\%$ of the network connections which are primarily established by interneurons and sensory neurons. Most of the peptides originate from the FLP and PDF family. About $31\%$ of the peptides are peptide transmitters and about $69\%$ are cotransmitters (\Cref{fig:network_frequencies}).
The number of peptides in humans is estimated to be over $1,000$, over $100$ have already been identified. Peptide transmission is involved in the perception of pain, modulation of emotions, and regulation of complex reactions to stress. Peptide cotransmission enhances or dampens synaptic activity and can influence many functions such as food intake, metabolism, social behavior, learning, and memory (\cite{Russo.2017}, \cite[Chapter 6]{Purves.2018}). In \elegans, peptides affect many behaviors including locomotion, dauer formation, egg-laying, sleep, learning, social behavior, mechano-, and chemosensation. Numerous  peptides of the FLP family are involved in feeding behavior \cite{Li.2008}. On the other side, peptides can fulfill a unique functions as \emph{nlp-22} is a regulator of \elegans\ sleep-like state (lethargus) during a larval transition stage \cite{Nelson.2013}. For this reason, neuropeptides should be considered as the third layer of information flow in neuronal communication next to chemical and electrical transmission. 
This study only uses a small subset of peptides found in \elegans, and many of their functions are still unknown. The actual number of neuropeptides in the worm exceeds $300$ \cite{Ringstad.2017,Chew.2018}.
Like monoamines, neuropeptides have a small wired network but a large wireless network \cite{Bentley.2016}. This raises the opportunity that neuropeptides could be involved in all \elegans\ behaviors.

\keyword{Electrical layer} - Electrical transmission accounts for $29.1\%$ of the network connections and is therefore the second largest layer behind ACh. Electrical synapses are found in all nervous systems. They enable a direct, passive flow of electrical current from one neuron to another. In contrast to chemical synapses, the current can flow in both directions (bidirectional) and the transmission is extraordinarily fast (virtually instantaneous). Communication is possible without delay which is not typical for chemical synapses. For this reason, they are found in places where quick actions are necessary and have the general purpose of coordinating and synchronizing network activity among neuron groups. For example, certain neurons in the brainstem are synchronized by electrical synapses to produce rhythmic breathing.
The same applies to populations of interneurons in the cerebral cortex, thalamus, and cerebellum \cite[Chapter 5]{Purves.2018}.
In \elegans, electrical transmission plays an important role in locomotion behavior and development \cite{Hall.2017,Simonsen.2014}.

\keyword{Neuromuscular layer} - Neuromuscular junctions are comparable which chemical synapses and have a proportion of $15.5\%$ of the network connections. They are made from motor neurons to body wall muscles. In \elegans, the transmitters ACh and GABA can be assigned to it. While ACh stimulates muscle contraction, GABA relaxes muscles cells \cite{Rand.2007, Jorgensen.2005}. In humans, it is well known that ACh is released by spinal motor neurons and results in contraction of skeletal muscles \cite[Chapter 5]{Purves.2018}.



\section{Logistic regression analysis}
\label{sec:Logistic regression analysis}

Logistic regression analysis is a statistical procedure used for classification problems. An application-oriented introduction to the methods of logistic regression can be found in \cite[Chapter 4]{Pruscha.2006}. In this study, we introduce the binary model.

\subsection{Binary logistic regression model}
\label{subsec:Binary logistic regression model}

The binary model considers a response variable (random number) which is denoted as $Y$ and accepts the two values $0$ and $1$. The expectation value $\mathbb{E}(Y)$ lies in the interval $[0,1]$. By applying a response function
%
%
\begin{align*}
F(z), \; z \in \mathbb{R}, \textrm{ with values in the interval } [0,1],
\end{align*}
the linear combination $z = c_0 + c{_1}x_1 + c{_2}x_2 + … + c{_m}x_m$ of $m$ explanatory variables $x_1, x_2, ..., x{_m}$ is restricted to the same interval whereby $c_0$ is a constant and $c_1, c_2, ..., c{_m}$ represent coefficients. This results in the following estimate 
%
%
\begin{align*}
\mathbb{E}(Y) = \mathbb{P}(Y = 1) = F(z).
\end{align*}
The binary model possesses as response function the logistic function (also known as sigmoid function)
\begin{eqnarray}\label{eq:logistic_function}
F(t)=\frac{1}{1+\exp(-t)}, \quad t \in \mathbb{R}.
\end{eqnarray} 
This function has a nonlinear s-shaped curve that asymptotically approaches the values $0$ and $1$ from above and below,
and its values can be interpreted as the probability that the response variable equals one.

For the observation $i$, the values of the response variable $Y$ ($0$ or $1$) and the explanatory variables $x_1, x_2, ..., x{_m}$ can be specified:
%
%
\begin{align*}
Y_{i},x_{1i},x_{2i},...,x_{mi},\quad i=1,2,...,n.
\end{align*}
The linear regression term for observation $i$ is 
%
%
\begin{eqnarray*}
z_i = c_0 + c{_1}x_{1i} + c{_2}x_{2i} + … + c{_m}x_{mi},\quad i=1,2,...,n. 
\end{eqnarray*}
The random variables $Y_1, Y_2, ..., Y{_n}$ are assumed to be independent. The $m + 1$ dimensional vector of the unknown parameters is designated as $\vecvar{c} = [c_{0},c_{1},c_{2},...,c_{m}]^{T}$. 
The probability for the occurrence of $Y_i = 1$ is abbreviated with 
\begin{eqnarray*}
\sigma_i(\vecvar{c}) = \mathbb{P}(Y_i = 1).
\end{eqnarray*}
Then, according to the above estimate, the binary logistic regression model reads as \cite[Chapter 4]{Pruscha.2006}
\begin{eqnarray}\label{eq:logistic_regression_model_Appendix}
\sigma_i(\vecvar{c})=F(z_i (\vecvar{c}))=\frac{1}{1+\exp(-z_i (\vecvar{c}))},\quad i=1,2,...,n. \;
\end{eqnarray} 
Since logistic regression predicts probabilities and not just classes, the unknown parameters $c_j$ can be estimated using the maximum likelihood method. This method maximizes the probability of the parameters for the given data. The response probability $p_i$ depending on observations ($0$ or $1$) of the response variable $y_i$ satisfies
\begin{align*}
p_i(y_i)=\begin{cases}
    \frac{1}{1+\exp(-z_i (\vecvar{c}))}, & \text{if $y_{i}=1$}\\
    1-\frac{1}{1+\exp(-z_i (\vecvar{c}))}, & \text{if $y_{i}=0$}
  \end{cases}
\end{align*} 
which can be summarized as 
\begin{flalign*}
\begin{split} 
p_i(y_i)=\left(\frac{1}{1+\exp(-z_i (\vecvar{c}))}\right)^{y_i}\left(1-\frac{1}{1+\exp(-z_i (\vecvar{c}))}\right)^{1-y_i}.
\end{split} &&
\end{flalign*} 
For all $i$ observations together, the probability theorem for independent events can be used to construct the likelihood function which needs to be maximized:
\begin{flalign*}
\hspace{3mm}
\begin{split}
L(\vecvar{c})={}& \prod\limits_{i=1}^{n}\left(\frac{1}{1+\exp(-z_i (\vecvar{c}))}\right)^{y_i} \\ 
& \quad\quad\quad \cdot \left(1-\frac{1}{1+\exp(-z_i (\vecvar{c}))}\right)^{1-y_i}\overset{!}{=}max.
\end{split} &&
\end{flalign*} 
Taking the natural logarithm yields the log-likelihood function
\begin{flalign}\label{eq:log-likelihood_function}
\hspace{0mm}
\begin{split}
\mathcal{L}(\vecvar{c})={}&\sum\limits_{i=1}^{n}\left[{y_i}\ln\left(\frac{1}{1+\exp(-z_i (\vecvar{c}))}\right)\right] \\
&  \quad +\left[({1-y_i})\ln\left(1-\frac{1}{1+\exp(-z_i (\vecvar{c}))}\right)\right]
\end{split} &&
\end{flalign} 
which has identical extreme values but is easier to calculate. In many program packages, the maximization is performed by the Newton-Raphson algorithm where the zero point is approximated by iteration \cite[Section 3.3]{Stein.2015}.

To evaluate the fitted model, the log-likelihood is often multiplied by $-2$. The negative twofold log-likelihood is approximately $\chi^2$ - distributed with $n - (m + 1) - 1$ degrees of freedom where $n$ is the number of observations and $m + 1$ the number of parameters. The expressiosn $-2\mathcal{L}$ is denoted as deviance and is comparable to the residual sum of squares of the linear regression. 
With a perfect fit, the deviance is zero. Under the null hypothesis "The model has a perfect fit", the fit is the better the lower the value $-2\mathcal{L}$ \cite[Section 3.5]{Stein.2015}. With the fitted model, the values $0$ or $1$ can be assigned to the response variable $Y$ in dependence on a threshold value. The binary logistic regression model is depicted in Figure \ref{fig:logistic_regression}. 

The requirement for creating a logistic regression model is that the explanatory variables should not be highly correlated with each other because this can cause estimation problems \cite{Bewick.2005}. 



\subsection{Power analysis to identify key factors}
\label{subsec:Power analysis to identify key factors}

The quality of a regression model is determined by its ability to distinguish correctly between observations ($0$ or $1$) of the response variable $y_i$. To figure out which explanatory variables $x_k$ have a high explanatory contribution to the model \eqref{eq:logistic_regression_model_Appendix}, the power $P$ is introduced. This measure lies in the interval $[-1,1]$ and can be defined as
\begin{eqnarray}\label{eq:power_calc_Appendix}
 \begin{array}{ll}
 P = \dfrac{1}{N_{Y=1}N_{Y=0}}\sum\limits_{l=1}^{N_{Y=1}}\sum\limits_{m=1}^{N_{Y=0}}S(x_{l,Y=1},x_{m,Y=0}),\\
\\
S(x_{l,Y=1},x_{m,Y=0})=\begin{cases}
    1, & \text{if $x_{l,Y=1}>x_{m,Y=0}$}\\
    0, & \text{if $x_{l,Y=1}=x_{m,Y=0}$}\\
    -1, & \text{if $x_{l,Y=1}<x_{m,Y=0}$}
  \end{cases}
 \end{array}
\end{eqnarray} 
where $x_{l,Y=1}$ and $x_{m,Y=0}$ are observations of the empirical distributions $X_{Y=1}$ and $X_{Y=0}$ with numbers of observations $N_{Y=1}$ and $N_{Y=0}$.

The power can be derived from the receiver operating characteristic (ROC) and the cumulative accuracy profile (CAP). Both ROC and CAP are important concepts to visualize the discriminative power (separation ability) of a model. They are applied in many scientific disciplines, for instance, in biology, information technology, and engineering sciences. The concepts convey the same message but present it in different ways. The information contained in a ROC or CAP curve can be aggregated into a single number, the area under the ROC curve (AUROC) or the accuracy ratio (AR). The AR can be interpreted as a simplified representation of AUROC since
\begin{align}\label{eq:power_rel}
AR = 2 \cdot AUROC-1
\end{align} 
%
and is also known as Gini coefficient or power statistics. 
Since the Mann-Whitney statistics can be introduced as an equivalent to the area under the ROC curve \cite[Chapter 13]{Engelmann.2011}, we obtain the power in \eqref{eq:power_calc_Appendix} using \eqref{eq:power_rel}. Equation \eqref{eq:power_calc_Appendix} indicates that $P$ can be expressed in terms of probabilities
\begin{eqnarray}\label{eq:power_prob}
\mathbb{E}(P)=\mathbb{P}(X_{Y=1}>X_{Y=0})-\mathbb{P}(X_{Y=1}<X_{Y=0})
\end{eqnarray}
which allows for an intuitive interpretation. 

\begin{figure}[!htbp]
	\centering
	    \begin{subfigure}{.235\textwidth}
  \centering
  \includegraphics[width=.95\linewidth]{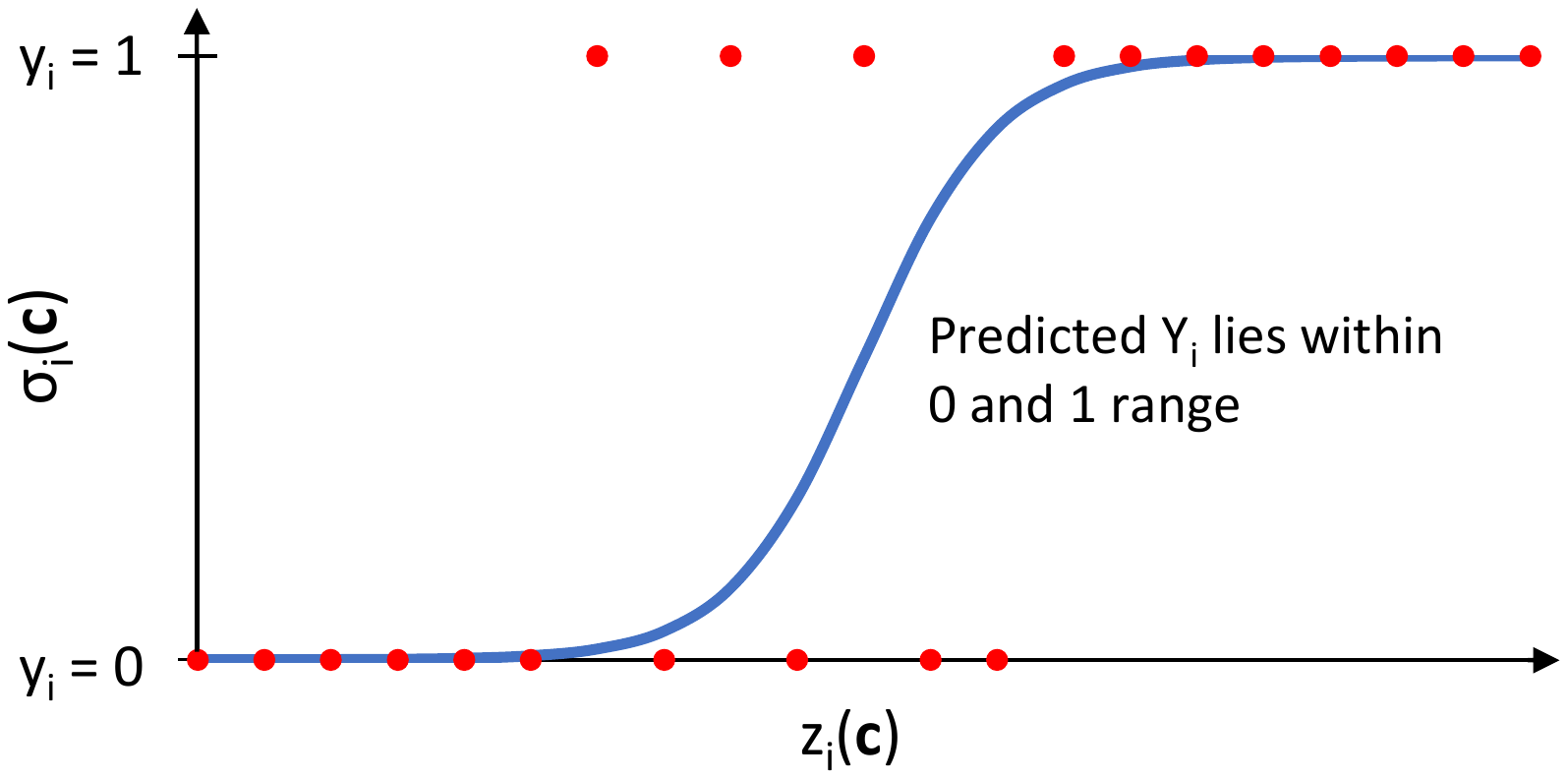}  
  \captionsetup{font=footnotesize}
  \caption{Binary logistic regression}
  \label{fig:sub-first}
\end{subfigure}
\begin{subfigure}{.235\textwidth}
  \centering
  \includegraphics[width=.95\linewidth]{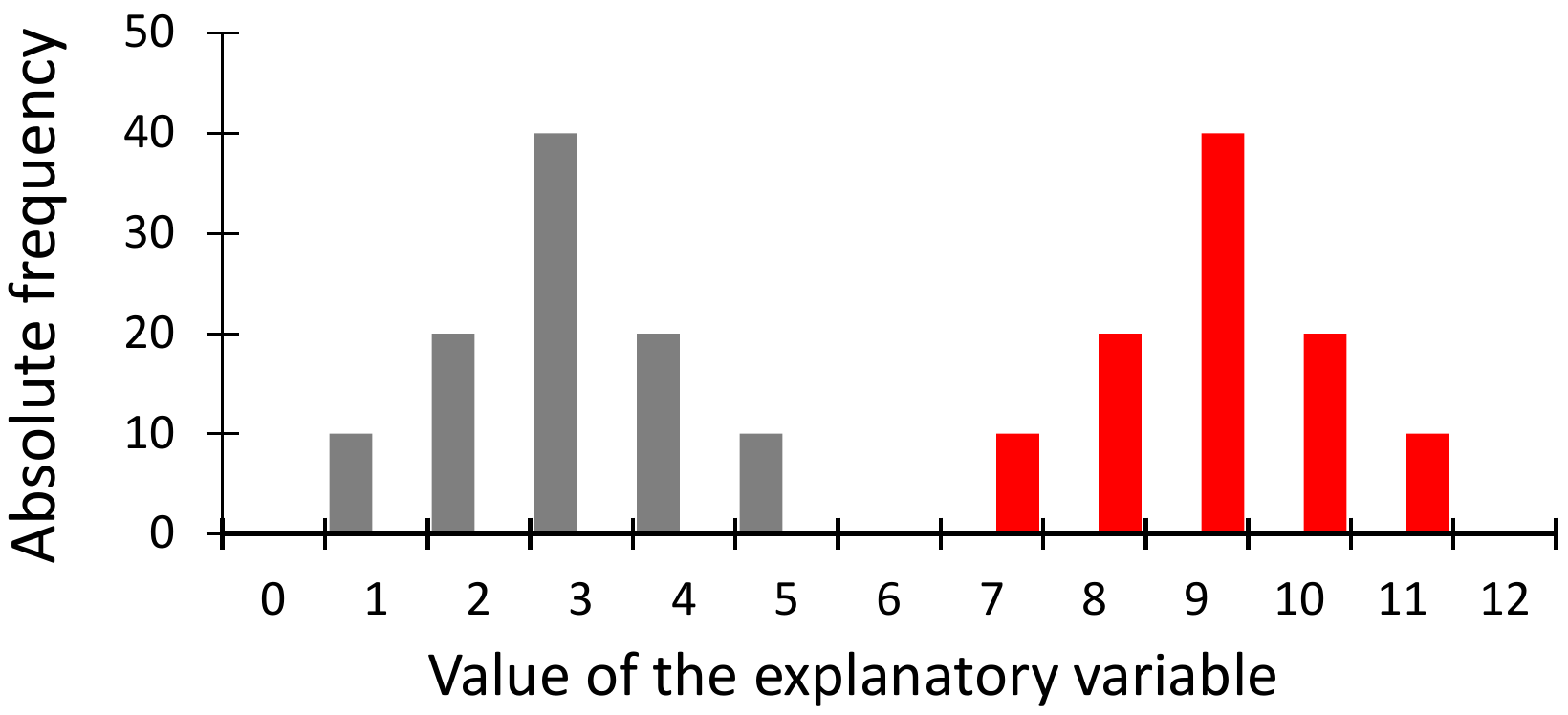}  
  \captionsetup{font=footnotesize}
            \caption{Perfect discrimination ($P = 100 \%$)}
  \label{fig:sub-second}
\end{subfigure}
\newline
\begin{subfigure}{.235\textwidth}
  \centering
  \includegraphics[width=.95\linewidth]{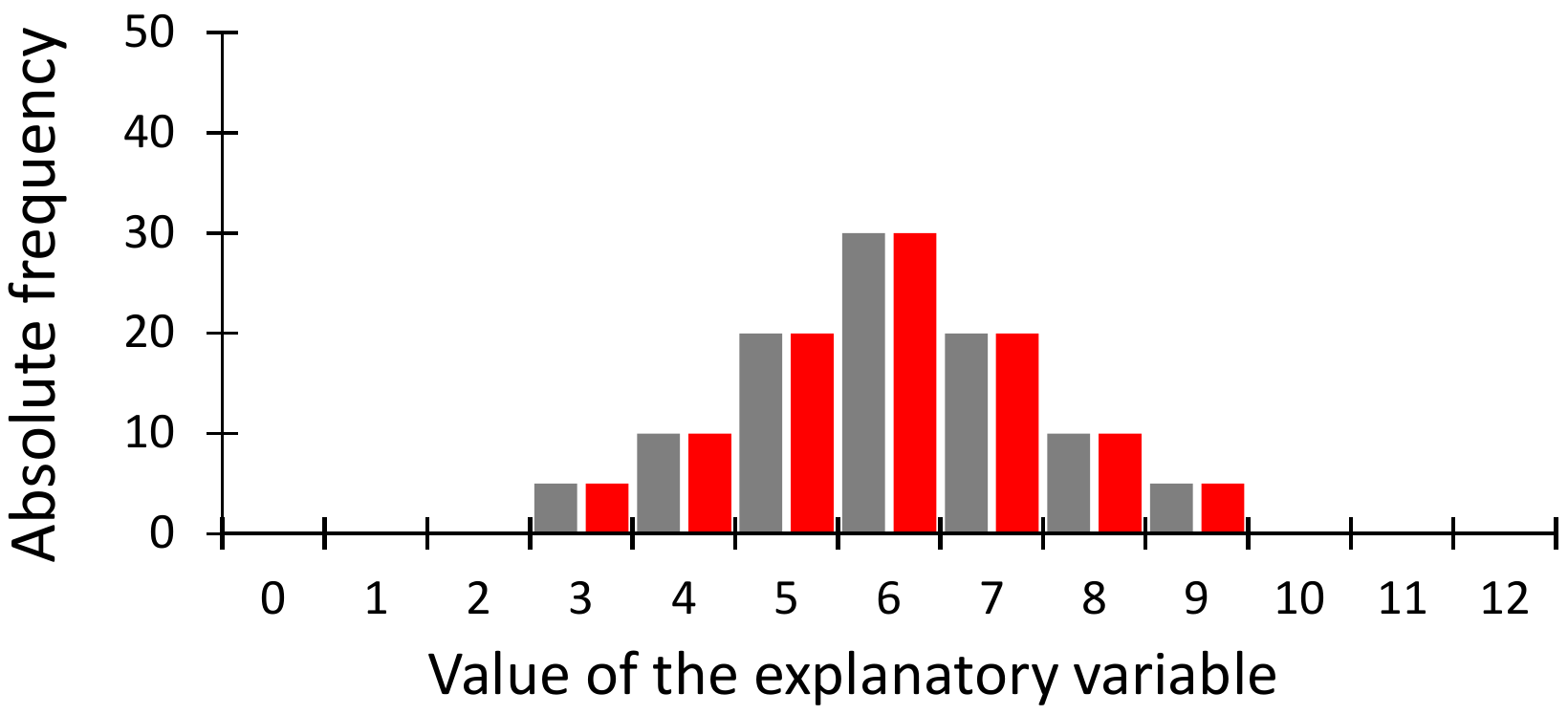}  
  \captionsetup{font=footnotesize}
  \caption{No discrimination ($P = 0\%$)}
  \label{fig:sub-third}
\end{subfigure}
\begin{subfigure}{.235\textwidth}
  \centering
  \includegraphics[width=.95\linewidth]{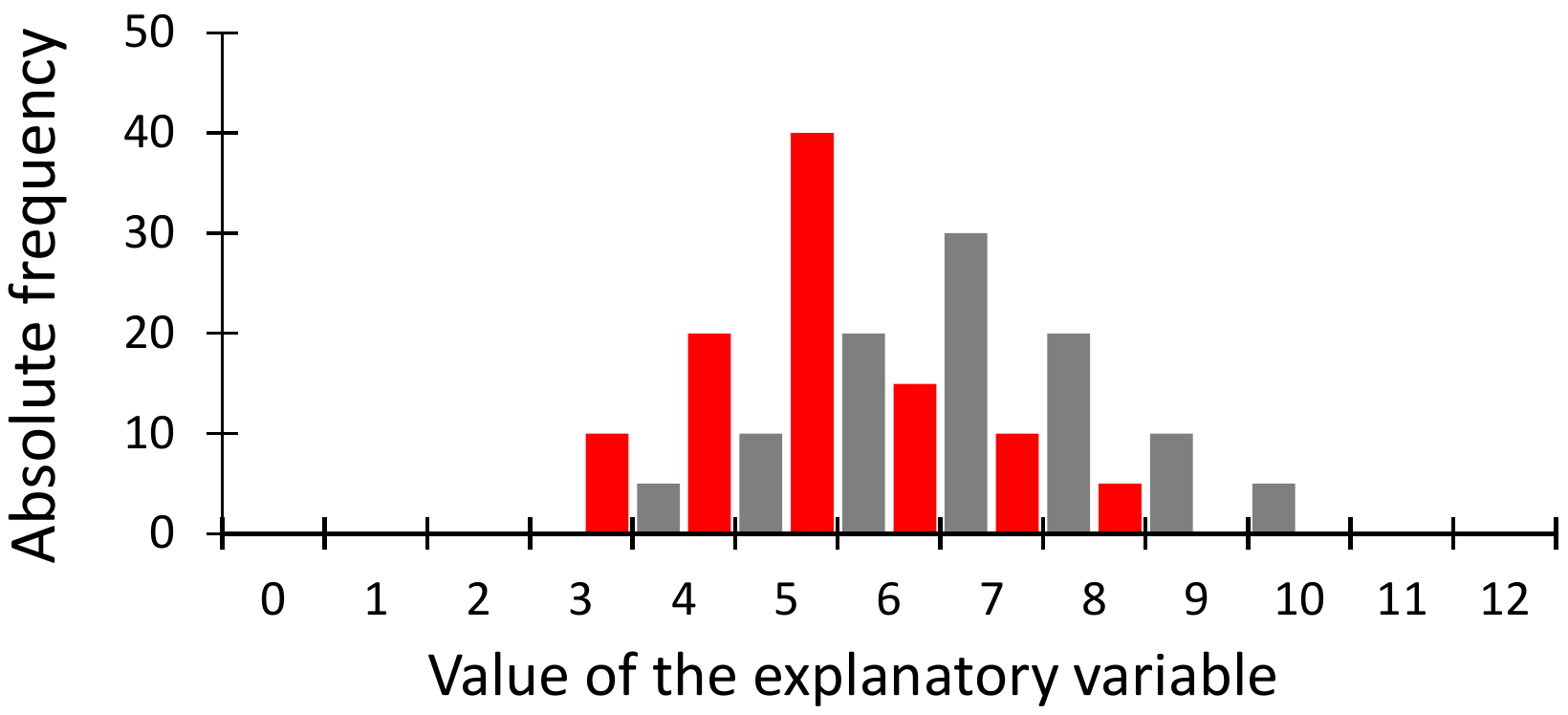}  
  \captionsetup{font=footnotesize}
  \caption{Excellent discrimination ($P = -66\%$)}
  \label{fig:sub-fourth}
\end{subfigure}
	\caption[Logistic regression model and example distributions with corresponding discriminative power.]{\keyword{Logistic regression model and example distributions with corresponding discriminative power.} \keyword{(a)} Binary logistic regression. The logistic function $\sigma_i$ is illustrated as a blue line. The observed values of the response variables $y_i$ are displayed as red circles. \keyword{(b)-(d)} Example distributions of an explanatory variable with the corresponding discriminative power. The empirical distribution $X_{Y=1}$ ($X_{Y=0}$) is indicated by the red (grey) bars.}
    \label{fig:logistic_regression}
	\end{figure}

If all observed values of the distribution $X_{Y=1}$ are larger than those of the distribution $X_{Y=0}$, the power $P$ is equal to $100\%$. The observations ($0$ or $1$) of the response variable $y_i$ can be perfectly separated. The same applies if both distributions switch the sides so that the power is now $-100\%$. 
If both distributions overlap completely, the power is zero, and the values of the response variable cannot be separated with logistic regression. Both examples are illustrated in \Cref{fig:logistic_regression}. As a third example, a case is shown where the distributions overlap slightly resulting in a power of $-66\%$. The minus sign indicates that the distribution $X_{Y=1}$ has more observations with lower values than the distribution $X_{Y=0}$. The absolute power value of $66\%$ indicates an excellent regression model. 

The power of a logistic regression model can be classified as follows \cite[Section 5.2.4]{Hosmer.2000}:
\begin{itemize}
\item If $P = 0 \%$, the model cannot separate between observed values ($0$ or $1$) of the response variable $y_i$ (predicted probabilities are pure random, like a coin flip).
\item If $0 \% > \left|P\right| < 40 \%$, the model is considered to have a poor separation ability.
\item If $40 \% \geq \left|P\right| < 60 \%$, the model is considered to have an acceptable separation ability.  
\item If $60 \% \geq \left|P\right| < 80 \%$, the model is considered to have an excellent separation ability.
\item If $\left|P\right| \geq 80 \%$, the model is considered to have an outstanding separation ability.   
\end{itemize}

It should be noted that in practice it is extremely unusual to observe absolute power values greater than $80\%$.
Since the power $P$ measures the discriminative power of potential explanatory variables, it is well suited to identify key factors for the regression model. In this case, it is extremely unusual to observe absolute power values greater than $40\%$. At the lower end, factors with $|P| > 5\%$ (on a significance level of $5\%$) can also be important for the regression model. Those factors can significantly increase the power of the regression model in combination with other uncorrelated factors. Please note that factors with a negative power value do not restrict the application of the logistic regression. For the regression model, they simply mean negative coefficients. However, it should be checked whether a negative relationship to the values of the response variable is plausible. 

From statistical point of view, the power $P$ can be interpreted as the probability to uncover a difference when there really is one. On a significance level of $5\%$, a power value of $\left|P\right| > 5\%$ indicates that the null hypothesis "There is no difference" must be rejected. 
To get a feeling for the accuracy of the calculated power \eqref{eq:power_calc_Appendix}, it is typical to specify a confidence interval. An efficient method to compute the $95\%$ confidence intervals based on the Mann-Whitney statistic is provided by \cite[Chapter 13]{Engelmann.2011}. 
This method becomes increasingly important for small sample sizes and a relatively small number of observations for $y_i=1$ or $y_i=0$. The uncertainty is considered more appropriately by significantly broader confidence intervals compared with other program packages.


\subsection{Complementary univariate power analyses for prediction of the locomotory subnetwork}
\label{subsec:Complementary univariate power analyses for prediction of the locomotory subnetwork}

\begin{figure*}[htbp] 
\centering
\begin{minipage}{.49\textwidth}
	\centering
    \includegraphics[width=.99\linewidth]{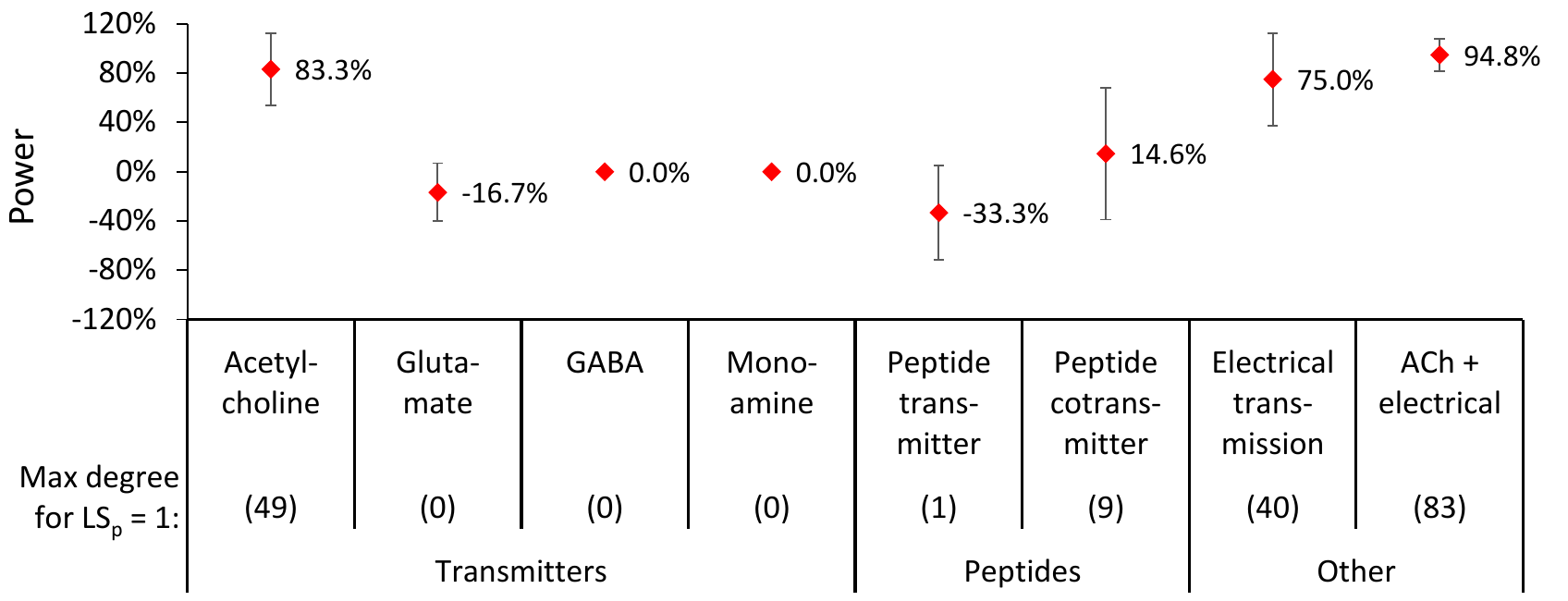}
	\caption[Outdegree power values for the interneurons.]{\keyword{Outdegree power values for the interneurons.}}
    \label{fig:Power_oudegree_IN}
\end{minipage}
\begin{minipage}{.98\textwidth}
\vspace{+3mm}
	\centering
        \begin{subfigure}{.49\textwidth}
                \centering
                \includegraphics[width=0.99\textwidth]{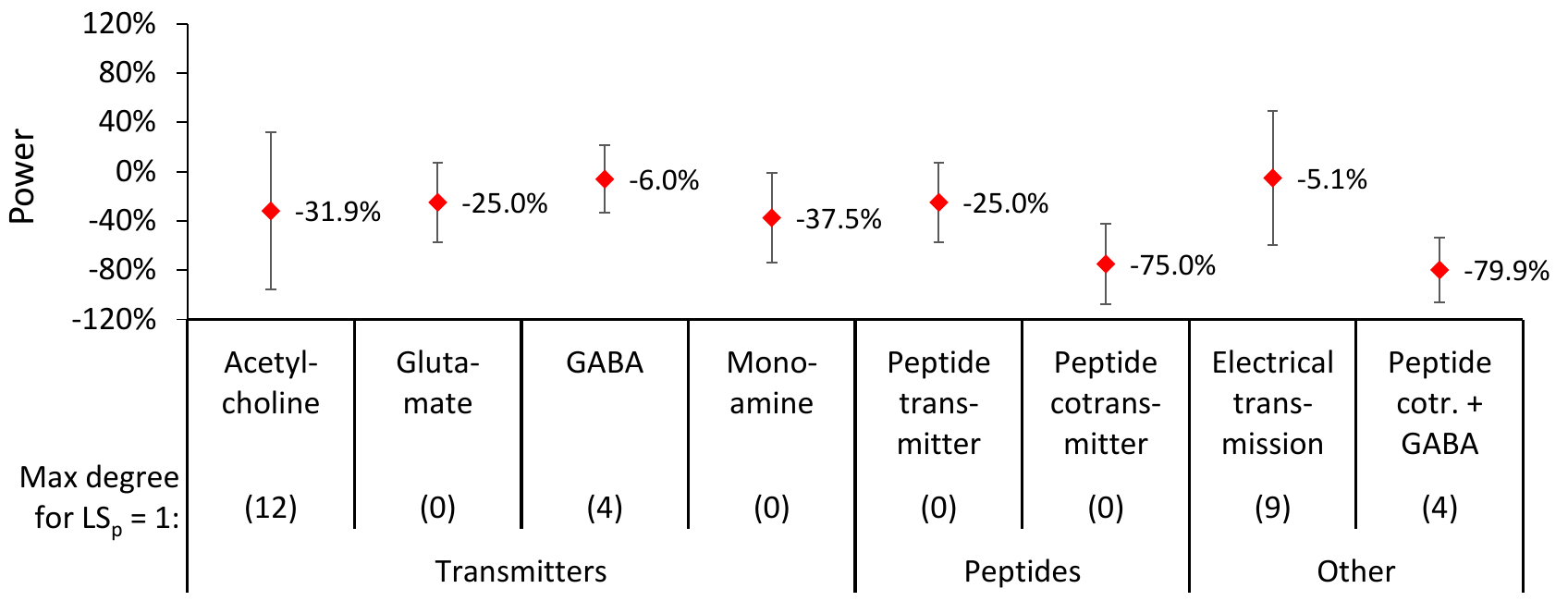}
                \caption{Outdegree of neurons}
                \label{fig:Power_outdegree_MO1}
        \end{subfigure}
        \hfill
        \begin{subfigure}{.49\textwidth}
                \vspace{+6mm}
                \centering
                \begin{subfigure}{.49\textwidth}
	                    \centering
	            \includegraphics[width=.99\linewidth]{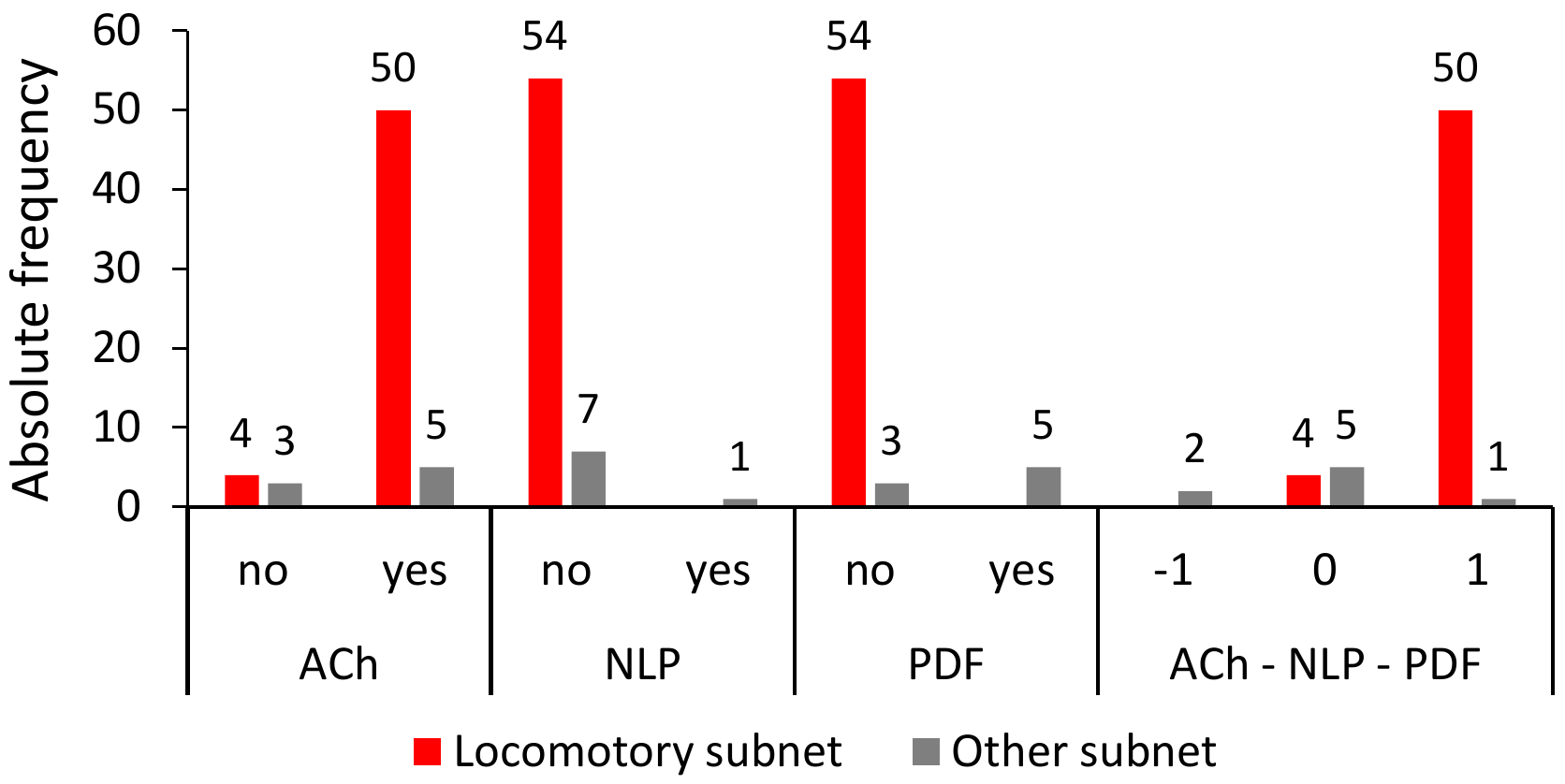}
	                    \caption*{\keyword{(1)} Absolute frequencies}
	                \end{subfigure}
	                \begin{subfigure}{.49\textwidth}
	                \vspace{+0.5mm}
	                    \centering
	            \includegraphics[width=.99\linewidth]{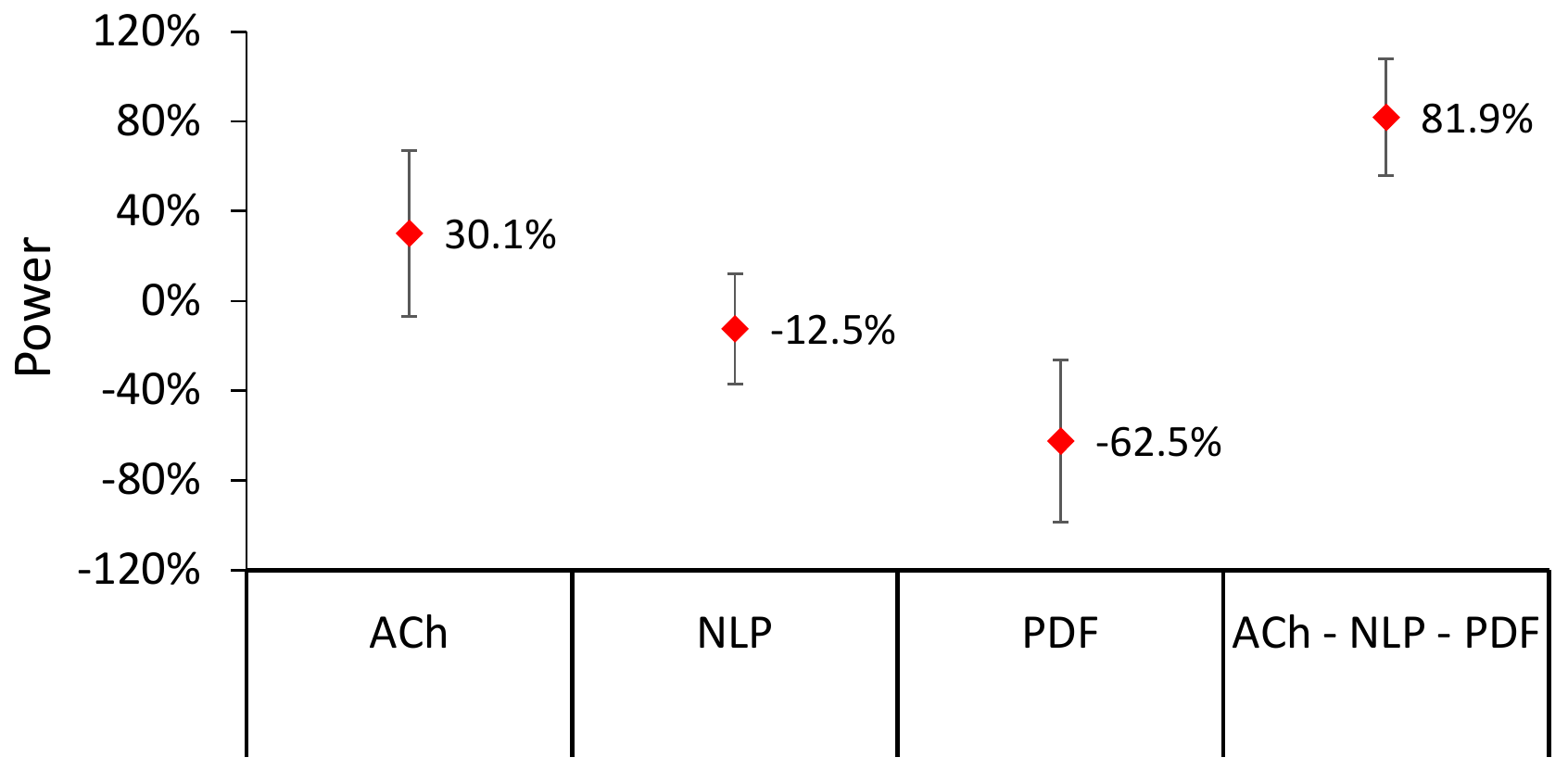}  
	                    \caption*{\keyword{(2)} Corresponding power}
	                \end{subfigure}
	       \caption{Existence of specific transmitters among neurons.}
	       \label{fig:Power_qual_MO1}
        \end{subfigure}
        	\caption[Power values for the first layer motor neurons.]{\keyword{Power values for the first layer motor neurons.}}
	\end{minipage}
\begin{minipage}{.98\textwidth}
\vspace{+3mm}
	\centering
	    \begin{subfigure}{.49\textwidth}
                \centering
                \includegraphics[width=0.99\textwidth]{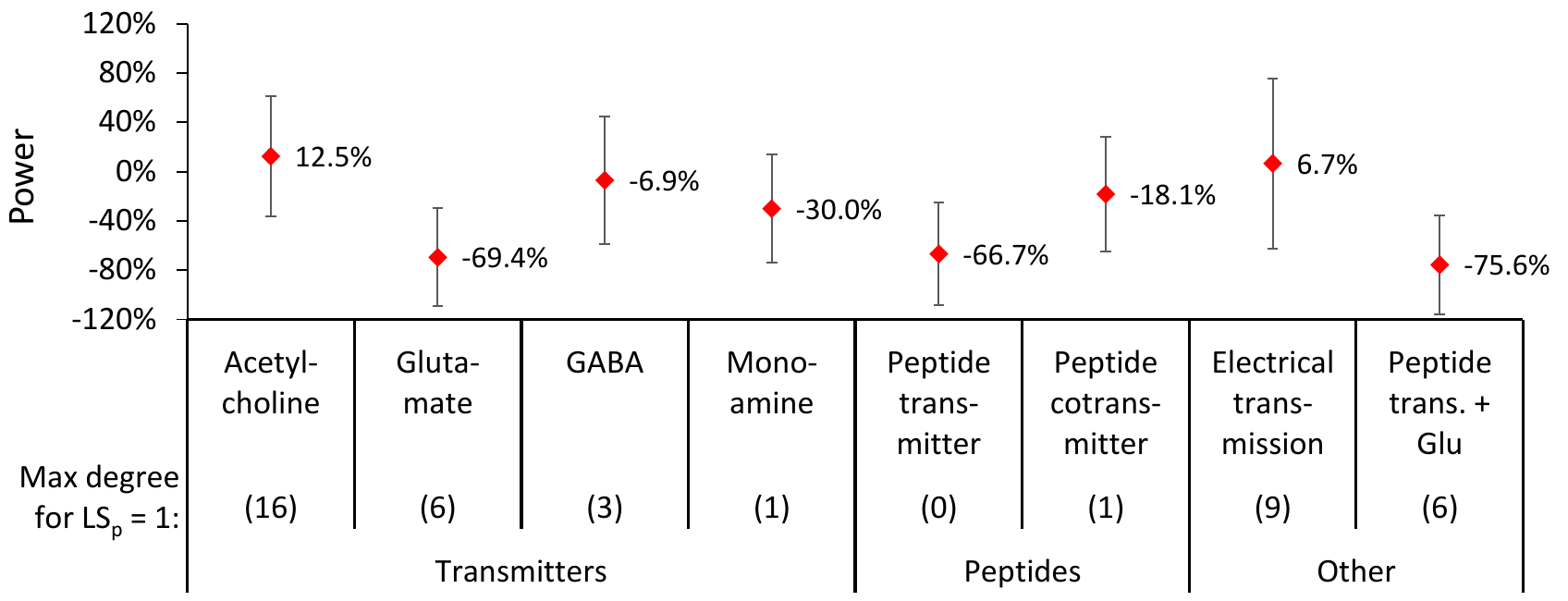}
                \caption{Indegree}
        \end{subfigure}
        \hfill
        \begin{subfigure}{.49\textwidth}
                \centering
                \includegraphics[width=0.99\textwidth]{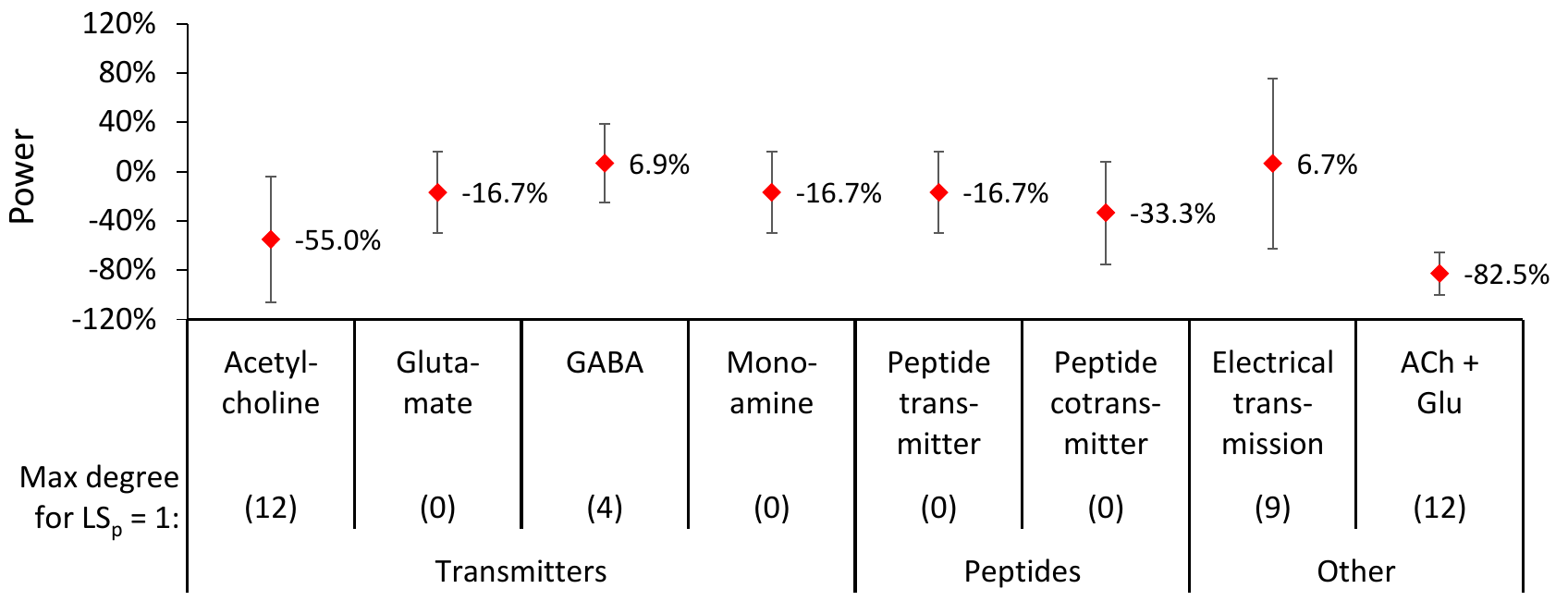}
                \caption{Outdegree}
        \end{subfigure}
	\caption[Power values for the second layer motor neurons.]{\keyword{Power values for the second layer motor neurons.}}
    \label{fig:Power_MO2}
	\end{minipage}
\end{figure*}

\keyword{Interneuron connectivity} - The results for considering the outdegree of the interneurons are depicted in \Cref{fig:Power_oudegree_IN}. For electrical connections, the power value of $75\%$ is the same as in the indegree analysis (\Cref{fig:Power_IN_a}) since they are bidirectional. Only ACh exhibits a higher power with $83.3\%$. The combined oudegree of both transmission types increases the power by $11.5\%$ to $94.8\%$ which is not far from a perfect model. For the other transmission types, no statement can be made due to low values or high uncertainty of the power.

\keyword{First layer motor neuron connectivity} - 
The outdegrees of the first layer motor neurons reveal one factor with excellent power of $-75\%$. The motor neurons of the locomotory subnetwork $LS_p = 1$ have no outgoing connections with peptide cotransmission. In combination with GABA transmission, the power can be increased to absolut $79.9\%$ (see \Cref{fig:Power_outdegree_MO1}).

\keyword{Qualitative view on the first layer motor neurons} - Instead of focusing on the connectivity of neurons in the network, it is also possible to target just the presence of specific transmitters. For example, the presence of certain transmitters among neurons can be binary coded. In the simplest case, $1$ stands for yes and $0$ for no.
This was done for ACh along with peptides of the families NLP and PDF. The results are plotted in \Cref{fig:Power_qual_MO1}.
The factor "PDF peptide or not" has already a good single power of $-62.5\%$. The neurons of the locomotory subnetwork do not have PDF peptides what can be seen in the distribution. The same is true for NLP peptides. 
If NLPs and PDFs are subtracted from ACh, the overlap of locomotory subnetwork neurons $LS_p = 1$ and other neurons $LS_p = 0$ can be minimized resulting in a power of $81.9\%$.  
This power value lies between the absolute values of the respective best factor from the indegree and outdegree analysis with $87.5\%$ and $79.9\%$ (see \Cref{fig:Power_IN_b,fig:Power_outdegree_MO1}).
Note that no qualitative consideration has been made for the interneurons since they can already be perfectly distinguished by their connectivity. However, this could also work fine for them.

\keyword{Second layer motor neuron connectivity} - Last, the second layer motor neurons are investigated (see exemplary \Cref{fig:shortest_paths}). 
It is assumed that the interneurons and the first layer motor neurons of the locomotory subnetwork are known. Only the motor neurons are considered which are postsynaptic to these of the first layer. In total, there are $66$ motor neurons of which $60$ belong to the locomotory subnetwork. 
New in this group are the motor neurons of the classes DD (6) and VD (13) with the exception of DD01, VD03, VD10, and VD13. In addition, many motor neurons of the first layer are also present here what is not reflected in \Cref{fig:shortest_paths} since it is only for illustrative purposes.
For example, motor neurons of the class VB (11) are all directly connected to muscles, but they have also postsynaptic partners in their class. The postsynaptic partners occur therefore also in the third group of neurons (second layer of motor neurons) in the shortest paths. 
Figure \ref{fig:Power_MO2} illustrates the resulting power values for indegree and outdegree distributions. For the indegree, most of the second layer motor neurons can be separated well because they do not have incoming connections with peptide transmitters. The absolute power value of $66.7\%$ is $20.8\%$ lower than for the first layer (see \Cref{fig:Power_IN_b}). On the other hand, Glu is a suitable factor for logistic regression with $-69.4\%$ power. If both factors are combined, the power rises to absolute $75.6\%$.
For the outdegree, no clear conclusion can be drawn at first glance for the individual factors due to high uncertainties. If the outdegree of ACh and Glu is combined, a power of $-82.5\%$ can be achieved which indicates a very good seperation result. 

\subsection{Prediction accuracy of logistic regression models with all sensory neurons}
\label{subsec:Prediction accuracy of logistic regression models with all sensory neurons}

The prediction accuracy of the logistic regression models on the test dataset with all sensory neurons is detailed in \Cref{tab:performance_test}.
In total, model 1 (2) predicts $99.2\%$ ($90.1\%$) of the shortest path correctly, and none of the paths through the locomotory subnetwork $y_i=1$ are misclassified. For paths that do not completely pass through the locomotory subnetwork $y_i=0$, there is a misclassification of $1.5\%$ ($18.5\%$). The latter value is significantly higher because model 2 does not take into account the outdegree of the interneurons.
However, this is less of a big problem since it can uncover additional potential neurons for the locomotory circuitry.

\subsection{Predictions with all sensory neurons}
\label{subsec:Predictions with all sensory neurons}

Using the predicted shortest paths $y_i=1$, model 1 (2) indicates 8 (29) further locomotory subnetwork neurons which are listed in \Cref{tab:Additional_locomotory_subnet_neurons}.
For all of these neurons except AVL, RMFL, and V03, it can be assumed on the basis of literature research and connectivity analysis that they affect the locomotion behavior of \elegans\ which is discussed in the following.

The interneurons are considered first. The neuron class AVE together with the classes AVA and AVD initiate the backward locomotion of \elegans\ (\cite[Section I]{Driscoll.1997}, \cite{Zhen.2015}). 
The neuron class AIB plays a role in turns. The classes AIB and AIZ promote turns whereas the classes AIY and AIA inhibit turns \cite{Garrity.2010}. The RIG neurons are involved in reversal behavior, and they are likely responsible for increased reversal rates \cite{Chao.2005}.
The AVJ and AVF neurons are suggested to function generally as modulators of the command interneurons that promote forward movement. Therefore, they are involved in the coordination of locomotion. Especially in periods of active egg-laying, they coordinate egg-laying and locomotion \cite{Hardaker.2001,Schafer.2005}.  

Next, the motor neurons are considered.
The AVL neuron is involved in the defecation. It controls together with the interneuron DVB the expulsion muscle contraction in the defecation motor program \cite[Section III]{Avery.1997}. Since this neuron has no apparent influence on the locomotion behavior of the worm, it is not considered to be part of the locomotory subnetwork.
The function of the PDA neuron is currently unknown, but it could be a counterpart to the PDB neuron. The latter generally ensures that ventral turns are preferred by the worm \cite{Yan.2017}. The PDA neuron is connected to the dorsal muscle MDL21 which lies opposite to the muscles that are connected to the PDB neuron. 
The PDB is not only connected with one but with two ventral muscles which could be the reason why ventral turns are preferred.
The RID neuron modulates the motor state of \elegans\ to maintain forward locomotion \cite{Lim.2016}.
The function of the PVN neurons in the locomotion of the worm is currently not investigated. These are presynaptic to command interneurons of the classes AVA, AVB, AVD, AVE, PVC, and to coordinating neurons of the classes AVJ and AVF, as well as to some motor neurons of the classes VD and DD. Therefore, they could contribute to the coordination of the locomotion. For the neurons of the DD and VD class, it is well established that they coordinate forward and backward movements \cite[Section I]{Driscoll.1997}. 
The RIM neurons play a role in the backward locomotion. For example, photostimulation of RIM causes the worm to reverse \cite{Guo.2009}. As a reaction to an anterior touch, head movements are suppressed and the backward movement is maintained \cite{Pirri.2012}. On the other hand, RIM is suggested to inhibit the initiation of reversals during locomotion \cite{Piggott.2011}. 
The RIV neurons are involved in the local foraging behavior consisting of reversals and deep omega-shaped turns. The ablation of the neurons reduces the frequency of omega turns \cite{Gray.2005}.
The neurons of the classes RME, SMD, and RMD are connected with head and neck muscles.  
The SMD and RMD motor neurons drive dorsoventral undulations \cite{Izquierdo.2018} and are needed for multiple navigation behaviors \cite{Gray.2005}, such as the local foraging behavior.
The RME neurons are linked with the head bending amplitude during forward locomotion. The undulatory activity of RME is regulated by SMD neurons via extrasynaptic neurotransmission which ensures optimal efficiency of forward locomotion \cite{Shen.2016}. The extrasynaptic neuromodulation between RIM and RME neurons allows deep head bending during omega turns and plays a role in the escape behavior of \elegans\ \cite{KagawaNagamura.2018}. 
Regarding the RMF neurons, no conclusions can be drawn without a detailed analysis. Therefore, RMFL is not counted to the locomotory subnetwork.
The role of the VC motor neurons in locomotion is also unclear. VC04 and VC05 innervate the vulva muscles and are involved in egg-laying \cite{Gjorgjieva.2014}. The first three VC neurons innervate motor neurons of the VD and DD class and could therefore contribute to the coordination of locomotion. This is not contradictory to the fact that the release of acetylcholine from these VC neurons inhibits egg-laying behavior. The neuron VC06 has no connections. It is hard to say whether the neurons VC01-VC03 are part of the locomotory subnetwork or belong to the egg-laying program. Therefore, VC03 is not counted.

In summary, it can be assumed that probably $26$ of the neurons in \Cref{tab:Additional_locomotory_subnet_neurons} are involved in locomotion behavior of \elegans\ except AVL, RMFL, and V03.
Model 1 only predicts $5$ of them because it only allows little margin regarding the interneurons. Model 1 includes the indegree and outdegree of the interneurons and model 2 only the indegree.

\begin{figure*}[htbp] 
\centering
\begin{minipage}{.98\textwidth}
\centering
\captionof{table}[Model performance on the test dataset (all sensory neurons).]{\keyword{Model performance on the test dataset (all sensory neurons).}}
  \small
\begin{tabular}{rrrcrrc}
\toprule
\multicolumn{1}{l}{\multirow{3}[2]{*}{\textbf{\specialcell{Locomotory\\subnetwork}}}} & \multicolumn{3}{c}{\textbf{Prediction Model 1}} & \multicolumn{3}{c}{\textbf{Prediction Model 2}} \\
    & \multicolumn{1}{c}{\multirow{2}[1]{*}{\textbf{correct}}} & \multicolumn{1}{c}{\multirow{2}[1]{*}{\textbf{incorrect}}} & \textbf{correct} & \multicolumn{1}{c}{\multirow{2}[1]{*}{\textbf{correct}}} & \multicolumn{1}{c}{\multirow{2}[1]{*}{\textbf{incorrect}}} & \textbf{correct} \\
    &     &     & \textbf{in \%} &     &     & \textbf{in \%} \\
\midrule
\multicolumn{1}{l}{yes} & 72,912 & 0   & 100.0 & 72,912 & 0   & 100.0 \\
\multicolumn{1}{l}{no} & 82,730 & 1,246 & 98.5 & 68,407 & 15,569 & 81.5 \\
\midrule
\textbf{Total} & \textbf{155,642} & \textbf{1,246} & \textbf{99.2} & \textbf{141,319} & \textbf{15,569} & \textbf{90.1} \\
\bottomrule
\end{tabular}%
  \label{tab:performance_test}%
\end{minipage}%
\newline
\begin{minipage}{.98\textwidth}
  \vspace{+5mm}
  \centering
  \captionof{table}[Additional predicted neurons for the locomotory subnetwork of \elegans.]{\keyword{Additional predicted neurons for the locomotory subnetwork of \elegans.} The assignment of the neuron type is arbitrary. The neurons are indicated as motor neurons if there are direct connections to muscle cells. Many of the listed neurons are multifunctional. The involvement of the neurons AVL, RMFL, and V03 in locomotion behavior is either unknown or cannot be clearly indicated.}
  \small
\begin{tabular}{lllcl}
\toprule
\multicolumn{1}{c}{\multirow{2}[2]{*}{\textbf{Neuron Type}}} & \multicolumn{2}{c}{\textbf{Model 3 additionally predicts}} & \textbf{Same for} & \multicolumn{1}{c}{\textbf{Suggested}} \\
    & \multicolumn{1}{c}{\textbf{neuron class}} & \multicolumn{1}{c}{\textbf{which neurons}} & \textbf{Model 1} & \multicolumn{1}{c}{\textbf{function}} \\
\midrule
\multirow{7}[2]{*}{Interneuron} & \multirow{2}[1]{*}{AVE (2)} &     &     & - Drive backward \\
    &     &     &     & \hspace{0.4em} movement \\
    & AIB (2) &     &     & - Promoting turns \\
    & \multirow{2}[0]{*}{RIG (1/2)} & \multirow{2}[0]{*}{RIGL} &     & - Involved in \\
    &     &     &     & \hspace{0.4em} reversal behavior \\
    & \multirow{2}[1]{*}{AVJ (1/2)} & \multirow{2}[1]{*}{AVJL} &     & - Coordination \\
    &     &     &     & \hspace{0.4em} of  movement \\
\midrule
\multirow{18}[2]{*}{Motor neuron} & \multirow{2}[1]{*}{AVF (2)} &     &     & - Coordination \\
    &     &     &     & \hspace{0.4em} of  movement \\
    & \multirow{2}[0]{*}{AVL (1)} &     & \multirow{2}[0]{*}{yes} & - Involved in \\
    &     &     &     & \hspace{0.4em} defecation process \\
    & PDA (1) &     & yes & - Promoting turns \\
    & \multirow{2}[0]{*}{RID (1)} &     & \multirow{2}[0]{*}{yes} & - Maintain forward \\
    &     &     &     & \hspace{0.4em} movement \\
    & \multirow{2}[0]{*}{PVN (1/2)} & \multirow{2}[0]{*}{PVNR} & \multirow{2}[0]{*}{yes} & - Coordination \\
    &     &     &     & \hspace{0.4em} of  movement \\
    & \multirow{2}[0]{*}{RIM (1/2)} & \multirow{2}[0]{*}{RIML} & \multirow{2}[0]{*}{yes} & - Maintain backward \\
    &     &     &     & \hspace{0.4em} movement \\
    & RIV (2) &     &     & - Promoting turns \\
    & RMD (6) &     &     & - Promoting turns \\
    & RME (2/4) & RMEL, RMEV &     & - Promoting turns \\
    & RMF (1/2) & RMFL & yes &  \\
    & SMD (4) &     & only SMDVR & - Promoting turns \\
    & \multirow{2}[1]{*}{VC (1/6)} & \multirow{2}[1]{*}{VC03} & \multirow{2}[1]{*}{yes} & - Inhibition of \\
    &     &     &     & \hspace{0.4em} egg-laying \\
\bottomrule
\end{tabular}%
  \label{tab:Additional_locomotory_subnet_neurons}%
\end{minipage}%
\end{figure*}%


\section{Supplementary information for simulation of dynamics}

\subsection{Estimated parameters for time-delayed feedback control and harmonic wave model}
\label{subsec:Additional parameters for time-delayed feedback control and harmonic wave model}

Time-delayed feedback control is applied in order to enhance the synchronicity of harmonic waves \eqref{eq:body_dynamics} which contributes to a coordinated locomotion of \elegans. The utilized parameters are given in \Cref{tab:time_delays,tab:feedback_strengths}. 
As a consequence, simulated extreme values in the time series of subtracted muscle activity $A^\text{ventral}_m(t) - A^\text{dorsal}_m(t)$ occur at different times (see exemplary \Cref{fig:feedback_control}).
For determining the initial phase in Eq.~\eqref{eq:body_dynamics}, the first observed extreme values after a certain transient time are required. For these, the time of occurrence is shown in \Cref{tab:extracted_extrema}. 

\begin{figure*}[htbp] 
\centering
\begin{minipage}{.98\textwidth}
\centering
  \captionof{table}[Time delays for motor neurons.]{\keyword{Time delays for motor neurons.}}
    \label{tab:time_delays}%
  \small
\begin{tabular}{rl}
\toprule
\multicolumn{1}{l}{\textbf{Time}} & \multicolumn{1}{c}{\multirow{2}[2]{*}{\textbf{Motor neurons}}} \\
\multicolumn{1}{l}{\textbf{delay (s)}} &  \\
\midrule
0.75 & AS11 \\
1.25 & AS05, AS06, DA04, DA05, DB03, DD03, VA05, VA06, VB03, VB04, VB05, VD05 \\
1.5 & AS02, DB01, VA01, VA02 \\
2   & VA10, VA12, VB08, VB11, VD09, VD13 \\
2.25 & AS08, DD04, VA07, VA08, VA09, VB06, VB07, VD07, VD08 \\
3   & DA09, DB07, DD06 \\
0   & Others \\
\bottomrule
\end{tabular}%
\end{minipage}%
\newline
\begin{minipage}{.98\textwidth}
\vspace{+5mm}
\centering
  \captionof{table}[Feedback strengths for muscle cells.]{\keyword{Feedback strengths for muscle cells.}}
    \label{tab:feedback_strengths}%
  \small
\begin{tabular}{rl}
\toprule
\multicolumn{1}{l}{\textbf{Feedback}} & \multicolumn{1}{c}{\multirow{2}[2]{*}{\textbf{Muscle cells}}} \\
\multicolumn{1}{l}{\textbf{strength}} &  \\
\midrule
0.25 & MDL11, MDR11, MVL11, MVR11 \\
0.5 & MDL12, MDR12, MVL12, MVR12 \\
\multirow{2}[0]{*}{0.75} & MDL07, MDL15, MDL16, MDR07, MDR15, MDR16, MVL07, MVL15, MVL16,  \\
    & MVR07, MVR15, MVR16 \\
1.5 & MDL18, MDR18, MVL18, MVR18 \\
1.75 & MDL13, MDL14, MDR13, MDR14, MVL13, MVL14, MVR13, MVR14 \\
2   & MDL19, MDR19, MVL19, MVR19 \\
\multirow{2}[0]{*}{2.75} & MDL21, MDL22, MDL23, MDL24, MDR21, MDR22, MDR23, MDR24, MVL21,  \\
    & MVL22, MVL23, MVR21, MVR22, MVR23, MVR24 \\
0   & Others \\
\bottomrule
\end{tabular}%
\end{minipage}%
\newline
\begin{minipage}{.66\textwidth}
  \vspace{+5mm}
  \centering
  \captionof{table}[Determining initial phase for harmonic waves.]{\keyword{Determining initial phase for harmonic waves.} The table provides the times of occurrence for first exploited extreme values in the time series of subtracted muscle activity $A^\text{ventral}_m(t) - A^\text{dorsal}_m(t)$. The calculation of the initial phase is then based on the opposite type of extreme value assumed for $t = 0$ and can be consequently calculated with the linear functions $\phi_{min}(m) = -0.3489m - 1.5753$ for a minimum and $\phi_{max}(m) = -0.3495m + 1.5774$ for a maximum. The given extremes are the first in the time series whose distance to the following extreme value of equal type is greater than 1.5 time units.}
    \label{tab:extracted_extrema}%
  \small
\begin{tabular}{ccccc}
\toprule
\multirow{2}[2]{*}{\textbf{m}} & \multicolumn{2}{c}{\textbf{Uncontrolled}} & \multicolumn{2}{c}{\textbf{With feedback control}} \\
    & \textbf{Extremum} & \textbf{Time} & \textbf{Extremum} & \textbf{Time} \\
\midrule
5   & min & 0.790 & min & 0.790 \\
7   & min & 0.445 & min & 1.865 \\
8   & max & 0.100 & max & 0.100 \\
9   & max & 0.275 & max & 0.275 \\
10  & min & 0.715 & min & 0.715 \\
11  & min & 0.620 & min & 0.615 \\
12  & min & 0.570 & max & 2.370 \\
13  & min & 0.795 & min & 0.625 \\
14  & min & 0.570 & min & 0.180 \\
15  & max & 0.310 & max & 0.205 \\
16  & max & 2.340 & min & 0.665 \\
17  & max & 0.155 & max & 0.155 \\
18  & max & 0.445 & max & 0.415 \\
19  & max & 0.205 & max & 2.565 \\
20  & max & 0.420 & max & 0.420 \\
21  & min & 0.840 & max & 0.805 \\
22  & min & 0.885 & max & 0.775 \\
23  & min & 0.840 & max & 0.805 \\
24  & min & 0.840 & max & 0.805 \\
\bottomrule
\end{tabular}%
\end{minipage}%
\end{figure*}%

\subsection{Complementary results for silencing of neuronal activity}
\label{subsec:Complementary results for silencing of neuronal activity}

\begin{figure*}[htbp] 
 \centering
 \begin{minipage}{.98\textwidth}
 	\centering
 	  	\begin{subfigure}{0.49\textwidth}
 	        \centering
 	        \includegraphics[width=1.0\linewidth]{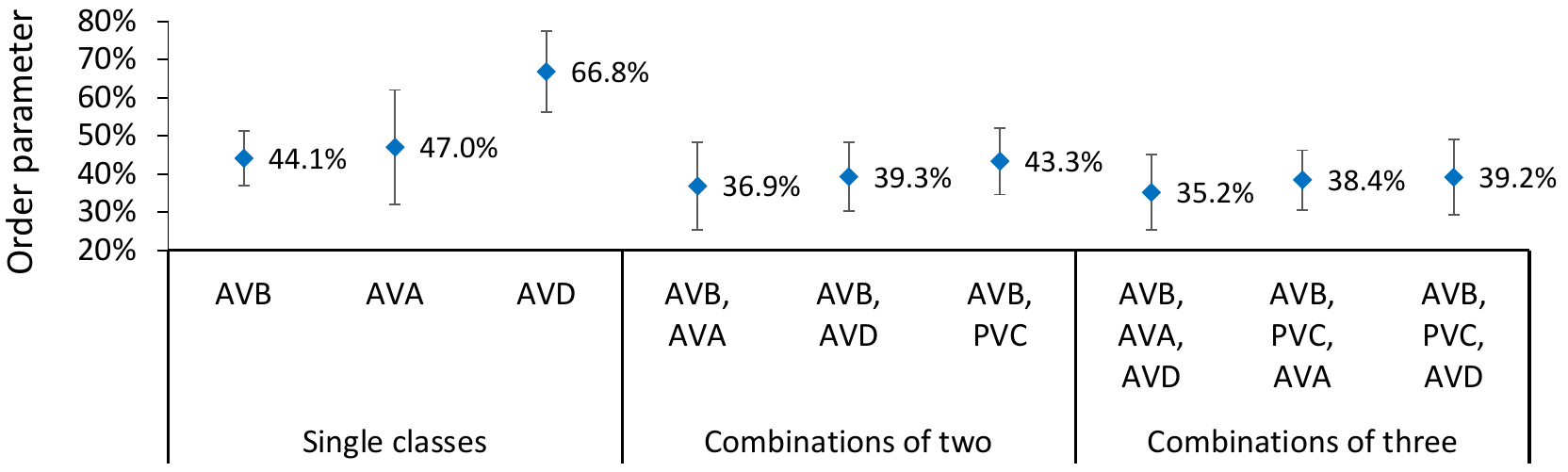}	  
 	    \caption{Silencing activity of interneuron classes}
 	    \end{subfigure}
 	    \begin{subfigure}{0.49\textwidth}
 	        \centering
 	        \includegraphics[width=1.0\linewidth]{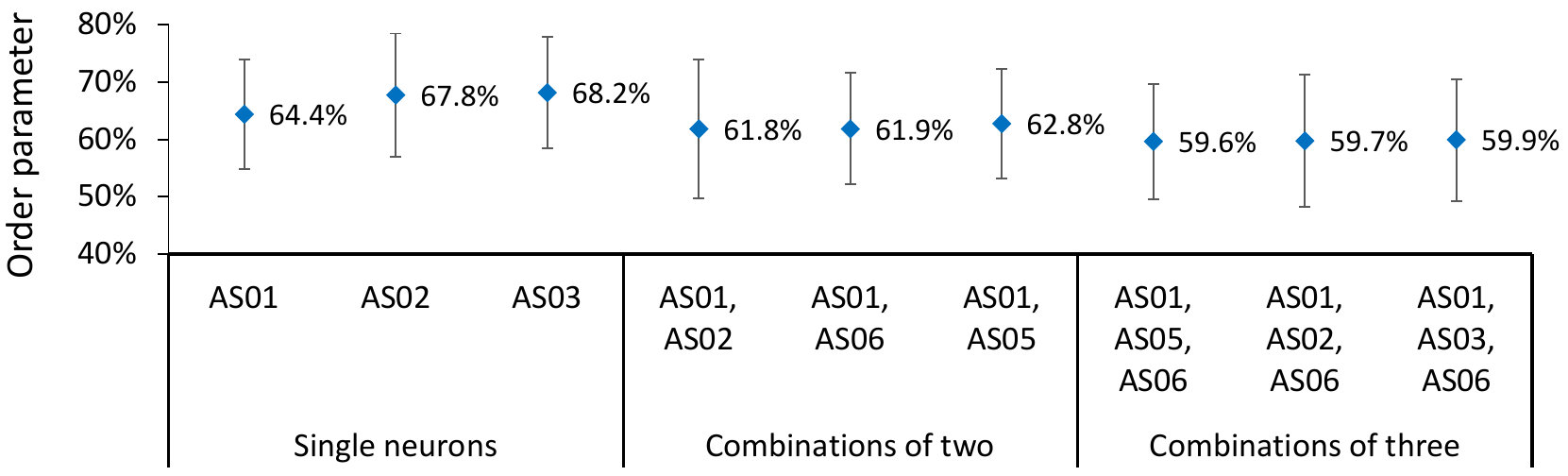}
 	        \caption{Silencing activity of AS neurons}
 	    \end{subfigure}
 	    \caption[Most significant interneuron classes and AS motor neurons for synchronicity of muscular waves.]{\keyword{Most significant interneuron classes and AS motor neurons for synchronicity of muscular waves.} Without silencing of neuronal activity, the reference value of the time-averaged order parameter is $70.1\%$.}
         \label{fig:Deletion_IN_AS}
 	\end{minipage}
 \begin{minipage}{.98\textwidth}
 \vspace{+3mm}
 	\centering
     \begin{subfigure}{0.49\textwidth}
 	    \centering
 	    \includegraphics[width=1.0\linewidth]{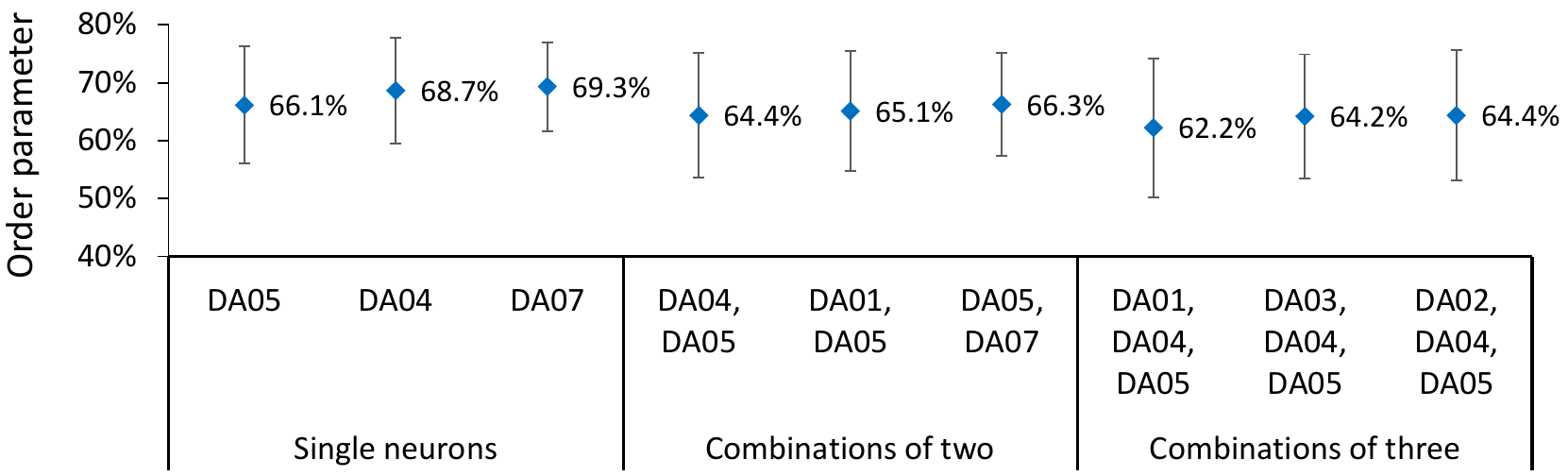}    
 	    \caption{Silencing activity of DA neurons}
 	\end{subfigure}
 	\begin{subfigure}{0.49\textwidth}
 	    \centering
 	    \includegraphics[width=1.0\linewidth]{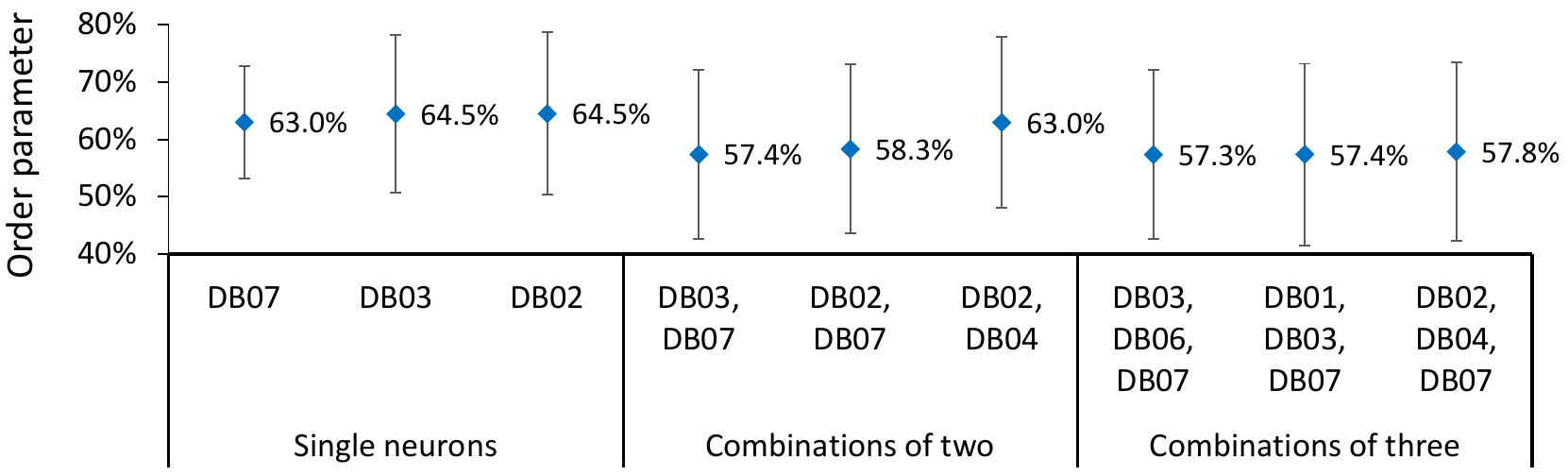}  
 	    \caption{Silencing activity of DB neurons}
 	\end{subfigure}
 	\caption[Most significant DA and DB motor neurons for synchronicity of muscular waves.]{\keyword{Most significant DA and DB motor neurons for synchronicity of muscular waves.} Without silencing of neuronal activity, the reference value of the time-averaged order parameter is $70.1\%$.}
 	\label{fig:Deletion_DA_DB}
 	\end{minipage}
 \begin{minipage}{.98\textwidth}
 \vspace{+3mm}
 	\centering
 	  	\begin{subfigure}{0.49\textwidth}
 	  \centering
 	  \includegraphics[width=1.0\linewidth]{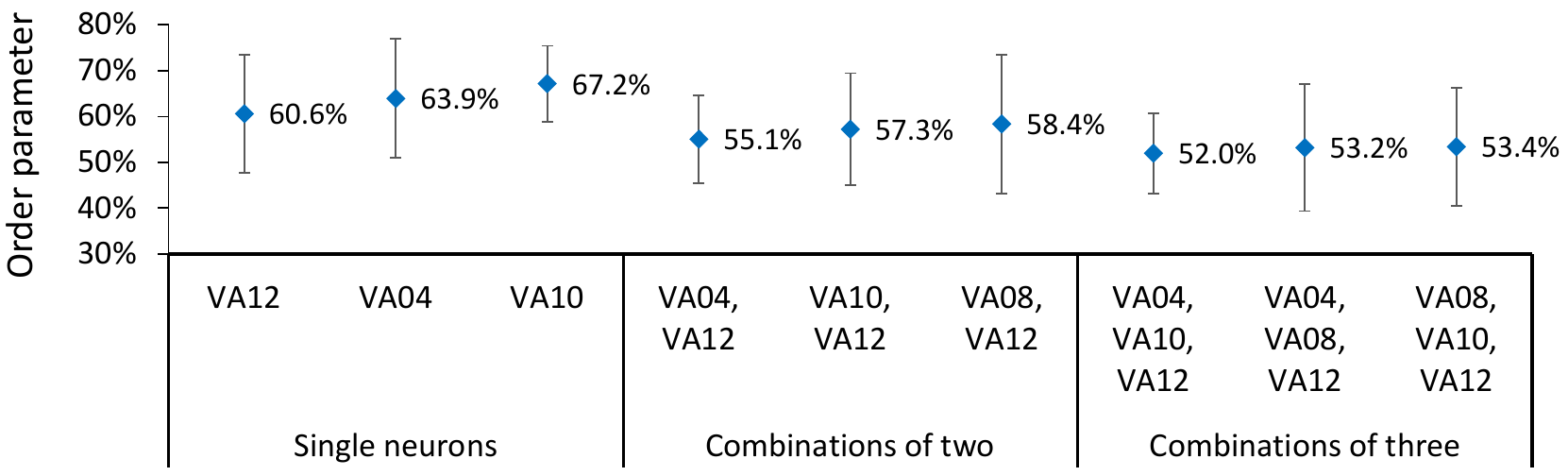}    
 	  \caption{Silencing activity of VA neurons}
 	\end{subfigure}
 	\begin{subfigure}{0.49\textwidth}
 	  \centering
 	  \includegraphics[width=1.0\linewidth]{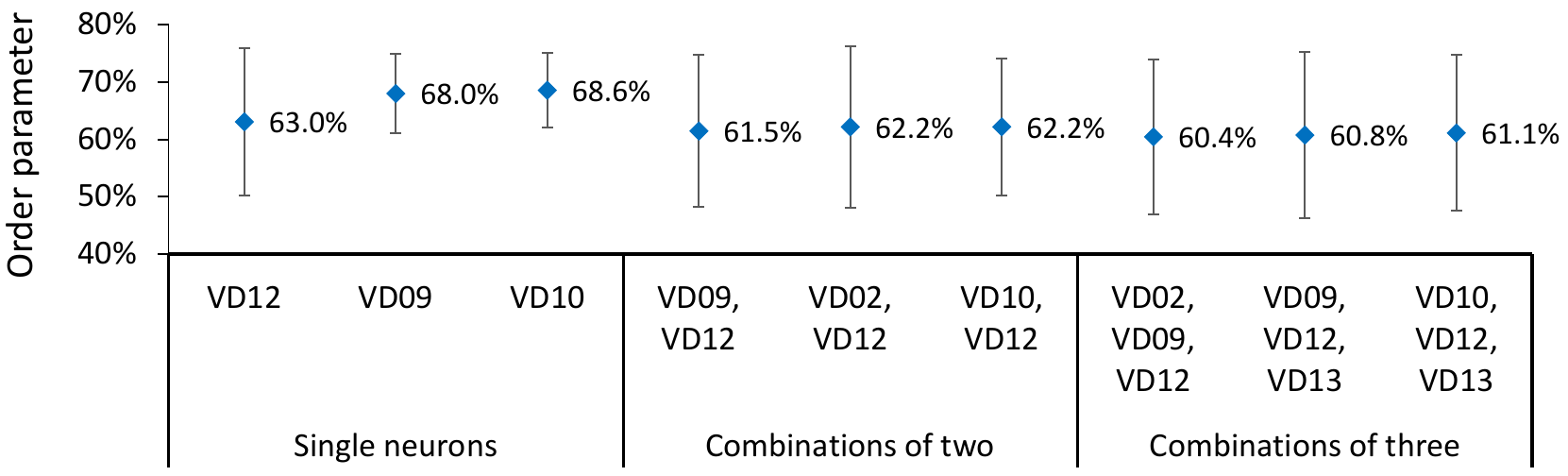} 
 	  \caption{Silencing activity of VD neurons}
 	\end{subfigure}
 	\caption[Most significant VA and VD motor neurons for synchronicity of muscular waves.]{\keyword{Most significant VA and VD motor neurons for synchronicity of muscular waves.} Without silencing of neuronal activity, the reference value of the time-averaged order parameter is $70.1\%$.}
     \label{fig:Deletion_VA_VD}
 	\end{minipage}
 \end{figure*}

The results in \Cref{fig:Deletion_IN_AS,fig:Deletion_DA_DB,fig:Deletion_VA_VD} supplement the synchronicity analyses when specific neurons are silenced (see Section~\ref{sec:Synchronicity_in_absence_of_neurons}).
For the different motor neuron classes, neurons are silenced individually, in combinations of two, and combinations of three. For these, the top three results with the lowest time-averaged order parameter ± its standard deviation are considered.
The lower this number is, the more important are the underlying neurons for a coordinated locomotion of \elegans.

\end{document}